\documentclass[twocolumn]{aastex62}
\newcommand{\msun}{$\rm M_\odot$}

\newcommand\nuk[2]{$\rm ^{\rm #2} #1$}

\graphicspath{{./}{figures/}}

\received{..........}
\revised{..........}
\accepted{..........}

\submitjournal{ApJS}

\shorttitle{Rotating massive stars}

\shortauthors{Limongi,Chieffi} \shorttitle{The evolution of rotating massive stars at various metallicities}

\begin{document}

\title{Presupernova evolution and explosive nucleosynthesis of rotating massive stars \\ in the metallicity range -3 $\rm \le[Fe/H]\le$ 0}
\author[0000-0003-0636-7834]{Marco Limongi}
\affiliation{Istituto Nazionale di Astrofisica - Osservatorio Astronomico di Roma, Via Frascati 33, I-00040, Monteporzio Catone, Italy; marco.limongi@oa-roma.inaf.it}
\affiliation{Kavli Institute for the Physics and Mathematics of the Universe, Todai Institutes for Advanced Study, the University of Tokyo, Kashiwa, Japan 277-8583 (Kavli IPMU, WPI)}       
\author[0000-0002-3589-3203]{Alessandro Chieffi}
\affiliation{Istituto Nazionale di Astrofisica - Istituto di Astrofisica e Planetologia Spaziali, Via Fosso del Cavaliere 100, I-00133, Roma, Italy; alessandro.chieffi@inaf.it}
\affiliation{Monash Centre for Astrophysics (MoCA),
School of Mathematical Sciences, Monash University, Victoria 3800, Australia}

\correspondingauthor{Marco Limongi}
\email{marco.limongi@oa-roma.inaf.it}

\begin{abstract} 
We present a new grid of presupernova models of massive stars extending in mass between 13 and 120\ \msun, covering four metallicities (i.e. [Fe/H]=0, -1, -2 and -3) and three initial rotation velocities (i.e. 0, 150 and 300 km/s). The explosion has been simulated following three different assumptions in order to show how the yields depend on the remnant mass - initial mass relation. An extended network from H to Bi is fully coupled to the physical evolution of the models. The main results can be summarized as follows.

a) At solar metallicity the maximum mass exploding as Red Super Giant (RSG) is of the order of 17\ \msun\ in the non rotating case, all the more massive stars exploding as WR stars. All rotating models, vice versa, explode as Wolf-Rayet (WR) stars. 

b) The interplay between the core He burning and the H burning shell, triggered by the rotation induced instabilities, drives the synthesis of a large primary amount of all the products of the CNO, not just \nuk{N}{14}. A fraction of them enriches enormously the radiative part of the He core (and is responsible of the large production of F) and a fraction enters the convective core leading therefore to an important primary neutron flux able to synthesize heavy nuclei up to Pb.
 
c) In our scenario, remnant masses of the order of those inferred by the first detections of the gravitational waves (GW150914, GW151226, GW170104, GW170814) are predicted at all metallicities for none or moderate initial rotation velocities. 
\end{abstract} 


\section{Introduction}

Massive stars play a pivotal role in the evolution of the Galaxies because of their influence on the environment: they contribute to the chemical enrichment of the gas clouds, eject an enormous amount of energy either as neutrinos {\color{black} or} kinetic energy, are the protagonists of some of the more spectacular explosions we do see in the sky, leave compact remnants that extend in mass from the neutron stars to black holes and are therefore also intimately linked to the spectacular detections of the collapse of compact remnants in binary systems by the LIGO-VIRGO collaboration. A proper understanding of the many evolutionary properties of these stars, including their distribution in the HR diagram, the relative numbers of stars in the various WR stages, the physical and chemical structure of the mantle at the onset of the collapse (responsible {\color{black} for} the different kinds of core collapse supernovae), the changes of the surface chemical composition during their evolution and, last but not least, the properties of the explosion including the explosive yields and the mass distribution of the remnants, demands the build up of an extended and homogeneous grid of models which may be used to study the contribution of these stars to the evolution of the galaxies since the formation of the first stars as well as to understand a large variety of objects we observe in the sky. Though it exists a huge literature addressing one or another aspect of the evolution of the {\color{black} massive stars \citep{hlw00,mm03,hws05,mm05,brott11,lc12,mm12,ekstroemetal12,cl13}}, an homogeneous and extended set (in mass, metallicity, initial rotation velocity and number of nuclear species followed) is still missing.

In this paper we present for the first time all the relevant properties of a wide set of rotating stellar models: we cover three different initial rotational velocities, namely, v=0, 150 and 300 km/s, and 4 different initial metallicities, namely, [Fe/H]=0, -1, -2 , -3.  The physical and chemical evolution of these models is fully coupled together and the number of nuclear species followed explicitly amounts to 338, from H to Bi. All the models were computed with the latest version of our stellar evolution code (FRANEC), improved with respect to the version used in \citet{cl13} {\color{black} (CL13 hereinafter)} in order to (1) refine the treatment of the angular momentum transport in the envelope of the star; (2) take into account the dynamical mass loss caused by the approach of the luminosity of the star to the Eddington limit; (3) refine the computation of the angular momentum loss due to the stellar wind and (4) increase the size of the adopted nuclear network. Compared to our previous study, we also adopted a different approach to calibrate the rotation induced mixing efficiency, that takes advantage now of the observations of the surface chemical composition of many B-stars in the LMC samples of the FLAMES survey \citep{untetal09}. The paper is organized as follows: the latest version of the FRANEC code is presented in Section \ref{code} while the calibration of the rotational mixing efficiency is discussed in Section \ref{calimix}; Section \ref{evol} is devoted to the presupernova evolution of all the models while the explosive yields are presented in Section \ref{yields}. A comparison with models computed by other authors is shown in Section \ref{compa} while the remnant mass - initial mass relation is discussed in Section \ref{inifi}. A final summary and conclusions follows.

\section{The models and the stellar evolution code}\label{code}

The results presented in this paper are based on a grid of models with initial masses 13, 15, 20, 25, 30, 40, 60, 80 and 120 $\rm M_\odot$, initial metallicities [Fe/H]=0, -1, -2, -3 and initial equatorial velocity v=0, 150, 300 km/s (at the beginning of the MS phase). 
The evolution of all these models has been {\color{black} followed} from the pre Main Sequence phase up to the presupernova stage, {\color{black} that is, when the integration of the equations does not converge anymore. Note that the central temperature of all the models at this stage is well above $\rm \sim 6\cdot 10^{9}~K$.
The evolutions have been computed by means of the latest version of the FRANEC code. }
The main features of this code, as well as all the input physics and assumptions, have been already extensively discussed in CL13 and {\color{black} are summarized in Appendix \ref{app:franec} for sake of completeness}. The improvements with respect to the version described in CL13 are the following: (1) a better treatment of the angular momentum transport in the envelope of the star; (2) the inclusion of the mass loss triggered by the approach to the Eddington limit; (3) a proper computation of the angular momentum loss due to the stellar wind and (4) the increase of the size of the adopted nuclear network. 

\textcolor{black}{(1) In the FRANEC the star is divided in two zones: the atmosphere and the inner region. In the atmosphere the luminosity is assumed to be constant so that only three equations, instead of four, are solved. In presence of rotation the transport of the angular momentum is ignored in the atmosphere and it is assumed that it rotates as a solid body together to the external border of the inner zone.  The mass fraction that we traditionally include in the atmosphere is fixed at 1\% of the current total mass of the star, and we adopted this value also in CL13. In order to increase the fraction of mass in which the angular momentum is properly transported we pushed forward the base of the envelope so that in this new grid of models only 1 per ten thousand of the mass is included in the atmosphere, i.e. 99.99 \% of the mass now being included in the inner zone. Such a choice cannot be maintained for the whole evolution because the dramatic increase  of the radius when the star turns redward would imply a prohibitive increase in the number of meshpoints and timesteps, so we adopted such a refined choice until a star is in central H burning. Beyond that, the border of the inner zone automatically shifts slowly down in mass until  it reaches 99\% of the current mass. Such a choice is partially justified by the fact that after the core H exhaustion the star quickly evolves toward the red supergiant (RSG) stage and therefore the surface rotation velocity reduces dramatically. (2) During the redward excursion in the HR diagram occurring after core H depletion, the radiative luminosity $\rm L=(16 \pi a c/3)(G m T^4)/(k P) \nabla$ (where $\nabla$ is the local effective temperature gradient, defined as $\rm \nabla=dLog~P/dLog~T$) may approach, and even overcome, the Eddington luminosity $\rm L=4 \pi c G m/k$. When this happens, all the zones above the region exceeding this limit become essentially unbound and one would expect a strong episode of mass loss. In order to treat such a phenomenon, we remove all the unbound zones with a maximum limit of $3\cdot 10^{-3}~\rm M_\odot$ lost per model. (3) When a star loses mass it also loses a certain amount of angular momentum. The determination of this amount is not trivial, it is subject to large uncertainties and is somewhat arbitrary.  In CL13 we arbitrarily made a {\it minimalist} choice in the sense that the amount of angular momentum removed per timestep was simply the one contained in the mass removed by the star as a consequence of the mass loss. In the present calculations, on the contrary, we compute explicitly the angular momentum loss $\rm J$ ($\rm =\dot{J} dt =j_{\rm surf} \dot{M} dt$, where $\rm j_{\rm surf}$ is the specific angular momentum at the surface and $\rm \dot{M}$ and dt the mass loss rate and the timestep, respectively) and remove such an  amount from the outer region of the star by requiring that no more than few per cent of the angular momentum may be removed from each layer.}
(4) The nuclear network adopted in CL13 included 293 isotopes, from H to \nuk{Mo}{98} coupled by all possibile links among them due to weak and strong interactions, for a total of about 3000 reactions in the various nuclear burning stages. Since one of the main issue related to the nucleosynthesis in rotating massive stars of low metallicity is the production of the s-process elements (see below), we extend our nuclear network in order to include as many elements as possibile between \nuk{Mo}{98} and \nuk{Bi}{209}. In order to save computer memory and computational time, we make the following assumption. Since in the neutron capture chain the slowest reactions are the ones involving magic nuclei, we explicitly follow, and includes into the nuclear network, all the stable and unstable isotopes around the magic numbers corresponding to N=82 and N=126, consider for these isotopes only neutron captures and beta decays, and assume all the other intermediate isotopes at the local equilibrium. In this way we are able to follow in detail the flux of neutrons through all the magic number bottlenecks. As a consequence the nuclear network adopted in the present calculations includes 335 isotopes (from neutrons to \nuk{Bi}{209}) and is reported in Table \ref{tabnetwork}

\begin{deluxetable}{lrrlrr}
\tablewidth{0pt}
\tablecaption{Nuclear network adopted in the present calculations\label{tabnetwork}}
\tablehead{
\colhead{Element} & \colhead{$\rm A_{\rm min}$} & \colhead{$\rm A_{\rm max}$} &
\colhead{Element} & \colhead{$\rm A_{\rm min}$} & \colhead{$\rm A_{\rm max}$}
}
\startdata
n........  &   1   &   1   & Co.......  &  54   &  61   \\  
H........  &   1   &   3   & Ni.......  &  56   &  65   \\  
He.......  &   3   &   4   & Cu.......  &  57   &  66   \\  
Li.......  &   6   &   7   & Zn.......  &  60   &  71   \\  
Be.......  &   7   &  10   & Ga.......  &  62   &  72   \\  
B........  &  10   &  11   & Ge.......  &  64   &  77   \\
C........  &  12   &  14   & As.......  &  71   &  77   \\
N........  &  13   &  16   & Se.......  &  74   &  83   \\
O........  &  15   &  19   & Br.......  &  75   &  83   \\
F........  &  17   &  20   & Kr.......  &  78   &  87   \\
Ne.......  &  20   &  23   & Rb.......  &  79   &  88   \\
Na.......  &  21   &  24   & Sr.......  &  84   &  91   \\
Mg.......  &  23   &  27   & Y........  &  85   &  91   \\
Al.......  &  25   &  28   & Zr.......  &  90   &  97   \\
Si.......  &  27   &  32   & Nb.......  &  91   &  97   \\
P........  &  29   &  34   & Mo.......  &  92   &  98   \\
S........  &  31   &  37   & Xe.......  & 132   & 135   \\
Cl.......  &  33   &  38   & Cs.......  & 133   & 138   \\
Ar.......  &  36   &  41   & Ba.......  & 134   & 139   \\
K........  &  37   &  42   & La.......  & 138   & 140   \\
Ca.......  &  40   &  49   & Ce.......  & 140   & 141   \\
Sc.......  &  41   &  49   & Pr.......  & 141   & 142   \\
Ti.......  &  44   &  51   & Nd.......  & 142   & 144   \\
V........  &  45   &  52   & Hg.......  & 202   & 205   \\
Cr.......  &  48   &  55   & Tl.......  & 203   & 206   \\
Mn.......  &  50   &  57   & Pb.......  & 204   & 209   \\
Fe.......  &  52   &  61   & Bi.......  & 208   & 209   \\
\enddata                                              
 \end{deluxetable}                                                                               

\begin{figure}
\epsscale{1.15}
\plotone{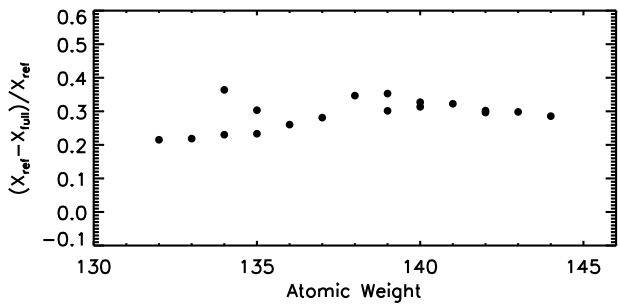}
\caption{Variation between the abundances obtained with our reference network (Table \ref{tabnetwork}), $\rm X_{ref}$, and the ones obtained with the larger network, $\rm X_{full}$, i.e., including in the reference network also the isotopes reported in Table \ref{tabnetcheck}}
\label{rationetwork}
\end{figure}

\begin{figure}
\epsscale{1.15}
\plotone{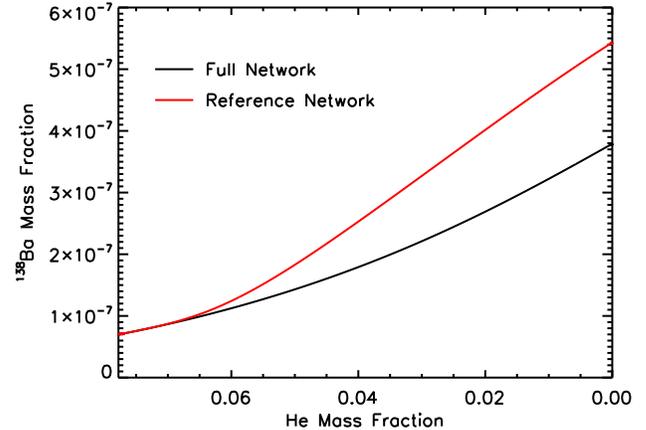}
\caption{Evolution of $\rm ^{138}Ba$ mass fraction as a function of $\rm ^{4}He$ mass fraction in the one-zone model calculation. The black line is obtained with the network in Table \ref{tabnetwork} while the red line is obtained with the network in Table \ref{tabnetcheck} (see text).}
\label{checknetwork}
\end{figure}

\textcolor{black}{In order to check the reliability of this assumption, we performed a one-zone calculation of a typical core He burning with two different nuclear networks, the one reported in Table \ref{tabnetwork} and another one in which we added also all the stable isotopes between \nuk{Mo}{98} and \nuk{Xe}{132} (Table \ref{tabnetcheck}). This test has been done by using the temporal evolution of the central temperature and density obtained for a $\rm 20~M_\odot$ star having [Fe/H]=-3 and initial equatorial velocity v=300 km/s. The continuous ingestion of fresh \nuk{N}{14} (driven by the rotation induced mixing, see below) that powers a steady production of primary neutrons is simulated by keeping constant the neutron mass fraction during the evolution: more specifically the neutron density is set equal to a mass fraction of $10^{-20}$ (.i.e. the value corresponding to the starting model, see below) at the beginning of each timestep . The computation starts when the \nuk{Ne}{22}($\alpha$,n) begins to be efficient, which means in this specific case when the central He mass fraction drops to $\sim 0.08$ and ends at the central He exhaustion. The result of this test shows clearly that at the end of the central He burning both run populate similarly the region $\rm 131<A<145$ (i.e. around the magic neutron closure shell n=82). Figure \ref{rationetwork} shows in fact  that final abundances of these nuclei vary between 20 and $40~\%$. Figure \ref{checknetwork} shows the temporal evolution of \nuk{Ba}{138} obtained in the two runs: the red line refers to the  reference network shown in Table \ref{tabnetwork} while the black line refers to the more extended network that includes also the elements reported in Table \ref{tabnetcheck}. Both these plots show that our approximation does not alter too substantially the flux of the matter through the neutron magic number bottlenecks because the increase of the abundances of the heavy elements induced by rotation is order of magnitude larger than the differences found above by comparing the two networks.}

\begin{deluxetable}{lrr}
\tablewidth{0pt}
\tablecaption{Isotopes included between $\rm ^{98}Mo$ and $\rm ^{132}Xe$ in 
the one-zone He burning calculation\label{tabnetcheck}}
\tablehead{
\colhead{Element} & \colhead{$\rm A_{\rm min}$} & \colhead{$\rm A_{\rm max}$}}
\startdata
Tc.......  &  99   &  99    \\  
Ru.......  & 100   & 102    \\  
Rh.......  & 103   & 103    \\  
Pd.......  & 104   & 104    \\  
Ag.......  & 109   & 109    \\  
Cd.......  & 110   & 114    \\
In.......  & 115   & 115    \\
Sn.......  & 116   & 120    \\
Sb.......  & 121   & 121    \\
Te.......  & 122   & 126    \\
I........  & 127   & 127    \\
\enddata
\end{deluxetable}                                                                               

The nuclear cross sections and the weak interaction rates have been updated whenever possible, with respect to the ones adopted in our previous version of the code. Most of them have been extracted from the STARLIB database \citep{starlib}. Tables \ref{tabreferences} and \ref{crosspecial} show the full reference matrix of all the processes taken into account in the network, together to its proper legend. As usual, for the weak interaction rates, $\beta^{+}$ and $\beta^{-}$ mean the sum of both the electron capture and the $\beta^{+}$ decay and the positron capture and the $\beta^{-}$ decay, respectively.

The initial composition adopted for the solar metallicity models is the one provided by \cite{agss09}, which corresponds to a total metallicity $\rm Z=1.345\cdot 10^{-2}$. For the models with initial metallicity lower than solar we assume the same scaled solar distribution for all the elements, with the exception of C, O, Mg, Si, S, Ar, Ca, and Ti for which we adopt an enhancement with respect to Fe derived from the observations of low metallicity stars, i.e., [C/Fe]=0.18, [O/Fe]=0.47, [Mg/Fe]=0.0.27, [Si/Fe]=0.37, [S/Fe]=0.35, [Ar/Fe]=0.35, [Ca/Fe]=0.33, [Ti/Fe]=0.23 \citep{cayreletal04,spiteetal05}.
As a result of these enhancements, the total metallicity corresponding to [Fe/H]=-1,-2,-3 is $\rm Z=3.236\cdot 10^{-3},~3.236\cdot 10^{-4},~3.236\cdot 10^{-5}$, respectively. The initial He abundances adopted at the various metallicities are: 0.265 ([Fe/H]=0), 0.25 ([Fe/H]=-1), 0.24 ([Fe/H]=-2 and [Fe/H]=-3). {\color{black} By the way, let us remind the reader that the abundance ratio [X/Y] is defined as $\rm [X/Y]=Log(X/Y)-Log(X/Y)_\odot$}.

\section{Calibration of the mixing efficiency}\label{calimix}

{\color{black} Since rotation is a multidimensional physical phenomenon, its inclusion in a 1D stellar evolution code implies a certain number of assumptions \citep[CL13]{mm00araa}. Therefore the calculation of the diffusion coefficients adopted to transport both the angular momentum and the chemical composition (see Appendix \ref{app:franec}) are intrinsically uncertain. For this reason the efficiency of the rotation induced mixing must be calibrated in some way.}
Different authors adopt different techniques to perform such a calibration, e.g. \cite{hlw00} require a surface enrichment of N of the order of 2-3 in solar metallicity models with initial mass in the range 10-20 $\rm M_{\odot}$, while  \cite{brott11}  try to reproduce the observed N abundance as a function of the projected rotation velocity (the Hunter diagram hereafter) in the LMC samples of the FLAMES survey \citep{untetal09}. In CL13 we computed only solar metallicity models and therefore we followed the same idea of \cite{hlw00}. In general, since we have two free parameters, namely $f_c$ and $f_\mu$ \citep[CL13]{hlw00}, and only one requirement, we cannot determine a single solution to this problem but only a family of possible choices. For this reason, in CL13, we chose a conservative approach by fixing $f_c=1.0$ and by calibrating $f_\mu$ in order to obtain an enhancement of the surface N abundance by a factor 2-3 in a $\rm 20~M_\odot$ star of solar metallicity at core H depletion. As a result of that calibration we found that the best choice of the two mixing parameters was $f_c=1.0$ and $f_\mu=0.03$. In this paper we present models of various metallicities and hence we decided to check whether the calibration obtained in CL13 is valid at subsolar metallicities. For this reason, we considered the LMC samples of the FLAMES survey that are centered toward the clusters NGC2004 and N11 \citep{untetal08}. The number of core H burning stars ($log~g\ge 3.2$), for which both the determination of the surface N abundance and the $\rm v\sin(i)$ are available, is $62$ and $30$ for NGC2004 and N11 respectively. Therefore, to obtain a sample which is as homogenous and populated as possible we decide to use only NGC2004. 

\begin{figure}
\epsscale{1.15}
\plotone{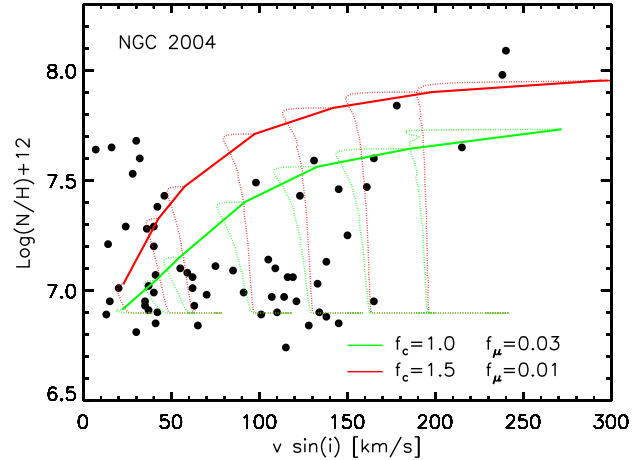}
\caption{Surface N abundance (black dots) as a function of the projected rotation velocity for a sample of stars in the LMC cluster NCG2004 \citep{untetal08}. Evolutionary tracks of a 13 $\rm M_{\odot}$ with different initial rotation velocities, computed with two different calibration of the rotation driven mixing efficiency, namely $f_c=1.0$ and $f_\mu=0.03$ (green line), and $f_c=1.5$ and $f_\mu=0.01$ (red lines).}
\label{fitngc2004}
\end{figure}

Figure \ref{fitngc2004} shows a plot of the surface N abundance as a function of the projected rotational velocity for the sample of stars in NGC2004. While there are stars that follow the general trend one would expect from the rotation induced mixing (i.e. the higher the initial rotation velocity the higher the surface N enhancement), there is also a conspicuous number of stars that do not follow such a general expectations: we will neglect these stars in the present calibration. We followed the same procedure adopted by \cite{brott11}, i.e. we aimed to reproduce what we expect it should be the main trend of the surface N enhancement as a function of the rotation velocity. The typical mass estimated for this sample is $\sim 13$ $\rm M_{\odot}$, therefore we computed a series of models of 13 $\rm M_{\odot}$ and different initial equatorial rotation velocities, assuming $f_c=1.0$ and $f_\mu=0.03$. The adopted initial metallicity is [Fe/H]=-.45, the abundances of most of the elements are assumed to be scaled solar, with the exceptions of  C, N, O, Mg and Si, for which we adopt the same scaling reported by \cite{brott11} in their Table 1. Figure \ref{fitngc2004} shows the evolutionary tracks of these models (green dotted lines) superimposed to the observed data. Note that the theoretical velocities have been multiplied by $\pi/4$ in order to take into account of the random inclination of the rotational axis. The Figure shows that the mixing obtained with $f_c=1.0$ and $f_\mu=0.03$ is not efficient enough to explain the highest N abundances observed for the fastest rotating models. Therefore we modified the two coefficients trying to get close to these highly enhanced stars. After a series of tests we decided (arbitrarily) to adopt $f_c=1.5$ and $f_\mu=0.01$. Figure \ref{fitngc2004} shows the evolutionary tracks computed with these choices as red dotted lines. The red and green solid lines are obtained by connecting the position of the various models at the central H exhaustion. As a final comment let us remark that with the present choices of  $f_c$ and $f_\mu$, the surface N enhancement obtained in a $\rm 20~M_\odot$ star of solar metallicity at core H depletion is of the order of 5-6, i.e., not much higher than the value obtained in CL13.


\section{Presupernova evolutions}\label{evol}

\begin{figure}
\epsscale{1.1}
\plotone{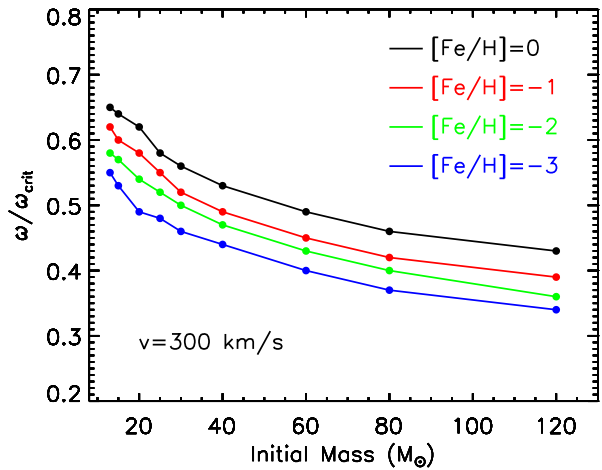}
\caption{Ratio between the angular and the critical velocities at the beginning of the main sequence for all the models of the present grid.}
\label{osuocritini300}
\end{figure}

We have already discussed extensively in CL13 the influence of rotation {\color{black} (see also \citet{hlw00,mm00})}, and of the associated instabilities, on the evolution of a generation of massive stars of solar metallicity and one initial rotation velocity v=300 km/s. Here we discuss the influence of the metallicity and two different initial equatorial velocities (150 and 300 km/s) on the evolution of a generation of massive stars. It is very important to note, and to remind through the reading of all the paper, that the role of rotation as a function of the mass largely depends on the way in which each grid of models is computed. More specifically, the dependence of any property (connected to rotation) on the mass is completely different if one compares models having the same initial rotation velocity or, e.g., the same fraction of the break out velocity. The reason is that the impact of rotation on the evolution of a star roughly scales directly with $\omega\over\omega_{\rm crit}$ but this parameter scales inversely with the mass if the initial rotation velocity is kept constant (Figure \ref{osuocritini300}). Hence a set of models of different masses and the same initial rotation velocity will obviously show a progressive reduction of the rotation induced effects as the initial mass increases. Note, also, that the setup of the stellar evolution code adopted for the present computations is different from the one adopted in CL13, therefore the solar metallicity models presented here are not identical to those discussed in that paper.  

\begin{figure}
\epsscale{1.1}
\plotone{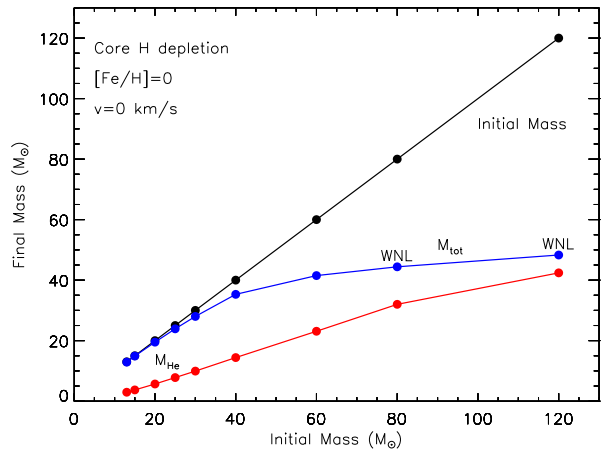}
\caption{Total mass (blue solid line) and He core mass (red solid line) at core H exhaustion as a function of the initial mass, for non rotating solar metallicity models. The label 'WNL' marks the models entering the Wolf-Rayet stage.}
\label{massHburncompare-a000}
\end{figure}

Tables \ref{tab0000}, \ref{tab1500}, \ref{tab3000}, \ref{tab0001}, \ref{tab1501}, \ref{tab3001}, 
\ref{tab0002}, \ref{tab1502}, \ref{tab3002}, \ref{tab0003}, \ref{tab1503} and \ref{tab3003} 
summarize the main evolutionary properties of all the computed models in the present grid, at the end of the main nuclear burning stages (e.g., "MS" refers to the end of the pre main sequence phase, "H" to the end of core H burning phase, "He" to the end of core He burning phase, "PSN" to the end of the evolution). For each burning stage, the various columns in these tables have the following meaning: (column 1) the evolutionary stage (as mentioned above);  (column 2) the star's lifetime in year; (column 3) the maximum extension of the convective core in solar masses; (column 4) the logarithm of the effective temperature in kelvin degrees; (column 5) the logarithm of the luminosity in solar luminosities; (column 6) the total mass of the star in solar masses; (column 7) the He core mass in solar masses; (column 8) the CO core mass in solar masses; (column 9) the equatorial velocity in km/s; (column 10) the surface angular velocity in $\rm s^{-1}$; (column 11) the ratio between the surface and the critical angular velocity; (column 12) the total star's angular momentum in unit of $\rm 10^{53}~g~cm^2~s^{-1}$; (columns 13 to 15) the surface H, He and N mass fractions; (columns 16 to 17) the N/C and N/O number ratios. 

\subsection{Core H burning}
\label{sect:hburning}
Mass Loss significantly affects the evolution of a massive star in central H burning and its influence increases with the initial mass because of the large dependence of the mass loss rate on the luminosity \citep{val00,val01}. With the presently adopted mass loss prescriptions, non rotating solar metallicity models with initial mass larger than $\rm 60~M_\odot$ lose a substantial fraction of their H-rich envelope and therefore enter the Wolf-Rayet (WR) stage already in this phase. In particular they become WNL stars during the late stages of core H burning. Note, however, that the minimum mass that becomes a WR star does not depend only  on the adopted mass loss rate but also on other uncertain properties like, e.g., the size of the H convective core. In fact, as it is well known,  the inclusion of some amount of convective core overshooting makes the evolutionary tracks cooler and brighter compared to the standard ones. This implies an overall higher mass loss and therefore a reduction of the minimum mass entering the WR stage.

Tables \ref{tabwr0000}, \ref{tabwr1500}, \ref{tabwr3000}, \ref{tabwr0001}, \ref{tabwr1501}, \ref{tabwr3001},\ref{tabwr0002}, \ref{tabwr1502}, \ref{tabwr3002}, \ref{tabwr0003}, \ref{tabwr1503} and \ref{tabwr3003} show the lifetimes during the various WR stages (see Chieffi \& Limongi 2013 for the definition of the various WR stages). Each table shows the following quantities: the  initial mass (column 1),  the lifetime during the O-type phase in years (column 2); the total WR lifetime in year (column 3); the lifetime during the WNL phase in years (column 4); the H or He central mass fraction at the time the star enters the WNL stage (column 5); the lifetime during the WNE phase in years (column 6); the H or He central mass fraction at the time the star enters the WNE stage (column 7); the lifetime during the WNC phase in years (column 8); the H or He central mass fraction at the time the star enters the WNC stage (column 9); the lifetime during the WC phase in years (column 10); the H or He central mass fraction at the time the star enters the WC stage (column 11). 

The He core mass $(\rm M_{He})$ at core H exhaustion increases, in general, with the initial mass $(\rm M_{ini})$ because it scales with the size of the H convective core that, in turn, increases with the mass of the star. Figure \ref{massHburncompare-a000} shows the $\rm M_{He}-M_{ini}$ (red dashed line) and the $\rm M_{tot}-M_{ini}$ (blue dashed line) relations at core H depletion, $\rm M_{tot}$ being the actual mass of the star. The bending of the $\rm M_{He}-M_{ini}$ relation is a consequence of the tremendous mass loss experienced by the more massive stars.

\begin{figure}
\epsscale{1.1}
\plotone{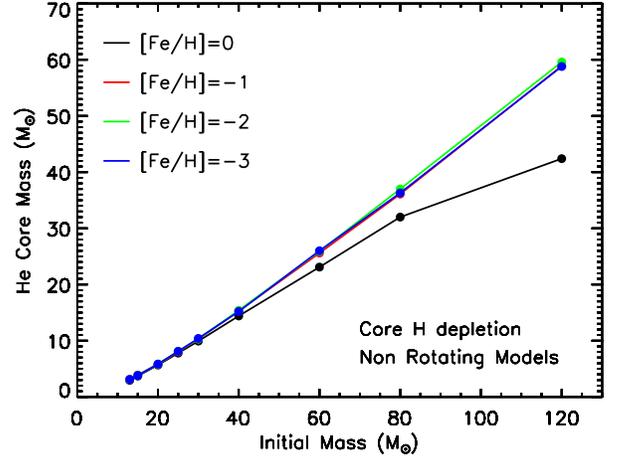}
\caption{He core mass at core H depletion as a function of the initial mass for non rotating models at various metallicities.}
\label{masshecore2}
\end{figure}

As the metallicity decreases, mass loss reduces significantly because it scales as $\rm \dot{M}\sim Z^{0.85}$ \citep{val00,val01}. As a consequence, all non rotating models with $\rm [Fe/H]\leq -1$ evolve essentially at constant mass in this phase. A lower initial metallicity implies also a reduction of the total abundance of the CNO nuclei and therefore an increase of the core H burning temperature. This leads, in principle, to more extended convective cores and hence to larger $\rm M_{He}$ at core H depletion \citep{tc86}. However, this effect is largely mitigated in our models by the inclusion of 0.2 $\rm H_P$ of overshooting in central H burning {\color{black} ($\rm H_P=-dLogr/dlogP$)}, occurrence that washes out most of the dependence of the convective core, and hence of the $\rm M_{He}$, on the initial metallicity. Figure \ref{masshecore2} shows, in fact, that stars with initial mass $\rm M<40~M_\odot$ develop He core masses essentially independent on the initial metallicity. Stars above this limiting mass show sizable  differences between models with solar and non solar metallicities but these differences are just the indirect effect of mass loss, as we have already discussed above.

\begin{figure*}
\plotone{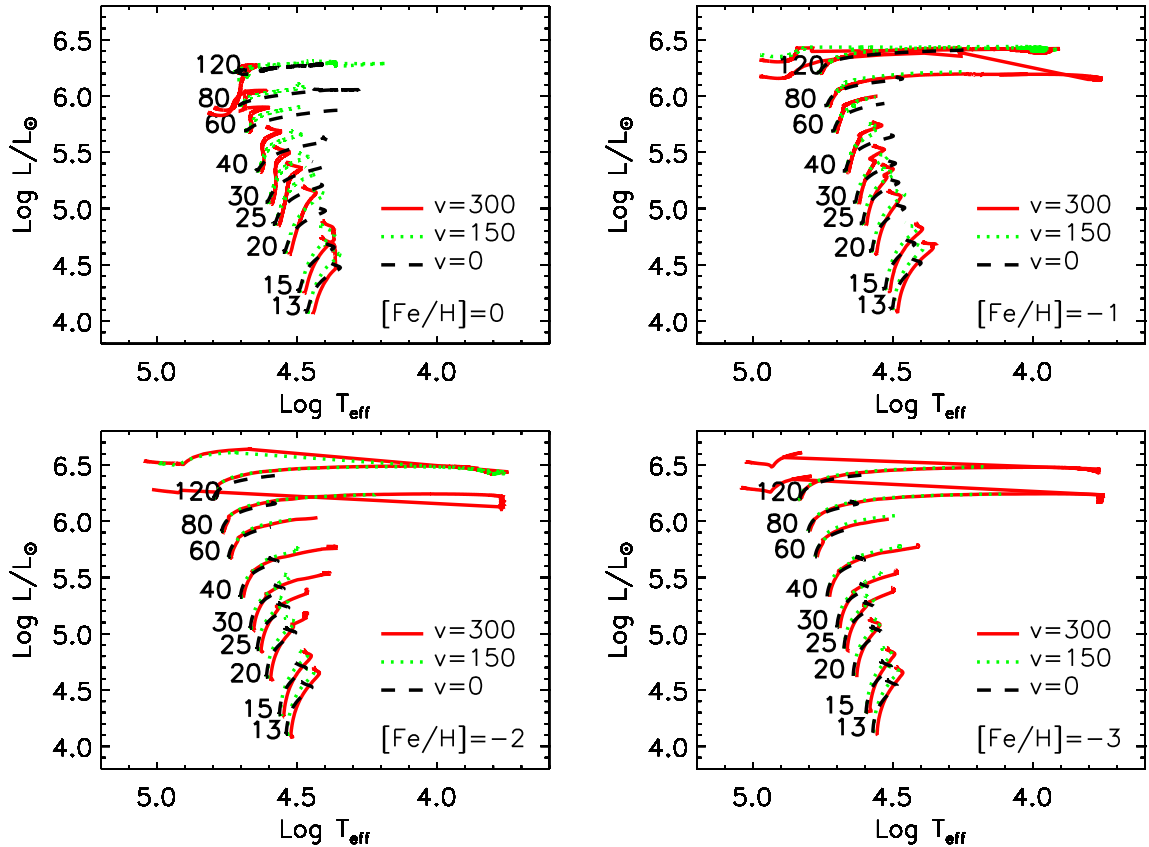}
\caption{Evolutionary tracks in the HR diagram of all the computed models during the core H burning phase at various metallicities. Black dashed lines refer to non rotating models, green dotted and red lines to models with initial velocity $\rm v=150~km/s$ and $\rm v=300~km/s$, respectively.}
\label{hrhburnallz} 
\end{figure*}

\begin{figure*}
\epsscale{1.1}
\plotone{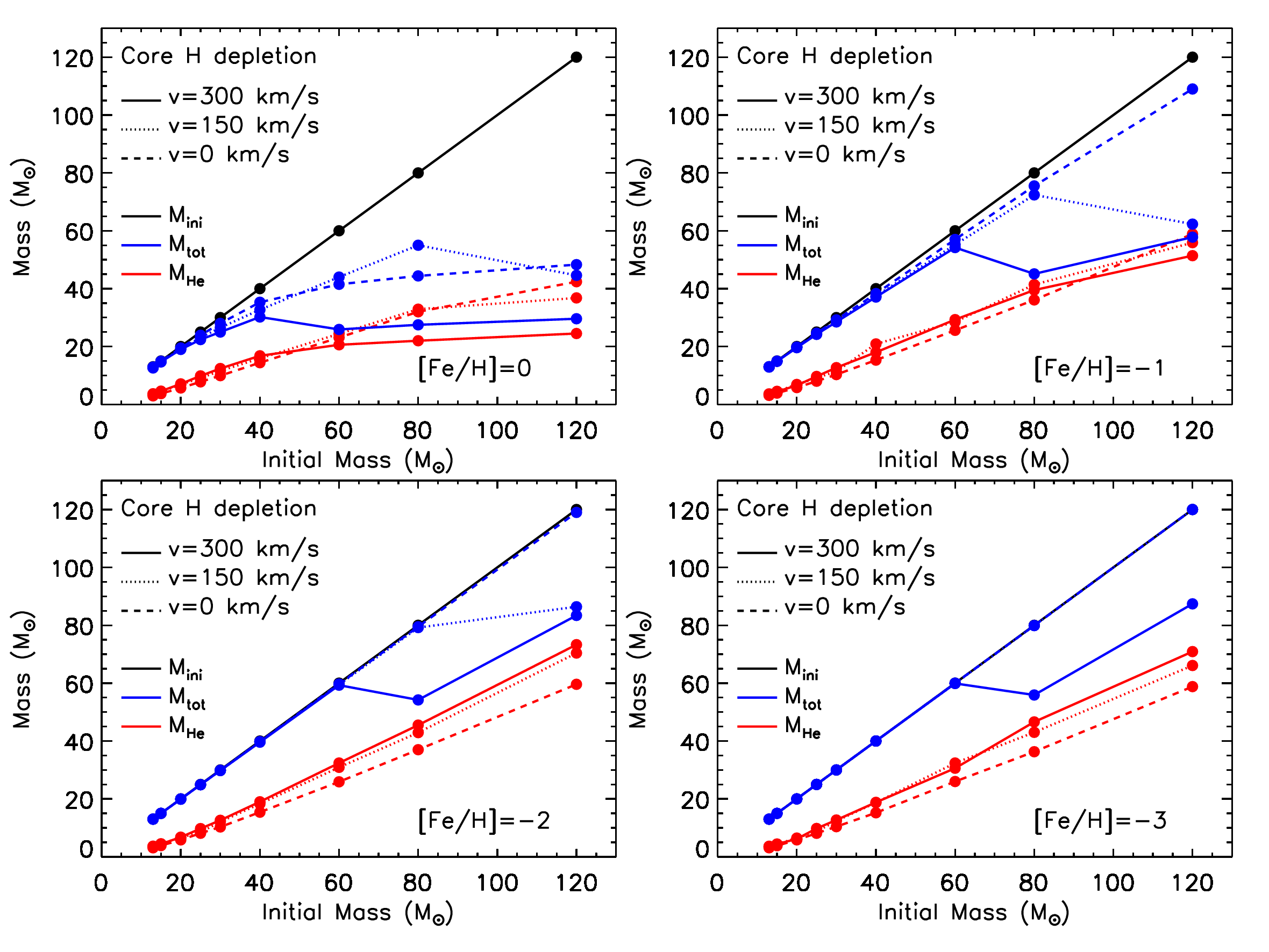}
\caption{Total mass (blue lines) and He core mass (red lines) at core H depletion as a function of the initial mass for various metallicities and different initial rotation velocities, i.e., $\rm v=0~km/s$ (dashed lines), $\rm v=150~km/s$ (dotted lines) and $\rm v=300~km/s$ solid lines.}
\label{massHdepletioncompare}
\end{figure*}

The effect of rotation on the evolutionary path of a massive star in the HR diagram in central H burning is twofold \citep[CL13]{mm00,mm00araa,mm01,mm12}.  
On one side, the lower gravity, due to the combined effects of the centrifugal force and the angular momentum transport, pushes the star towards lower effective temperatures. On the other side, the increase of the mean molecular weight in the radiative envelope, due to the rotationally driven mixing, has two key consequences: a) the mass of the H convective core increases, making therefore the track brighter and cooler (like the effect of the convective core overshooting); b) the opacity in the H rich mantle lowers, making the star more luminous and more compact and favoring therefore a blueward evolution. Depending on the initial mass, initial metallicity and initial rotation velocity, one of these effects may prevail over the others. Figure \ref{hrhburnallz} shows that, during core H burning, the evolutionary tracks of rotating solar metallicity stars are, on average, brighter and hotter than those of the non rotating ones and hence that the increase of the mean molecular weight in the H rich mantle is the dominant effect. On the contrary, as the metallicity decreases, the evolutionary tracks of rotating stars become brighter and cooler than those of the non rotating ones and hence that the reduction of the effective gravity, due to the centrifugal force and the angular momentum transport, mainly controls the evolution.

A change in the evolutionary path of a star in the HR diagram obviously reflects on the mass loss rate. At solar metallicity, the amount of mass lost in H burning increases significantly with the initial rotation velocity. As a consequence, the minimum mass entering the WNL stage in this phase decreases from $\rm M>60~M_\odot$, in the non rotating models, to $\rm M>40~M_\odot$, in the rotating ones (Tables \ref{tabwr0000}, \ref{tabwr1500} and \ref{tabwr3000}). At sub solar metallicities, viceversa, the effect of rotation on the mass loss rate is negligible because of its steep dependence on the metallicity. Hence at sub solar metallicities also the rotating models evolve essentially at constant mass, with the exception of the two most massive ones. These models experience a pronounced redward excursion in the HR diagram, approach the Eddington limit (when the effective temperature drops below $Log~{\rm T_{eff}}\sim 3.9$), enter a phase of very strong mass loss that drives the ejection of a substantial fraction of their H-rich envelope and eventually become WNL stars (Figure \ref{hrhburnallz}). Thus, in general, the minimum mass entering the WNL stage during core H burning decreases with increasing the metallicity and with increasing the initial rotational velocity (see Tables \ref{tabwr0000}-\ref{tabwr3003}).

In absence of rotation, the mixing of the matter occurs only within the regions where the thermal instabilities (convection) grow. On the contrary, in presence of rotation an additional mixing occurs also in the thermally stable (radiative) regions. The two main {\it engines} that drive such a mixing are the meridional circulation and the secular shear. The former instability dominates the diffusion of the chemical composition in the inner part of the radiative mantle, i.e., close to the outer edge of the H convective core, while the secular shear controls the mixing in the outer layers (see, e.g., Figure 5 in Chieffi \& Limongi 2013). The main consequences of this rotation driven mixing are: (1) the increase of the core H burning lifetime, (2) the increase of the He core mass at core H exhaustion and (3) the surface enhancement of the \nuk{N}{14} abundance.

\begin{figure*}
\epsscale{1.0}
\plottwo{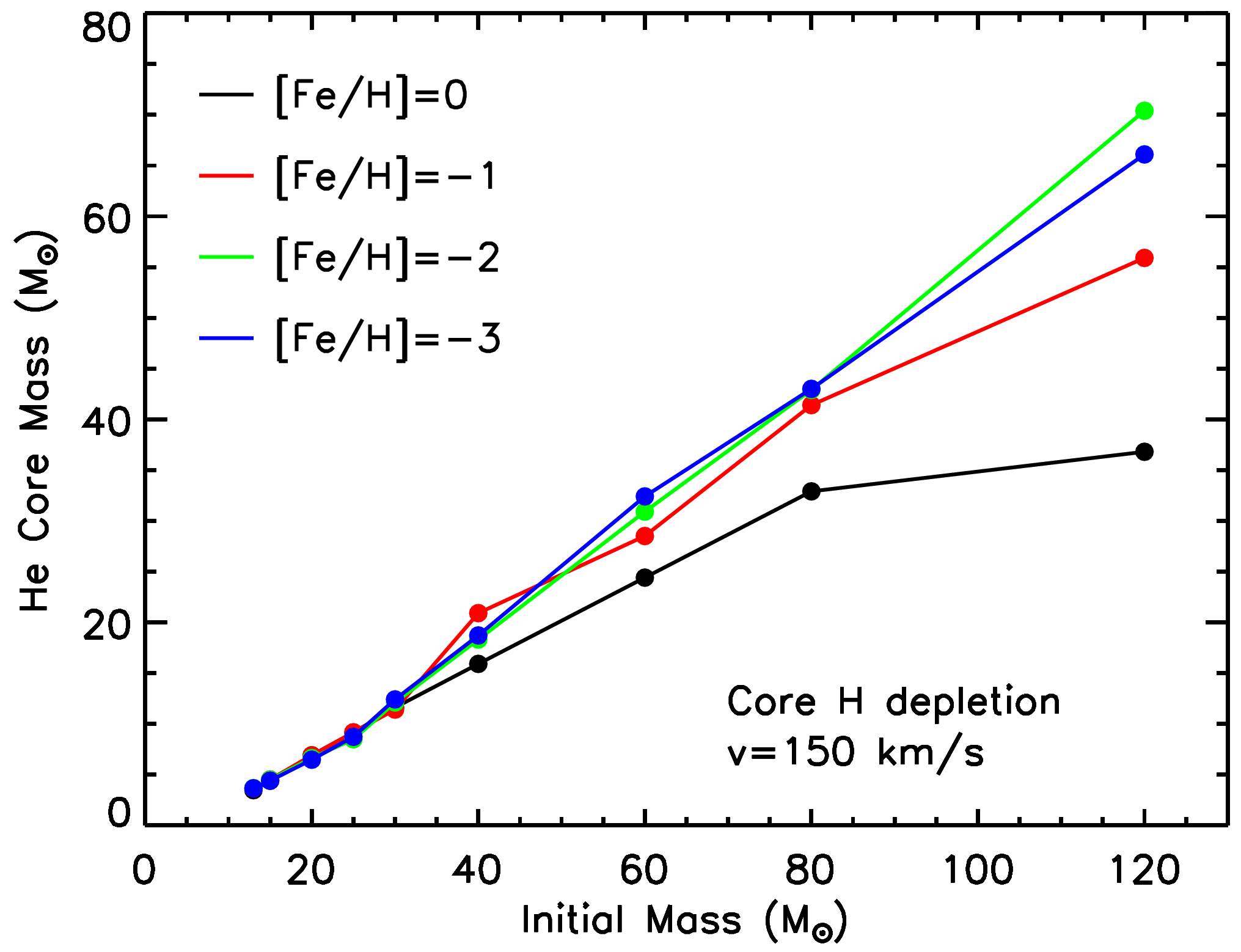}{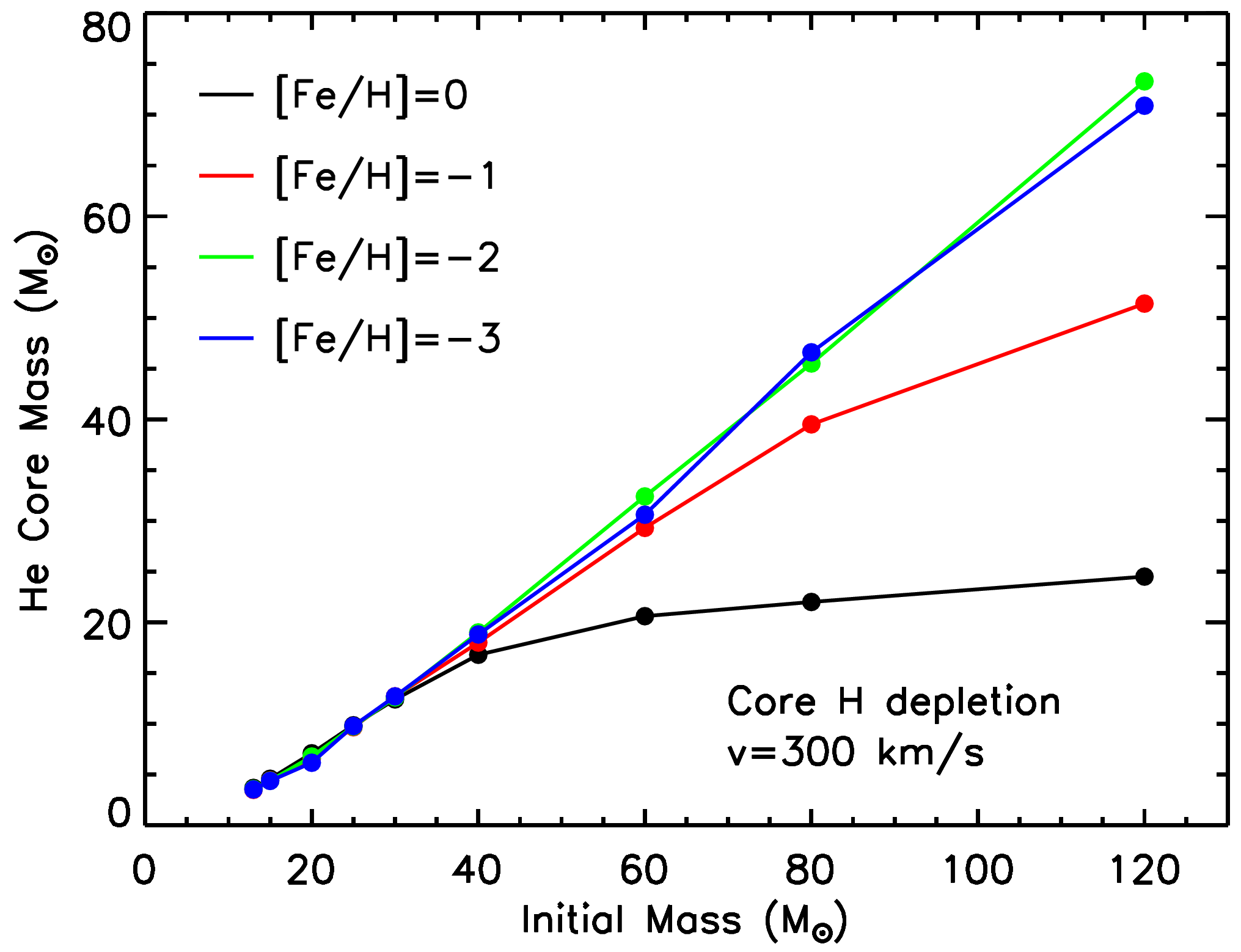}
\caption{He core mass at core H depletion as a function of the initial mass at various metallicities for models with initial rotation velocity $\rm v=150~km/s$ (left panel) and $\rm v=300~km/s$ (right panel).}
\label{massHdepletioncompareZ}
\end{figure*}

Figure \ref{massHdepletioncompare} shows the effect of the rotation induced mixing on the size of the He core at core H depletion as a function of the initial rotation velocity, for the various metallicities. The general trend is that, for any fixed initial mass, the higher the initial rotation velocity, the greater the $\rm M_{He}$ at core H depletion. This general trend fails for those models in which mass loss is efficient enough to reduce substantially the total mass of the star. In particular the fast rotating solar metallicity models with $\rm M > 40~M_\odot$ as well as the fast rotating ones with [Fe/H]=-1 and mass $\rm M > 60~M_\odot$. Figure \ref{massHdepletioncompareZ} shows the effect of the metallicity on the size of the $\rm M_{He}$ at core H depletion, for a fixed initial rotation velocity. It is worth noting that stars with initial mass $\rm M < 25~M_\odot$ do not show a significant dependence of the $\rm M_{He}$ on the initial metallicity (at least in the range of initial rotational velocities studied in this paper). More massive stars, vice versa, show an increase of the $\rm M_{He}$ as the metallicity decreases (for both rotational velocities) because the lack of a strong mass loss allow these stars to keep up to the end of the H burning a convective core bigger than that present in star that have lost a consistent amount of mass. 

\begin{figure*}
\epsscale{1.0}
\plotone{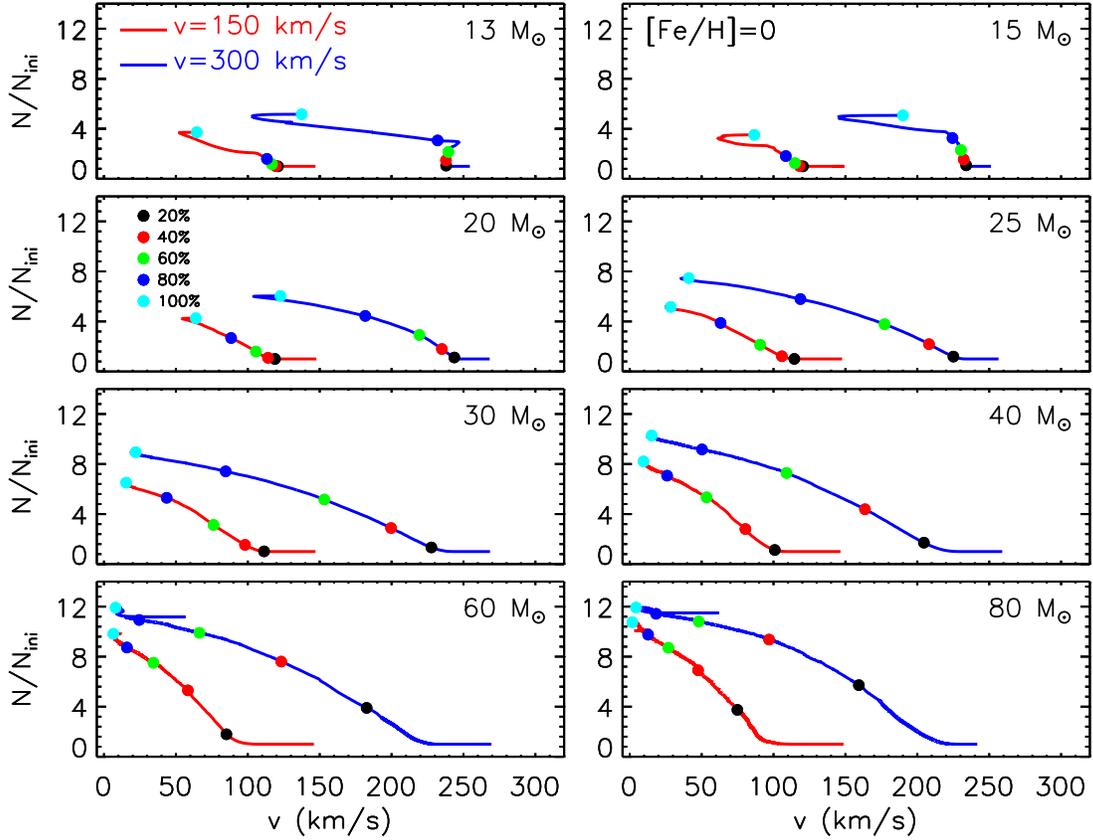}
\caption{Ratio between the surface N abundance and the initial one as a function of the equatorial velocity during the core H burning phase for solar metallicity rotating models. The dots with different colors mark the locations corresponding to a given percentage of the total H burning lifetime (see the legend in the figure corresponding to the $\rm 20~M_\odot$ star model).}
\label{hunterA}
\end{figure*}

\begin{figure*}
\epsscale{1.0}
\plotone{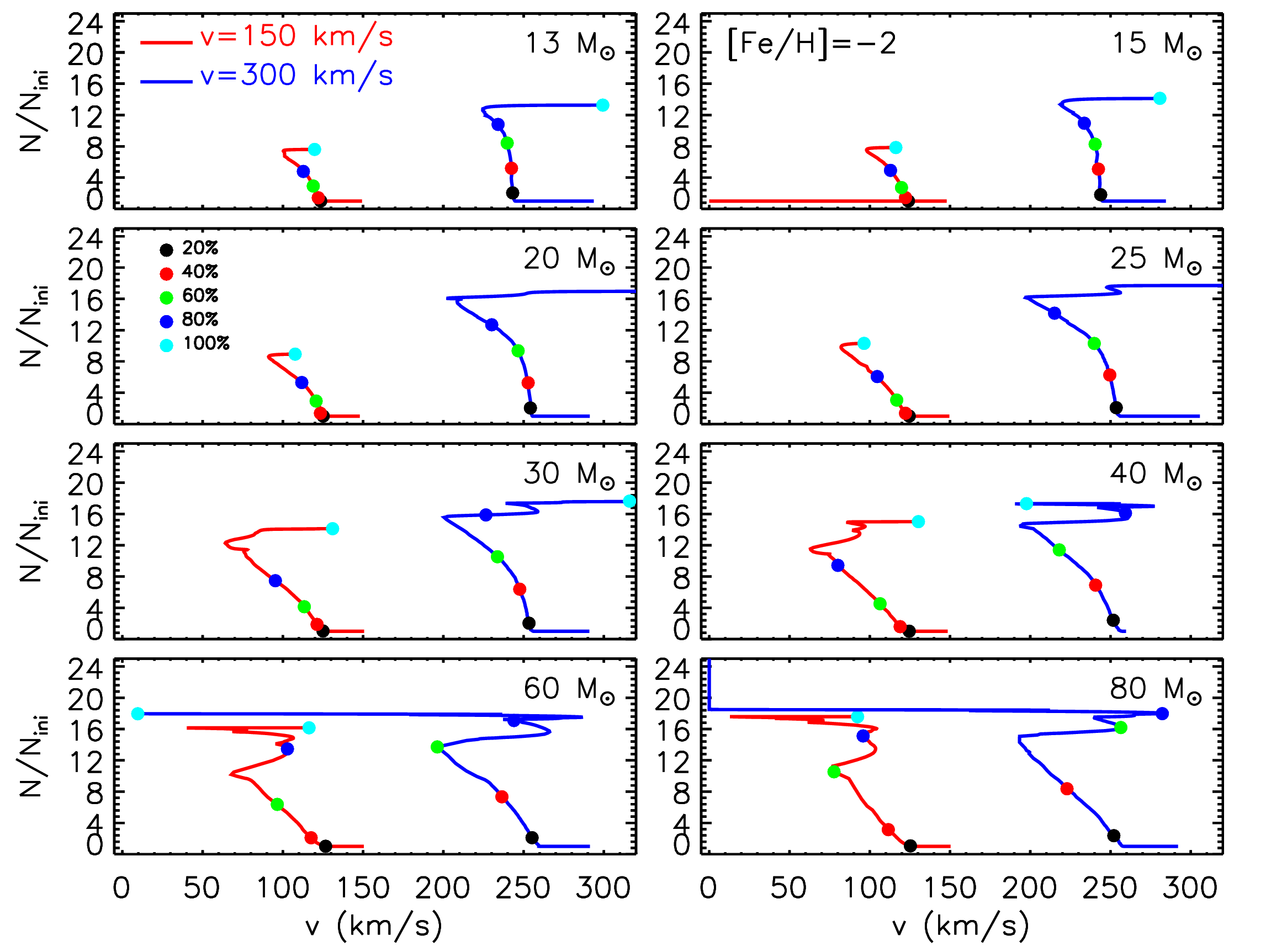}
\caption{Same as Figure \ref{hunterA} but for metallicity [Fe/H]=-2.}
\label{hunterC}
\end{figure*}

An important aspect of rotating models that is worth discussing is the temporal variation of the surface chemical composition during the main sequence phase. Figures \ref{hunterA} and \ref{hunterC} show the effect of the metallicity on the trend of the surface \nuk{N}{14} abundance versus the current equatorial rotational velocity during the core H burning phase (the so called Hunter diagram). At solar metallicity most of the stars show a quite smooth increase of the surface \nuk{N}{14} abundance coupled to a corresponding decrease of the surface equatorial rotational velocity. This behavior is the consequence  of the efficient loss of angular momentum triggered by the strong stellar winds. The only exceptions to this behavior are the two smaller masses, namely the $\rm 13~M_\odot$ and the $\rm 15~M_\odot$, that show an increase of \nuk{N}{14} at almost constant equatorial rotational velocity because of the modest loss of angular momentum from the surface due to the weaker stellar wind.

Low metallicity models behave essentially like the 13 and the 15 $\rm M_\odot$ stars of solar metallicity. In fact, due to the strong reduction of the mass loss with the metalliticy, in this case the increase of the surface \nuk{N}{14} is coupled to a modest reduction of the surface equatorial velocity; a substantial change of the surface velocity occurs only towards the end of the H burning, when the surface abundance of \nuk{N}{14} does not change any more. It goes without saying that these different behaviors may play a crucial role in the interpretation of the Hunter diagrams. A more detailed study of this issue will be addressed in a forthcoming paper. 

\begin{figure*}[htb!]
\epsscale{1.0}
\plottwo{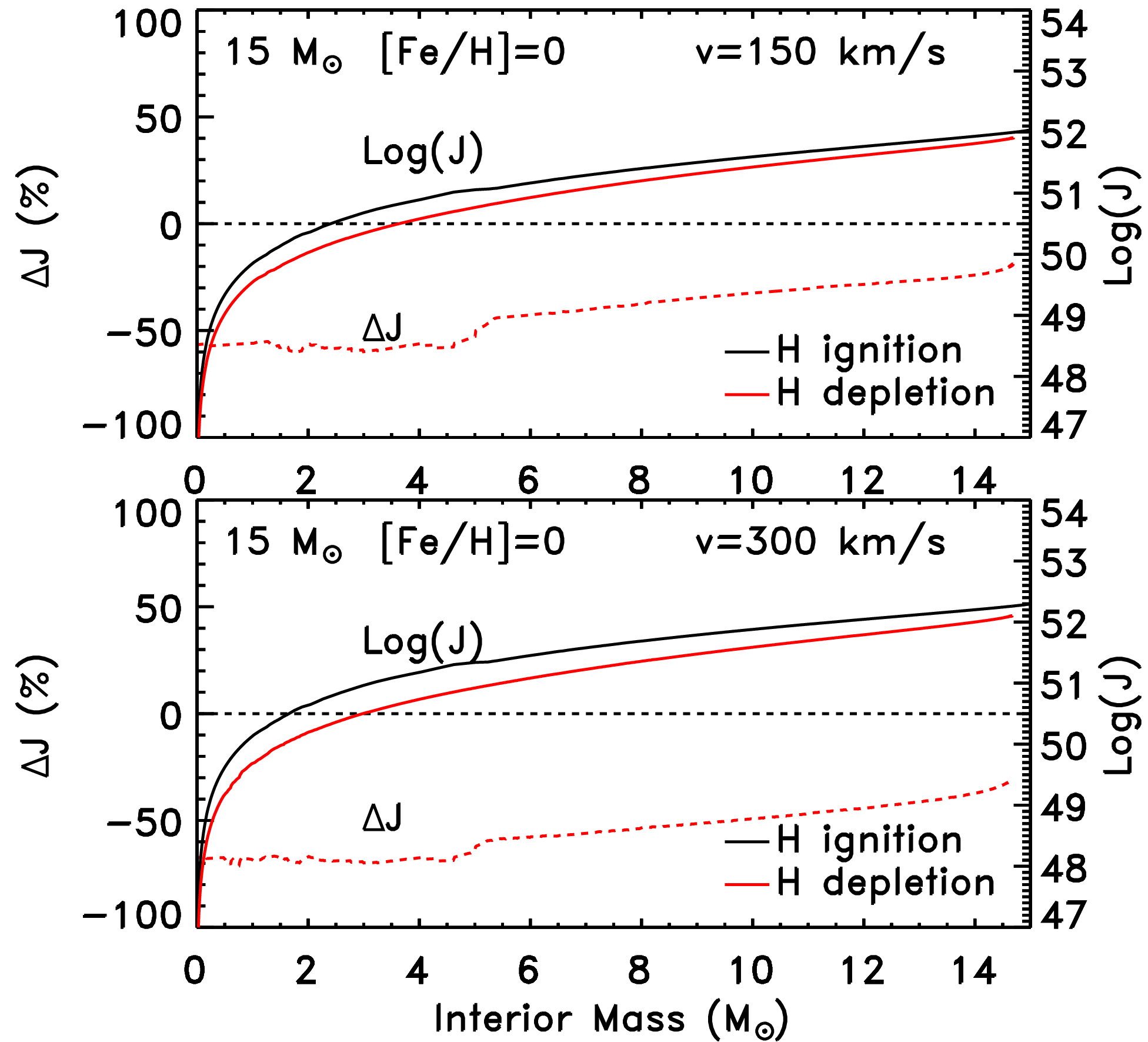}{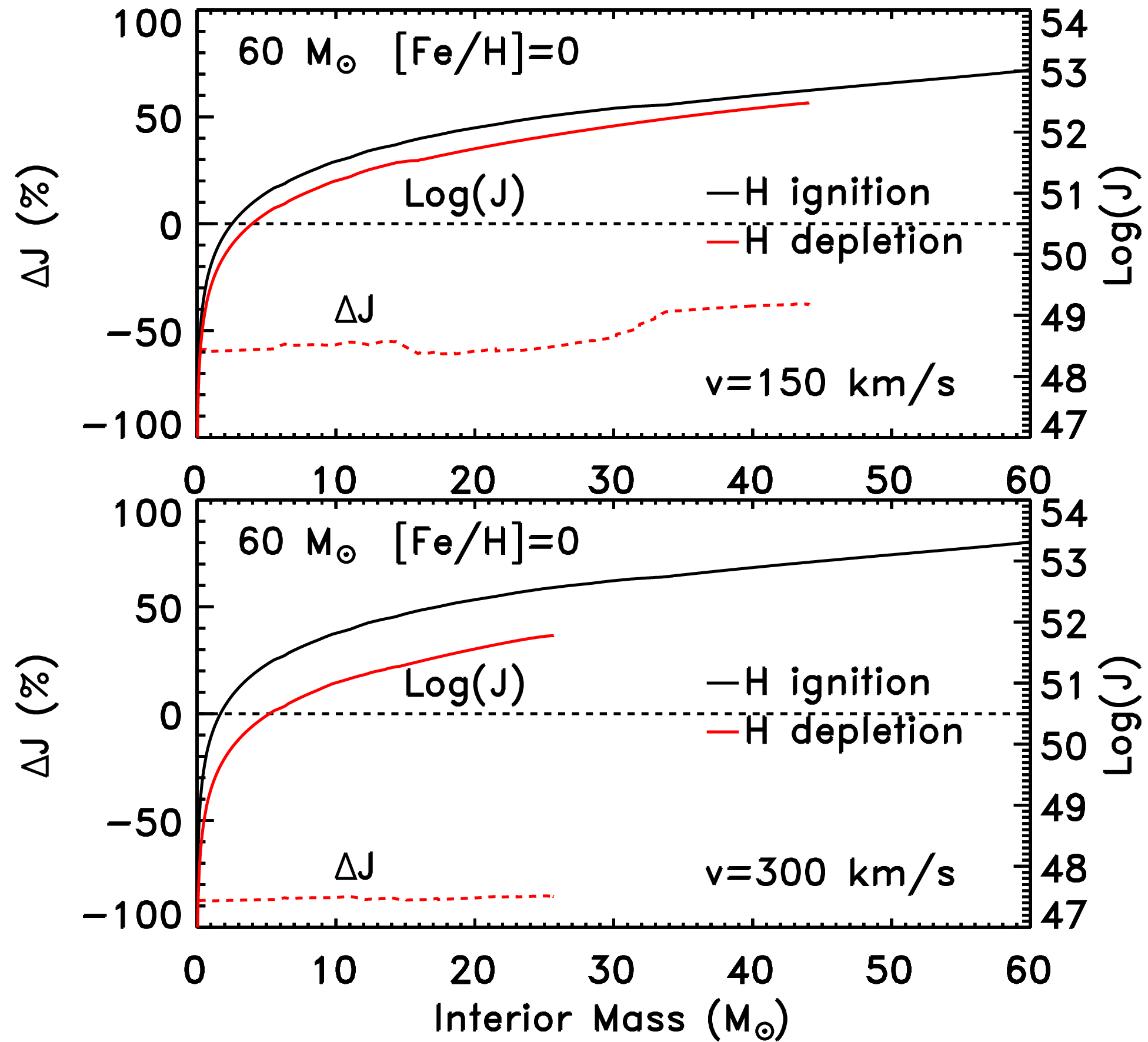}
\caption{Total (cumulative) angular momentum (J) at core H ignition (black solid line) and at core H depletion (red solid line) as a function of the interior mass (secondary {\em y-axis}) for solar metallicity models of $\rm 15~M_\odot$ (left panels) and $\rm 60~M_\odot$ (right panels) with initial rotation velocities $\rm v=150~km/s$ (upper panels) and $\rm v=300~km/s$ (bottom panels). Also shown is the difference, in percentage, between the total angular momentum at core H depletion and at core H ignition ($\rm \Delta~J$, red dotted line) as a function of the interior mass (primary {\em y-axis}).}
\label{momanghburn0}
\end{figure*}

\begin{figure*}[htb!]
\epsscale{1.0}
\plottwo{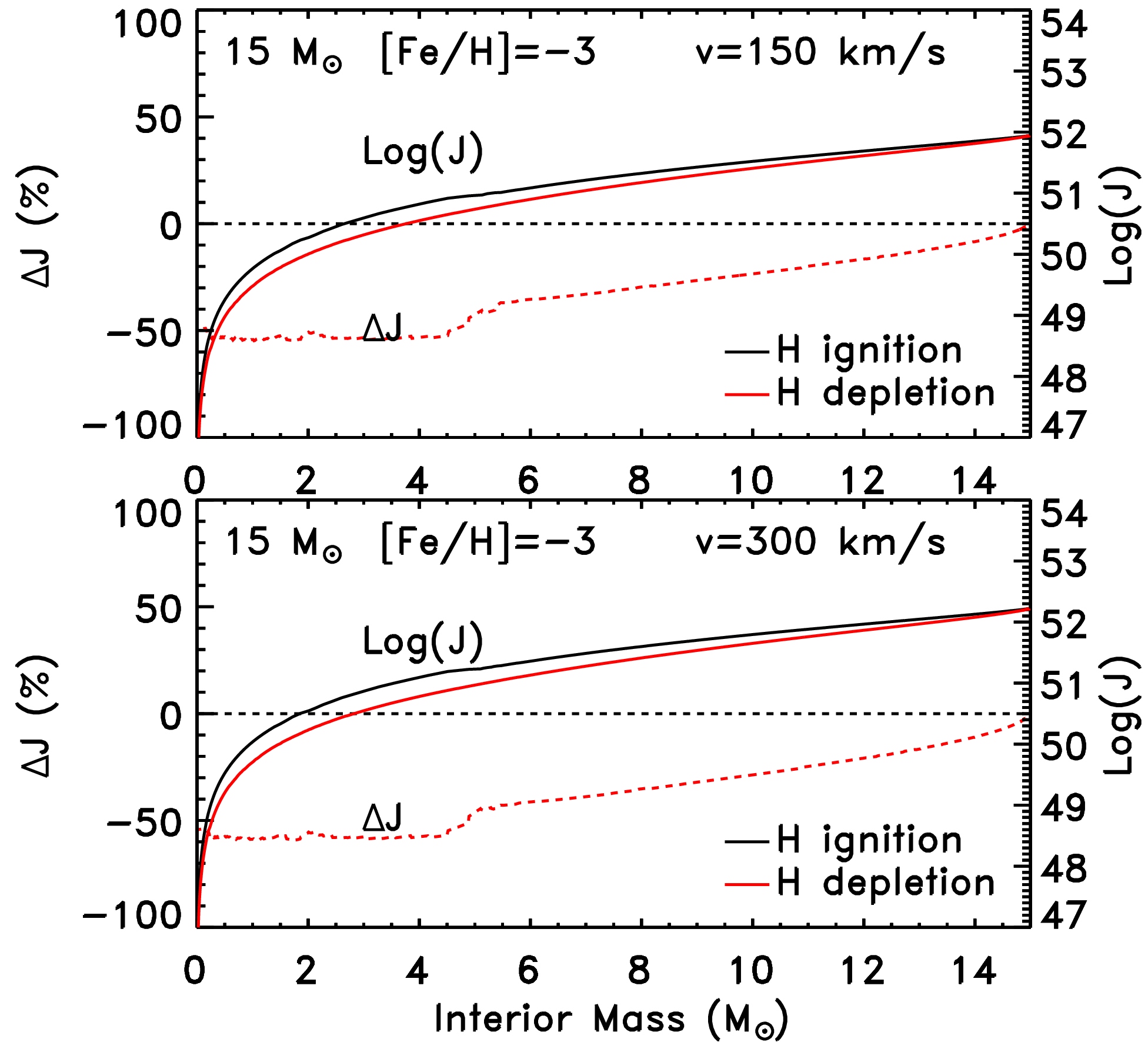}{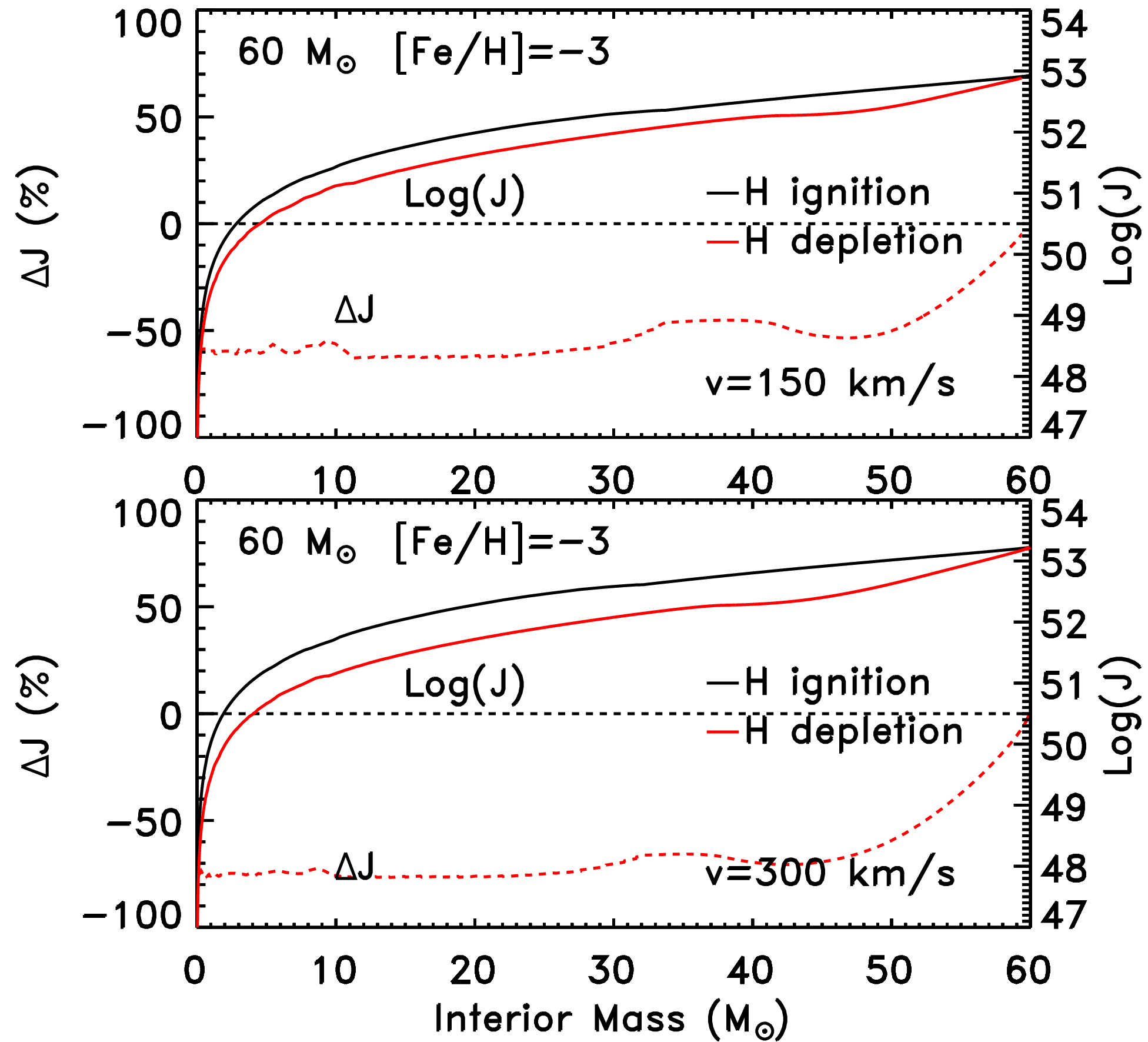}
\caption{Same as Figure \ref{momanghburn0} but for models with initial metallicity [Fe/H]=-3.}
\label{momanghburn3}
\end{figure*}

Another key feature worth being discussed is the variation of the total angular momentum and its internal distribution at the end of the core H burning phase. Both these properties are the result of the combined effects of either the transport and the loss of angular momentum. The present solar metallicity models lose between $\sim 30\%$ and $\sim 90\%$ of their initial angular momentum during core H burning (see Figure \ref{momanghburn0}). The solid black and red lines show the internal (cumulative) run of the angular momentum at H ignition and exhaustion, respectively, for both the 15 and the 60\ \msun. As expected, the larger the mass and the larger the initial equatorial velocity the more efficient the angular momentum loss. The red dashed lines in the same Figure show the run of the (cumulative) change (with respect to the H ignition) of angular momentum (in percentage) at the central H exhaustion. A comparison between the two red dashed lines in the left panels (that refer to the 15\ \msun) shows that the amount of angular momentum lost by the H exhausted core is almost independent on the initial rotation velocity. This is easily understood by reminding that in the convective zones we assume a flat omega profile which implies the maximum possible transport of angular momentum. Also the 60\ \msun\ with v=150 km/s behaves similarly to the two 15\ \msun, while the 60\ \msun\ rotating initially at 300 km/s shows a much more pronounced reduction of angular momentum in the H exhausted core. This is due to the very efficient mass loss that this star experiences in H burning. 
At lower metallicities the strong decrease of the stellar wind inhibits the loss of angular momentum, as a consequence these models do not lose a substantial amount if angular momentum (see Figure \ref{momanghburn3}). Note, however, that also in this case the presence of a convective core forces the angular momentum present in the H exhausted core to drop by an amount quite similar to that lost by the more metal rich stars.

\begin{figure*}
\epsscale{1.19}
\plotone{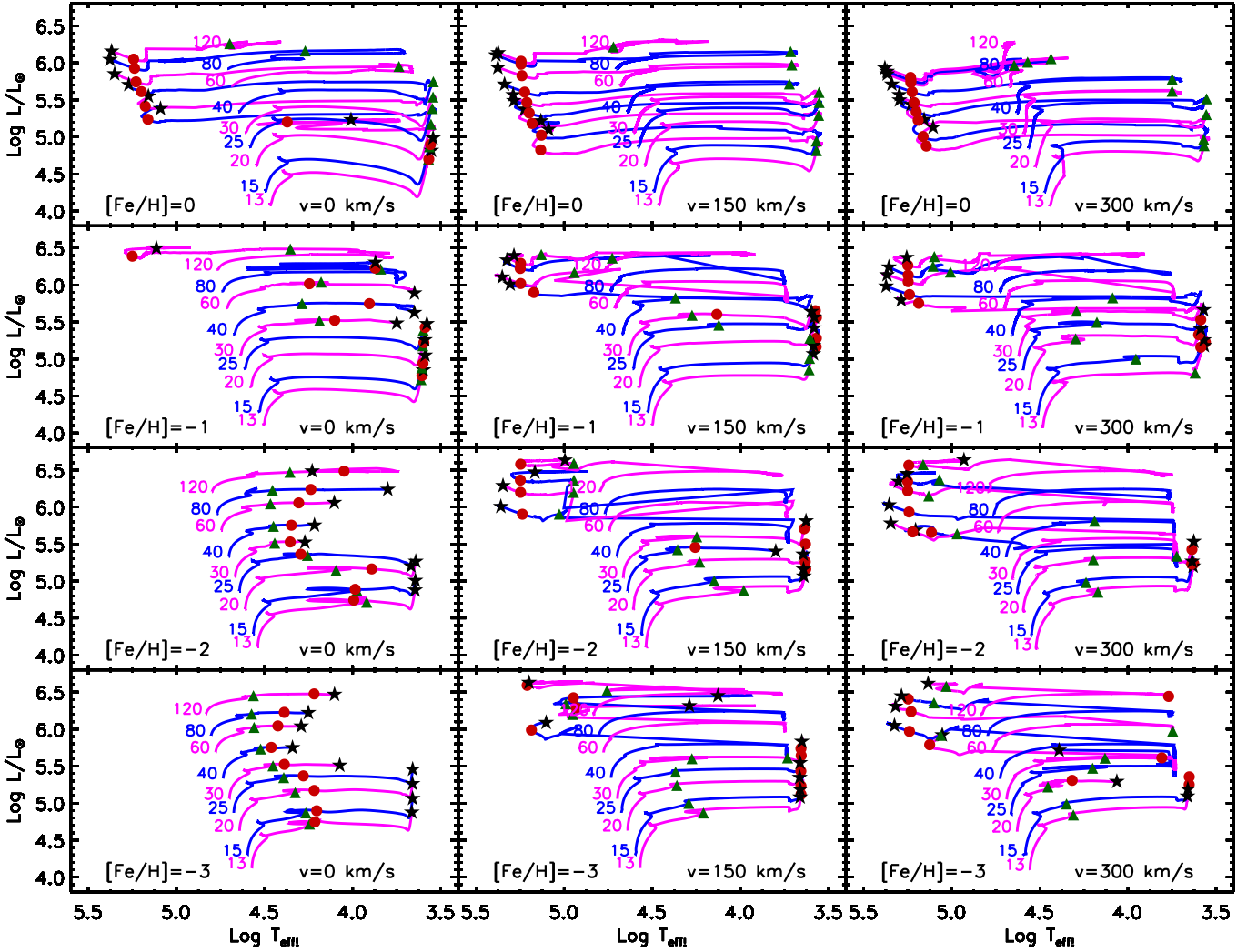}
\caption{Evolutionary tracks of all our models in the HR diagram. The various symbols mark: the central He ignition (green triangles), the central He exhaustion (red dots) and the final position at the time of the core bounce (black star).}
\label{hrpresnallz}
\end{figure*}

\subsection{Core He burning}
\label{sect:heburning}
The evolutionary track of a massive star beyond the central H burning depends on a complex interplay among several factors: (1) the He core that contracts towards a new quasi equilibrium configuration powered by the $3\alpha$ nuclear reactions, (2) the H rich mantle that expands towards a Red SuperGiant (RSG) configuration, (3) the actual He core mass and (4) the amount of mass the star loses during this phase.
Figure \ref{hrpresnallz} shows the full HR diagram for all our models: the green triangles and red filled dots mark the beginning and the end of the central He burning phase, respectively.
Before discussing Figure \ref{hrpresnallz} let us remark that the capability of the mantle of the star to expand (and therefore to become red giant) on a thermal or on a nuclear timescale depends on many factors including, e.g., the adoption of the Schwarzchild or the Ledoux criterion in the region of variable H abundance left by the receding H convective core, the opacity of the mantle, and so on. As a consequence, at present, this behavior is still poorly understood.

At [Fe/H]=0 all the non rotating models evolve toward their Hayashi track on a thermal timescale and therefore they start the core He burning as RSGs, the only exception being the 120\ \msun\ that loses enough mass to become a WR star already during core H burning. During this redward excursion stars with $\rm M\gtrsim 40~M_\odot$ approach the Eddington luminosity, at $\rm Log(T_{eff})\sim 3.7$), lose a substantial fraction of the H-rich envelope, evolve to a Blue SuperGiant (BSG) configuration and become WRs (Table \ref{tabwr0000} shows the time spent by each mass in the various WR subclasses). Less massive stars, on the contrary, reach their Hayashi track and cross the critical temperature for the dust driven wind to become efficient. However, within this mass interval, only stars with $\rm M\gtrsim 15~M_\odot$ enter this stage with a central He abundance high enough to have time to lose a consistent amount of mass. When a substantial fraction of the H-rich envelope has been lost, these stars deflate towards a BSG configuration and become WR stars. Stars with $\rm M\lesssim 15~M_\odot$ remain RSGs during the whole core He burning phase. Therefore, at solar metallicity, we predict a population of RSGs up to an initial mass of $\rm M\sim 40~M_\odot$ (corresponding to a maximum luminosity $\rm Log(L/L_\sun)\sim 5.7$) and a minimum mass for the WR stars of $\rm M\sim 20~M_\odot$.

At [Fe/H]=-1 non rotating stars with $\rm M\gtrsim 60~M_\odot$ quickly expand after the central H exhaustion but never reach their Hayashi track because they lose an enormous amount of mass when they exceed their Eddington luminosity (at $\rm Log(T_{eff})\sim 3.7-3.8$) and then become WR stars (see Table \ref{tabwr0001}). Stars in the range $\rm \sim 30-60~M_\odot$ ignite and burn He as BSGs without entering at all the WR phase because of the modest mass loss. Stars with $\rm M\lesssim 30~M_\odot$ ignite and burn He as RSGs, but none of them crosses the threshold temperature for the condensation of dust, hence they remain RSGs during the whole core He burning phase. Therefore, at this metallicity, we predict a population of RSGs up to an initial mass of $\rm M\sim 25~M_\odot$ (corresponding to a maximum luminosity $\rm Log(L/L_\sun)\sim 5.5$) and a minimum mass that enters the WR stage of $\rm M\sim 80~M_\odot$.


At [Fe/H]=-2 and -3 all the non rotating stars ignite and burn He in the core as BSG. For these metallicities, therefore, we do not expect neither RSGs nor WRs during this phase.

Summarizing the results discussed so far, we predict the core He burning non rotating models to populate the RSG branch with stars of mass $\rm M\lesssim 40~M_\odot$ ($\rm Log(L/L_\sun)\lesssim 5.7$) at [Fe/H]=0 and $\rm M\lesssim 25~M_\odot$ ($\rm Log(L/L_\sun)\lesssim 5.5$) at [Fe/H]=-1. 
No star becomes a RSG at [Fe/H]=-2 and -3. The minimum mass that enter the WR stage during this phase is $\rm M\sim 20~M_\odot$ at [Fe/H]=0. No WR star is expected at lower metallicities. 

\begin{figure}
\epsscale{1.1}
\plotone{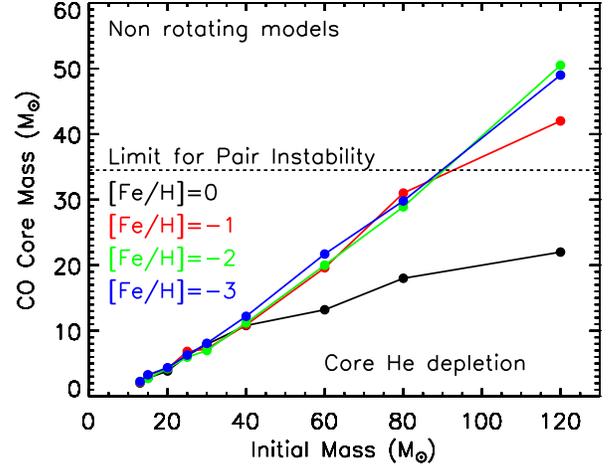}
\caption{$\rm M_{\rm CO}-M_{\rm INI}$ relation for the four metallicities. The horizontal dashed line marks the mass limit above which a star enters the pair instability regime ad explodes as Pair Instability Supernova}
\label{masscocoreZ}
\end{figure}

Turning to the evolution of the interior, core He burning occurs, as it is well known, in a convective core that advances progressively in mass until it vanishes at core He depletion. As a consequence a very steep He profile forms at a mass coordinate corresponding to the maximum extension of the convective core. Such a "typical" behavior fails when mass loss is strong enough to drive the complete ejection of the H rich mantle and to erode part of the He core. Since the properties of the He burning depend mainly on the actual He core mass, if the He core reduces, while the central burning is still active, (1) the He convective core progressively shrinks in mass leaving a region of variable chemical composition; (2) the surface luminosity progressively decreases; (3) the core He burning lifetime increases; (4) the CO core mass at the end of the He burning reduces while the $\rm ^{12}C$ mass fraction increases. Since all these effects are driven by mass loss, they tend to progressively disappear as the initial metallicity decreases. Figure \ref{masscocoreZ} shows the CO core mass as a function of the initial mass at core He exhaustion for the four metallicities.  As expected, the CO core scales directly with the initial mass at all four metallicities. Similarly to the trend shown by the He core (see Figure \ref{massHdepletioncompareZ}), the $\rm M_{CO}-M_{ini}$ relation is basically independent on the initial metallicity when mass loss does not erode the He core mass. Therefore only the solar metallicity stars of mass $\rm M>40~M_\odot$ show an evident bending due to the decrease of the He core mass.

The inclusion of rotation makes the picture discussed so far even more complex because its effect on the evolution of a star depends on the mass and the metallicity. 
As in the non rotating case, at [Fe/H]=0 we can identify ranges of models that (1) become WR during the core H burning phase and hence burn He in the core as BSGs; (2) approach their Eddington luminosity during the redward excursion, evolve toward a BSG configuration and burn He in the core as WR stars; (3) approach their Hayashi track, enter the dust driven wind stage and then turn again to the blue, sometime during the He burning phase. The basic rule is that the higher the initial rotation velocity the lower 
the limiting masses that divide these three mass intervals (Tables \ref{tab1500}, \ref{tab3000}). Note that, for this metallicity, all rotating models end their He burning lifetime as WR stars since in this case even the two smaller masses are pushed beyond the threshold temperature for dust formation early enough to have time to lose a large part of their mantle and turn again towards the blue (Tables \ref{tabwr0000}, \ref{tabwr1500}, \ref{tabwr3000}).


At metallicities $\rm [Fe/H]\leq -1$ the quite complex interplay between metallicity and rotation does not lead any more to a strictly monotonic trend with the mass. We can identify models that (1) move redward toward the Hayashi track on a thermal timescales, lose a substantial amount of mass because they approach their Eddington luminosity, turn to the blue and burn He as WR stars (Tables \ref{tabwr1501}, \ref{tabwr3001}); (2) ignite He as BSGs, move redward while core He burning goes on, become RSGs, turn to the blue sometime during core He burning and become WR stars; (3) ignite He as BSG, move redward on a nuclear timescale, reaching eventually the RSG phase when the central He abundance is more than halved; (4) ignite and burn He as RSGs. The basic rule in this case is that, on average, the limiting masses that divide the above mentioned mass intervals decrease by increasing the initial rotation velocity and by decreasing the initial metallicity (Tables \ref{tab1501}, \ref{tab3001}). Note that at these metallicities no model crosses the temperature threshold {\color{black} \citep{vanloonetal05})} for the activation of the mass loss due to the dust formation.


As in core H burning, also in core He burning the interplay among convection, meridional circulation and shear turbulence drives the outward transport of angular momentum and mixing of the chemicals. 

\begin{figure}
\epsscale{1.18}
\plotone{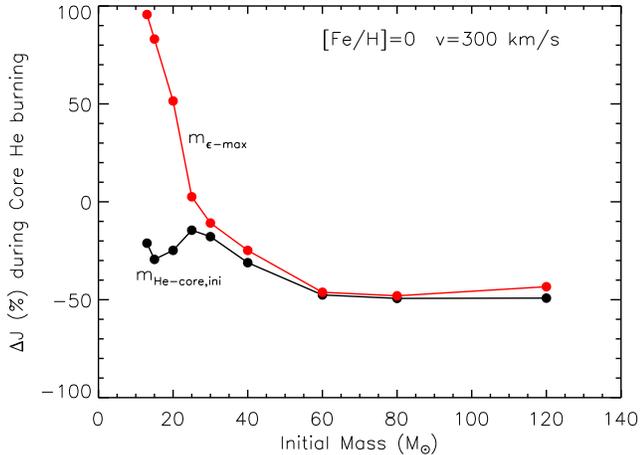}
\caption{Variation (in percentage) of the total amount of angular momentum stored in the He core, during core He burning, for the solar metallicity case, computed for two different definitions of the He core mass. The red line refers to the case in which the He core mass is defined at each time at the mass location where the maximum of the H nuclear burning occurs while the black one refers to the case in which the He core mass is fixed at the beginning of the central He burning and kept constant in time.}
\label{momaHevar}
\end{figure}

A quantitative determination of the variation of the angular momentum contained in the He core in this phase depends on the definition of the He core mass. Figure \ref{momaHevar} shows the variation of the amount of angular momentum in the He core for the solar metallicity case and initial velocity 300 km/s. The two lines correspond to two different choices for the He core mass. If we choose as the He core the amount of mass contained within the H burning shell, we obtain the red line in the Figure. In this case, the progressive advance of the burning shell adds continuously new mass, and hence angular momentum, to the He core and this increase is much larger than the amount of angular momentum that flows from the center outward. However stars more massive than 20\ \msun\ lose a consistent fraction of their He core mass through the wind and this phenomenon prevails in determining the amount of angular momentum left in the He core. If, vice versa, we fix the He core mass just at the beginning of the He burning and compute during the whole He burning phase the total amount of angular momentum contained within this mass, the scenario changes completely (black line in Figure \ref{momaHevar}). In this case the angular momentum always reduces because the mass is fixed and the angular momentum fluxes outward. Note that above 20\ \msun\ the red and black lines converge because in both cases the angular momentum present in the He core is dictated by the mass loss. At subsolar metallicities the amount of angular momentum that fluxes outward through a fixed He core mass is quite similar to the solar case, so that the decrease of angular momentum varies between 40\% and 10\% in the range 13 to 25\ \msun. The decrease of the angular momentun in the more massive stars, similarly to what happens in the solar case, depends on the efficiency of the mass loss. Though rotating massive stars enter the instability region where L/Ledd$>$1 and therefore lose a large fraction of their envelope, they do not lose as much mass as their solar counterparts so that also the total amount of angular momentum left in the He core at the central He exhaustion increases moderately as the initial metallicity reduces.

The mixing of the chemicals due to the rotation induced instabilities has essentially two basic consequences: (1) the increase of the CO core mass and (2) the exchange of matter between the two active burning regions, i.e., the He convective core and the H burning shell.

\begin{figure}
\epsscale{1.18}
\plotone{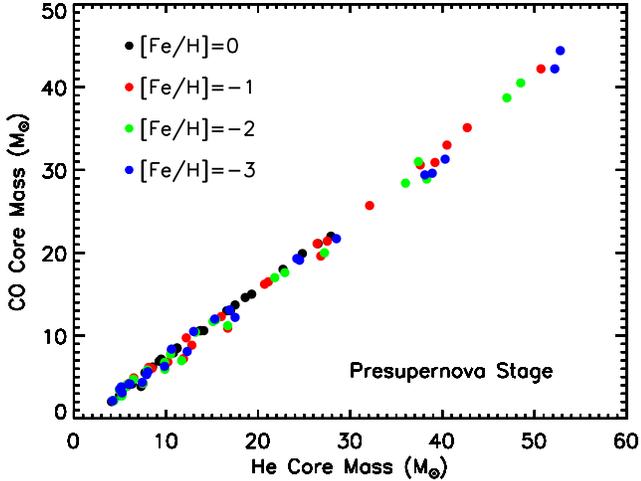}
\caption{\color{black} CO core mass as a function of the He core mass for all the computed models.}
\label{masshecocore}
\end{figure}

\begin{figure*}
\gridline{\fig{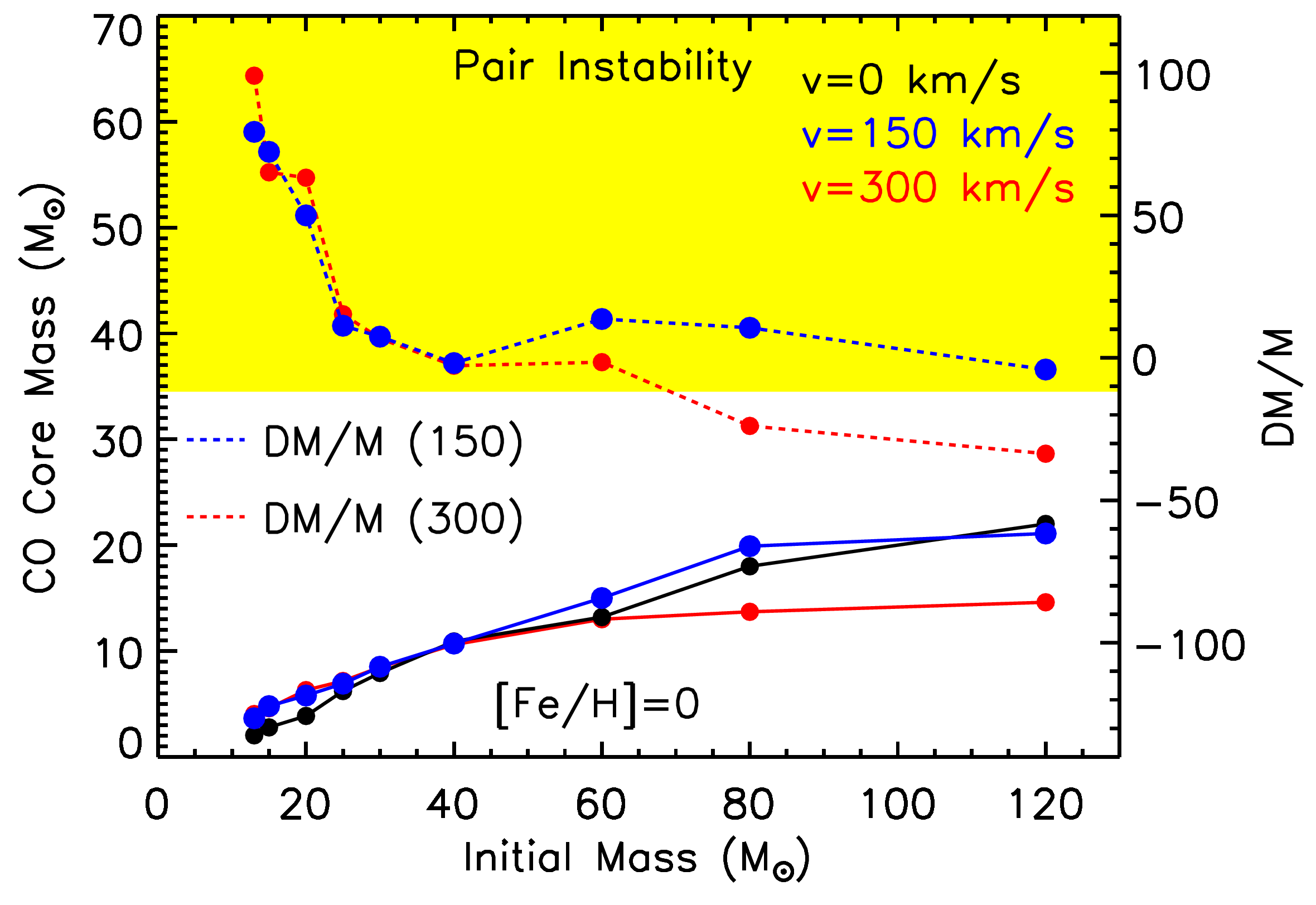}{0.5\textwidth}{(a)}
          \fig{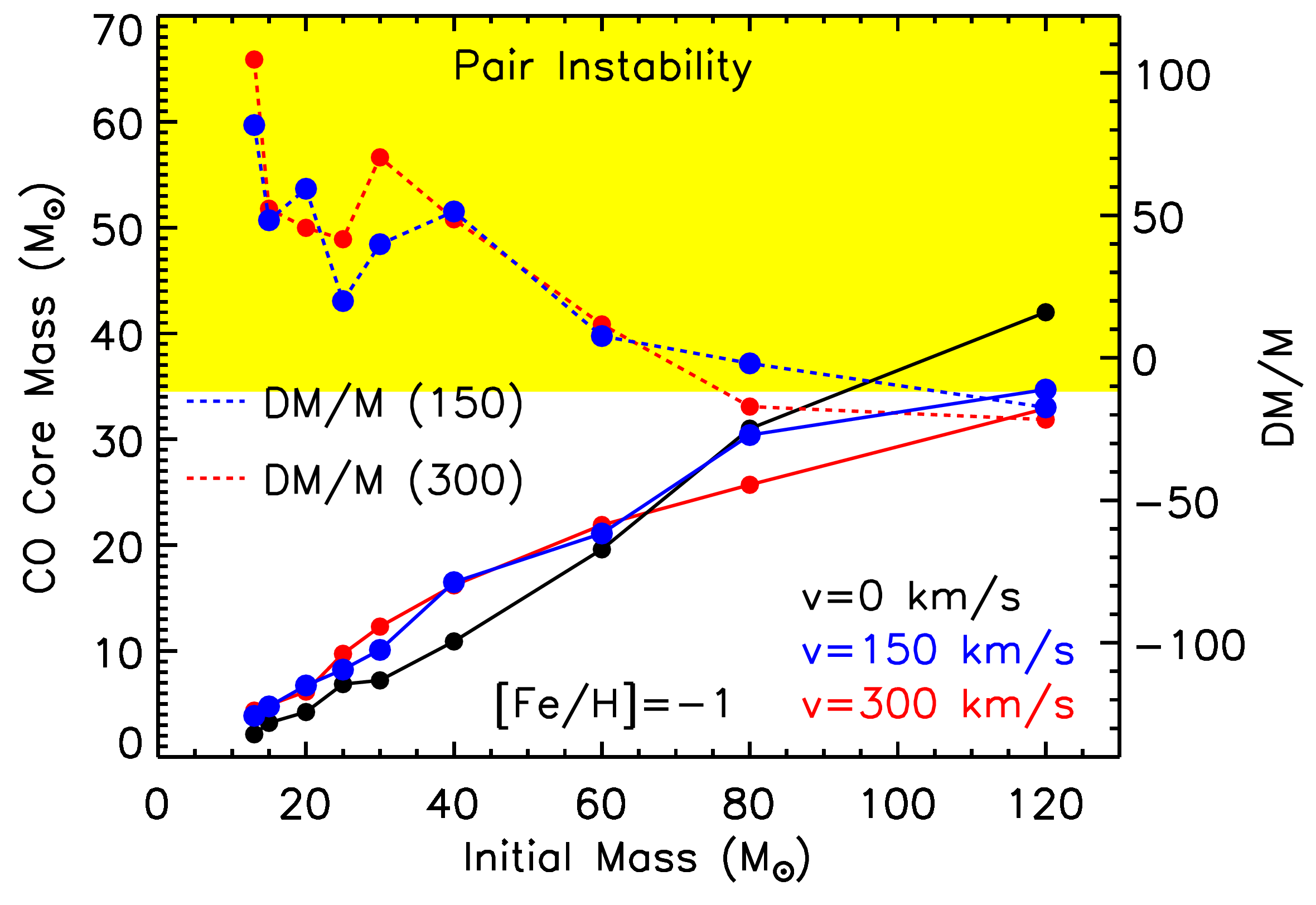}{0.5\textwidth}{(b)}
         }
\gridline{\fig{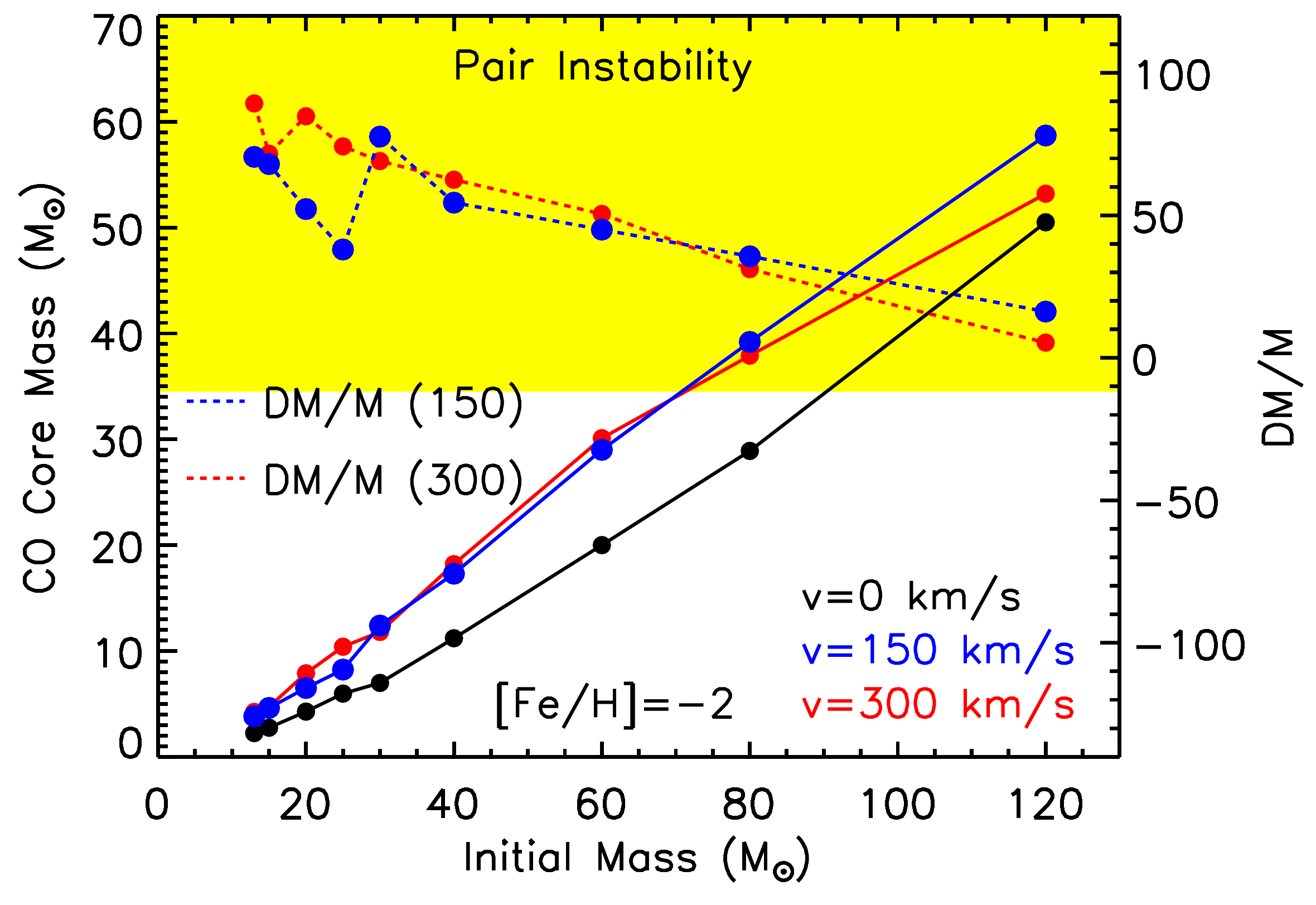}{0.5\textwidth}{(c)}
          \fig{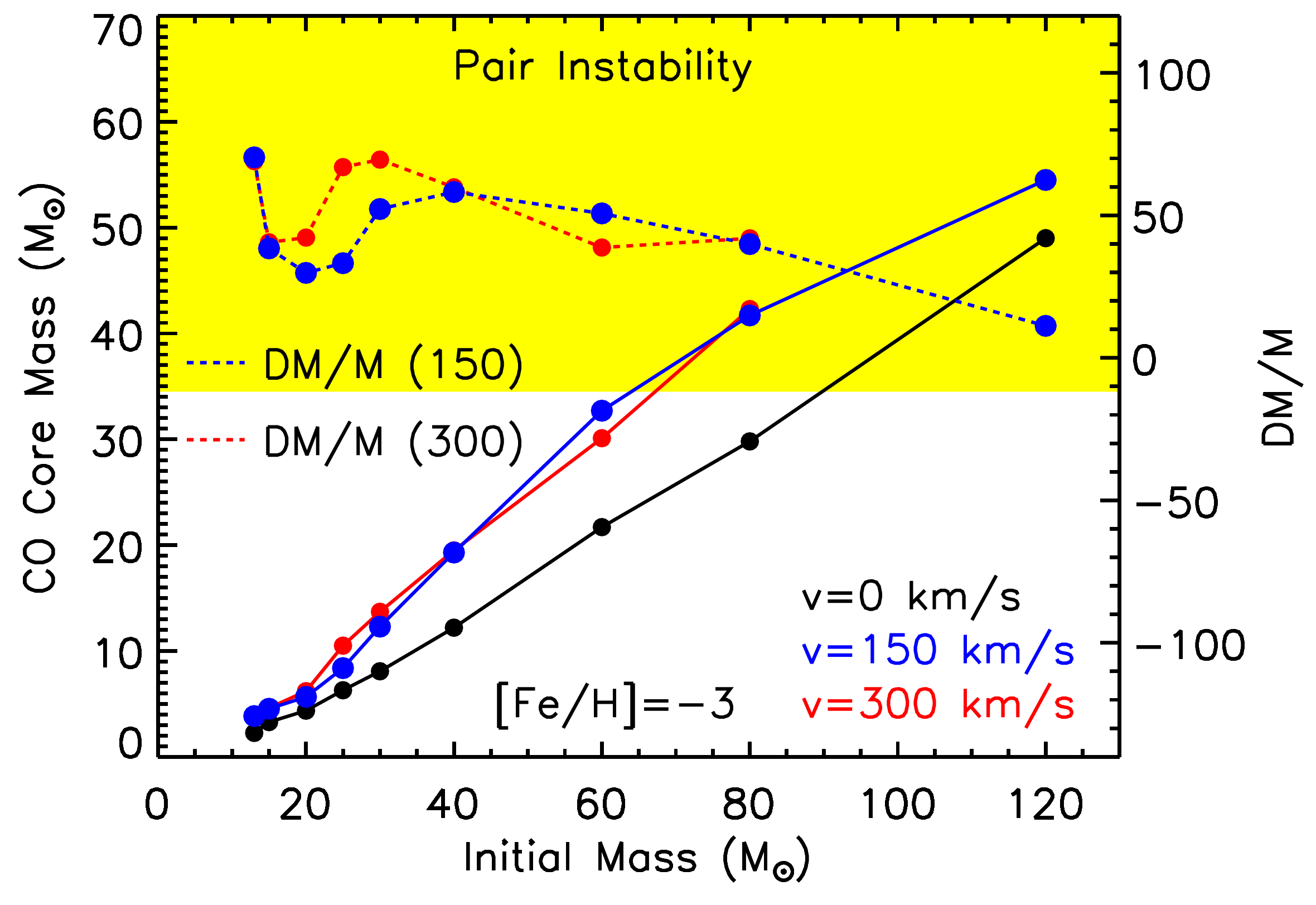}{0.5\textwidth}{(d)}
         }
\caption{The four panels show, for each initial metallicity, the $\rm M_{\rm CO}-M_{\rm INI}$ relation obtained for the non rotating (black) and the rotating cases, 150 km/s (blue) and 300 km/s (red), as solid lines (left Y axis). The dashes lines show the percentage difference {\color{black} (DM/M)} between rotating and non rotating $\rm M_{\rm CO}$ (right y axis).}         
\label{cocorehedepletion}         
\end{figure*}

Figure \ref{cocorehedepletion} shows the trend of the CO core mass (left y-axis) as a function of the initial mass for the three initial velocities as solid lines. Each of the four panels refers to a specific [Fe/H]. The dotted lines in the same Figure show the percentage difference between non rotating and rotating models, i.e. $\rm (M_{CO}^{rot}-M_{CO}^{norot})/M_{CO}^{norot}$ (right y-axis). The blue lines refer to $\rm v_{\rm ini}$=150 km/s while the red ones to $\rm v_{\rm ini}$=300 km/s. The two dotted lines clearly show that, in almost all cases, rotation increases the CO core mass and that the smaller the mass the larger the increase. Such a trend is due to the combination of basically two effects: a) the He burning lifetime scales inversely with the He core mass (and hence, in most cases, the initial mass) so that the smaller the He core mass the longer the secular instabilities may operate, and b) the trend with the mass is performed on models computed with a constant initial equatorial rotation velocity (see the comment at the beginning of section \ref{evol}). The general direct scaling of the CO core with the initial rotation velocity fails when the He core mass is eroded by mass loss. In these cases the core feels a smaller He core mass and the convective core shrinks accordingly. At [Fe/H]=0 and [Fe/H]=-1 the most massive rotating stars lose much more mass than their non rotating counterparts and this explains why in these cases the final CO core scales inversely with the initial rotation velocity. At solar metallicity, rotation roughly doubles the CO core mass of the less massive stars but its influence on $\rm M_{\rm CO}$ progressively reduces as the mass increases, becoming almost negligible for stars with $\rm M\leq 40~M_\odot$. In the more massive models with initial rotation velocity $\rm v=300~km/s$, the global effect of mass loss overcomes the effect of rotation, resulting in a reduction of CO core mass in rotating models up to $\sim 30-40~\%$ for the $\rm 120~M_\odot$. At metallicities lower than solar, mass loss reduces dramatically and therefore its effect on the CO core mass becomes progressively negligible. {\color{black} The spread in the CO core mass-initial mass relations evident in the four panels of Figure \ref{cocorehedepletion} (due the variation of both the initial metallicity and velocity) vanishes if the CO core mass is ranked as a function of the He core mass (Figure \ref{masshecocore}). The reason is obviously that the evolution of the star after core He depletion is essentially driven by the mass of the He core.  
Figure \ref{masshecocore} shows that CO core masses larger than 35 $\rm M_\odot$ correspond to He core masses larger than 45 $\rm M_\odot$, which is roughly the minimum mass entering the pulsation pair instability regime, as reported by \citep{HW02}. Let us note that this value has been adopted also by \citep{chaz12,yoon12,georgy17} for their works on pair instability supernovae.}

An obvious consequence of the increase of the CO core with rotation is that the minimum mass entering the pair instability regime reduces as the initial rotation velocity increases (yellow area in Figure \ref{cocorehedepletion}).

The diffusion of the chemicals between the He convective core and the H burning shell, induced by the rotation driven mixing, changes profoundly the chemical composition of the He core. In fact fresh $\rm ^{12}C$ synthesized in the core He burning is diffused up to the H burning shell where it is quickly converted not just into \nuk{N}{14} but in all the CNO nuclei, whose relative abundances are dictated by the temperature of the H shell. This means that all the nuclei involved in the CNO cycle are actually increased by this interplay. The fresh CNO nuclei, and in particular \nuk{N}{14}, plus {\it fresh He}, are brought back towards the center. The \nuk{N}{14} diffused back in the center is quickly converted into \nuk{Ne}{22} first and then into \nuk{Mg}{25,26}, becoming therefore an efficient primary neutron source. It must not be ignored that also the He brought towards the center plays an important role since it favors the conversion of $\rm ^{12}C$ into $\rm ^{16}O$, lowering therefore the final $\rm ^{12}C/^{16}O$ ratio in the CO core. 


\begin{figure}
\center{\includegraphics[scale=0.22]{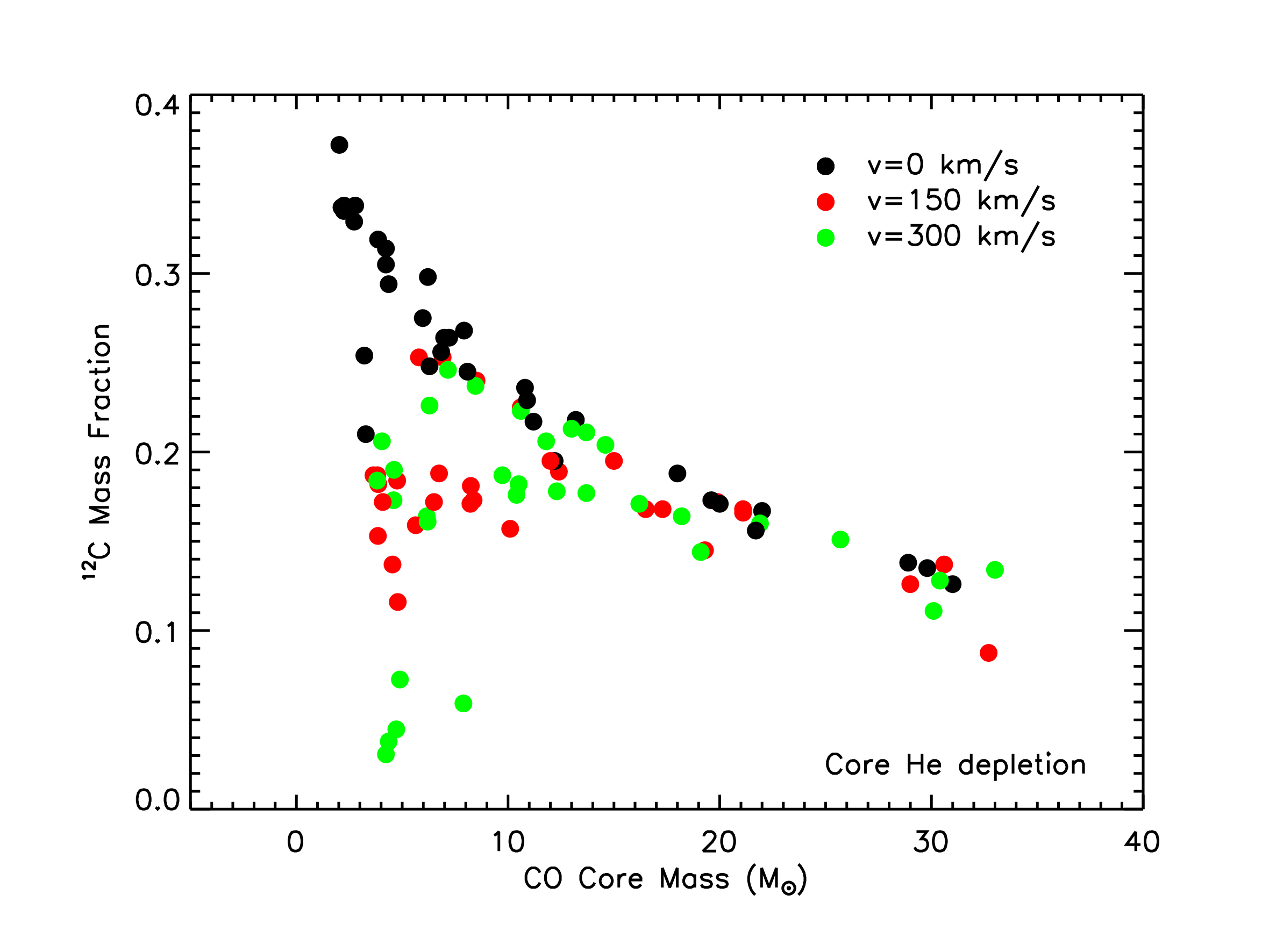}}
\caption{Central $\rm ^{12}C$ abundance at the core He exhaustion for all the models of our grid as a function of the CO core mass}
\label{c12mco}
\end{figure}

Figure \ref{c12mco} shows the central abundance of $\rm ^{12}C$ at the central He exhaustion for all the models of our grid as a function of the CO core mass. The effect of the rotational induced mixing on the amount of \nuk{C}{12} left by the He burning is readily visible in the large spread of abundances present at the lower CO core masses. The spread reduces progressively as the CO core mass increases because both the timescale over which the instabilities may operate reduces (because the He burning lifetime scales inversely with the CO core mass) and because we chose a constant initial rotation velocity as a function of the mass (see section \ref{evol}).


An easy way to quantify the efficiency of the rotation induced mixing in increasing the initial global abundance of the CNO nuclei, is that of defining the following quantity \citep[see, e.g.,][]{cls98}:
$$
\begin{array}{ll}

\chi(\rm{N,Mg})= & \rm{ \displaystyle {X(^{14}N) \over 14} + {X(^{18}F) \over 18} + {X(^{18}O) \over 18}} + \\
& \\
            &  \rm{ \displaystyle {X(^{22}Ne) \over 22} + {X(^{25}Mg) \over 25} + {X(^{26}Mg) \over 26}} 

\end{array}
$$

Such a total abundance (by number) remains constant in non rotating stars during the central He burning because the \nuk{N}{14} (the most abundant of the CNO nuclei at the central He ignition) left by the H burning may only evolve through the sequence that leads at most to the synthesis of \nuk{Mg}{25,26}, keeping unaltered the total sum (by number) of these nuclei all along the central He burning phase. 

\begin{figure*}
\epsscale{1.0}
\plotone{ntomg2.eps}
\caption{Scaling of $\Delta \chi(\rm{N,Mg})_{\rm He-burn}$ (see text for the definition of this parameter) with the initial metallicity for models of initial rotation velocity $\rm v=300~km/s$.}
\label{ntomg}
\end{figure*}

Vice versa, if fresh $\rm ^{14}N$ is brought in the convective core, the $\chi(\rm{N,Mg})$ parameter necessarily increases. Figure \ref{ntomg} shows the variation of $\chi(\rm{N,Mg})$ during the core He burning ($\Delta \chi(\rm{N,Mg})_{\rm He-burn}$) for all the models with initial rotation velocity $\rm v=300~km/s$ as a function of the initial metallicity. The figure clearly reveals the existence of two different behaviors: (1) in stars with mass $\rm M<40~M_\odot$, $\Delta \chi(\rm{N,Mg})_{\rm He-burn}$ does not show a monotonic dependence on the metallicity but the maximum variation remains confined within a factor of $\sim 7$ (the only exception being the $\rm 13~M_\odot$ models, left panel in Figure \ref{ntomg}), and (2) in stars with mass $\rm M\ge 40~M_\odot$, on the contrary, $\Delta \chi(\rm{N,Mg})_{\rm He-burn}$ scales inversely with the initial metallicity and shows a larger variation up to a factor of $\sim 20$ (right panel in Figure \ref{ntomg}). It is worth noting, however, that the range of metallicity over which the variation of $\Delta \chi(\rm{N,Mg})_{\rm He-burn}$ is evaluated, spans 3 orders of magnitude. The origin of the different behavior shown in Figure \ref{ntomg} is difficult to understand because the efficiency of the rotation driven mixing depends, in general, on the diffusion coefficients for the shear instabilities and the meridional circulation (see eqs. 2 and 5 in Chieffi \& Limongi 2013 and eq. 4.3 in Maeder \& Zahn 1998) that, in turn, depend in a very complex way on many interior properties of the He core among which, the local value of the angular velocity and its gradient, the mass and the extension of the He core, the mean molecular weight gradient, the difference between the radiative and the adiabatic gradients and so on. In addition to that, mass loss may play a crucial role in the primary $\rm ^{14}N$ production. In fact, if mass loss is efficient enough to remove all the H-rich envelope, it cancels out the engine (i.e. the H burning shell) needed to convert $\rm ^{12}C$ into $\rm ^{14}N$. Such an occurrence does not imply that the central value of $\chi(\rm{N,Mg})$ cannot increase anymore (because the radiative part of the He core is in any case still very $\rm ^{14}N$-rich) but simply that its increase is reduced with respect to the one the star would have without the removal of the H shell. Of course the earlier such a removal occurs the earlier the conversion of $\rm ^{12}C$ into $\rm ^{14}N$ stops. This phenomenon becomes progressively more important as the mass and metallicity increase because of the direct scaling of the efficiency of mass loss with these two quantities. As we have mentioned above, the primary $\rm ^{14}N$ production may have important consequences on the synthesis of the s-process nuclei at various metallicities - we will address this issue in section \ref{nfsprocess}. 

Once He is exhausted in the center, the CO core starts contracting again while the He burning shifts in a shell where a He convective shell forms. If mass loss did not erode the He core during the central He burning, or if the star did not rotate, the He convective shell forms above the He discontinuity left by the progressive advance of the convective core, in a region with a flat He profile where no products of the He burning are present.  Vice versa, if mass loss was efficient enough to erode the He core mass while the star was still in central He burning or rotation induced mixing modified the chemical composition of the intershell, the He convective shell forms in a region that has a pronounced He profile left over by either the receding He convective core (if the He core is reduced by mass loss) or determined by the continuous exchange of matter between the convective core and the H shell burning (in rotating models).

\subsection{Advanced nuclear burning stages}\label{advanced}
In the previous section we showed that all models that develop a CO core at core He depletion more massive than $\rm \sim 35~M_\odot$ enter the pair instability regime. This occurs when the adiabatic index $\Gamma_1$ drops below $\sim 4/3$ in a substantial fraction of the core that therefore becomes unstable. This instability is usually encountered in core O burning, but it may also occur either in core C burning or core Si burning, depending on the physical conditions of the interior of the star. Once the instability sets in, we stop the calculations; in the following we will describe the evolutionary properties of all the models that do not enter the pair instability regime.

Rotation (at least in the range of initial rotation velocities considered in this paper) does not affect significantly the evolutionary properties of a star beyond the He burning because: (1) the timescales over which the rotation instabilities operate are much longer than the advanced evolutionary timescales, so that they do not have enough time to operate; (2) the local gravity in the core dominates over the centrifugal force preserving therefore a quasi spherical shape (see CL13). Also the initial metallicity does not play any direct role in the physical evolution of a star in the advanced burning phases because all the relevant nuclear reactions involve primary nuclei. In other words the evolution of a star in the advanced burning phases is not linked any more to its initial mass, metallicity and initial rotation velocity, but it is controlled essentially by the mass of the CO core (that to all effects plays the role of the "total" mass) and the amount of $\rm ^{12}C$ left by the core He burning (the main fuel that powers both the C and the Ne burning). The role of the initial metallicity and rotation velocity influence the advanced burning only through their capability of modifying the mass of the CO core as well as the abundance of $\rm ^{12}C$ at the end of core He burning.

As already discussed in the previous section, there is a quite tight correlation between $\rm M_{\rm CO}$ and the $\rm ^{12}C$ left by core He burning for $\rm M_{\rm CO}\geq~15~M_\odot$ or so, the reasons being either the fact that in any case the $\rm ^{12}C/^{16}O$ ratio scales inversely with the size of the He core mass and also because the choice of an initial rotation velocity independent on the mass implies a progressive reduction of the effects induced by rotation as the initial mass increases. For these core masses, therefore, the only parameter that controls the advanced burning phases is the CO core mass.
Stars that develop $\rm M_{\rm CO}\leq~15~M_\odot$, vice versa, cannot be ranked just in term of the mass of the CO core but they constitute a family of models each of which depends on two {\it not} much correlated parameters.

Since the compactness of a star at the beginning of the collapse scales directly with the CO core mass and inversely with the $\rm ^{12}C$ abundance left by the He burning, and since both Figures \ref{cocorehedepletion} and \ref{c12mco} show that in most cases the  CO core mass scales inversely with the initial metallicity and directly with the initial rotation velocity, we can conclude that a reduction of the initial metallicity and/or an increase of the initial rotation velocity produce, in general, more compact cores.

\begin{figure}
\center{\includegraphics[scale=0.22]{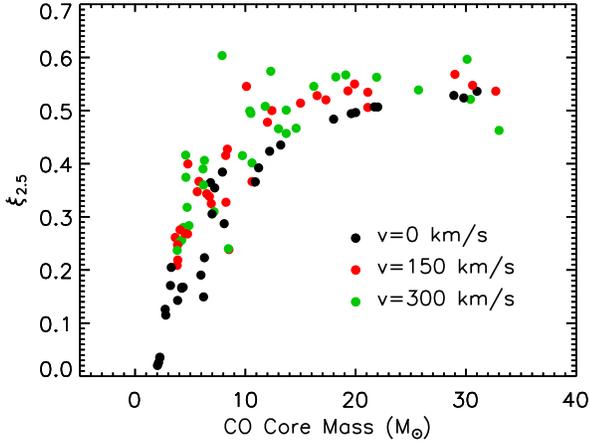}}
\caption{Plot of the compactness parameter $\rm \xi_{2.5}=M_i(M_\odot)/R_i(10^3~ km)\vert_{i=2.5~M_\odot}$ as a function of $\rm M_{\rm CO}$ for all models, rotating or not.}
\label{csimco}
\end{figure}

\begin{figure}
\center{\includegraphics[scale=0.22]{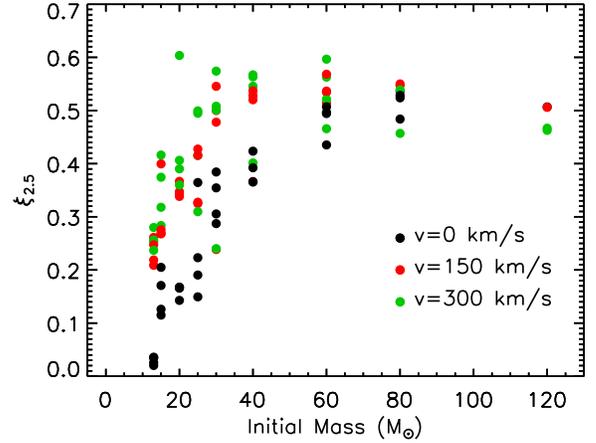}}
\caption{Same as Figure \ref{csimco} but now as a function of the initial mass.}
\label{csimass}
\end{figure}

{\color{black}
The compactness of a star at the presupernova stage is a fundamental property that influences the dynamics of the core collapse and the following explosion. In the past we usually discussed the compactness of a star in terms of the mass-radius relation (or density profile) at the presupernova stage. In the last years it has become very popular the use of a single or few parameters to describe such a property. For example \citet{oconnorott2011} define the compactness by means of the parameter $\rm \xi_{2.5}=M_i(M_\odot)/R_i(10^3~ km)\vert_{i=2.5~M_\odot}$, while \citet{ertl16} by means of the two parameters $M_4$ and $\mu_4$ (their equations 2 and 3). Just as an example we show in Figure \ref{csimco} how the $\rm \xi_{2.5}$ parameter, evaluated at the presupernova stage, scales with the CO core mass.} 
It is interesting to note that this Figure \ref{csimco} shows that a rather tight relation between the $\xi_{2.5}$ and the CO core mass.
Such a tight relation becomes much more scattered if the $\xi_{2.5}$ parameter is ranked as a function of the initial mass (Figure \ref{csimass}). 

While the dramatic shortening of the lifetimes of the advanced evolutionary stages, determined by the enormous neutrino energy losses due to the pair production, largely inhibits any transport of the angular momentum in the radiative zones, in the convective regions we assume that the angular momentum transport is so efficient that we impose a flat profile of the angular velocity. Note that, in all models of the present grid, no convective region crosses the outer edge of the CO core, therefore the total angular momentum stored in the CO core will remain constant up to the onset of the iron core collapse.


The surface properties (luminosity, radius and chemical composition) of a massive star at the time of the explosion is an important theoretical prediction since they are an "observable", in the sense that they control the way in which the supernova will appear to the observer (as a Type II, IIb, Ib etc). The black stars in Figure \ref{hrpresnallz} mark the positions of the models at the presupernova stage, in the HR diagram. Note that in some cases stars move, even substantially, from their position at core He exhaustion (red dots). Table \ref{propsn} summarizes the main properties of all the presupernova models of the present grid. The various columns refer to: the initial mass (column 1); the amount of H in the envelope (column 2); the amount of He in the envelope (column 3);  the mass of the iron core (column 4), defined as the mass coordinate corresponding to the sharp drop of the electron profile below $\sim 0.49$; the binding energy of the mass above the iron core (column 5); the compactness parameter $\xi_{2.5}$ (column 6); the angular momenta contained within the iron core (column 7), the CO core (column 8), the He core (column 9), the inner $\rm 1.5~M_\odot$ (column 10) and the inner $\rm 2.0~M_\odot$ (column 11); the expected SN type, according to the classification suggested by \citet{{hachietal12}}. 

On the basis of all the results discussed so far we predict that at solar metallicity and $\rm v_{\rm ini}=0$ stars less massive than 20\ \msun\ explode as RSGs while the more massive ones as WR stars. The relative numbers among the WNL, WNE, WNC and WC obviously depend on the total amount of mass lost: the present set of models predicts a much larger number of WNE and WC with respect to the WNL and no WNC. According to the supernova classification of \citet{hachietal12}, we predict that the maximum mass exploding as SNIIP ranges between $\rm 15<M_{\rm IIP}<20~M_\odot$, while stars more massive than $\rm M_{\rm Ib}\sim 20~M_\odot$ explode as SNIb. We cannot determine (for any initial metallicity) the range of masses that explode as Type IIb because they are intermediate between the SNIIP and the SNIb and our grid is not refined enough to allow us to determine their mass interval. At [Fe/H]=-1 mass loss reduces and hence the maximum mass that explodes as a RSG increases up to a mass in the range $\rm 60<M_{\rm IIP}<80~M_\odot$, the more massive stars reaching the collapse as either WNL or WC and exploding as a SNIb. At this metallicity the minimum mass that becomes a PISN ranges between $\rm 100<M_{\rm PISN}<120~M_\odot$. At $\rm [Fe/H]\leq -2$, mass loss reduces so much that no star loses a large fraction of its H rich mantle, and hence no star becomes WR. Stars less massive than $\rm \sim 25-30~M_\odot$ explode as RSGs while stars above this limiting value explode as BSGs. At these low metallicities all stars explode as SNIIP. The minimum mass exploding as a PISN ranges now between 60 and 80\ \msun. 

Rotation favors the redward evolution at the end of the central H burning and in many cases it also pushes stars above the Eddington luminosity, strongly enhancing mass loss. Hence on one side it increases the maximum mass that explodes as a RSG but it also lowers the minimum mass that enters the WR stage, squeezing therefore the range of masses that reach the core collapse as BSG. Note, however, that the interplay between rotation and mass loss has a complex and non monotonic impact on the expected number of the various types of WR supernovae and that, in any case, at variance with the non rotating case, it allows some progenitors to explode as WNC stars. As a general trend, at any given initial metallicity both $\rm M_{\rm IIP}$ and $\rm M_{\rm Ib}$ reduce as the initial rotation velocity increases. Also the 
minimum mass that explodes as PISN decreases as the rotation velocity increases because of the direct scaling of the CO core mass with the rotation velocity (see above).

\section{The yields}\label{yields}
The chemical composition of the ejecta of each stellar model after the explosion has been computed as described in CL13. Since this approach is not based on first principles, a proper calibration of the explosion is necessary. Such a calibration is obtained by requiring the fit to some observable, typically the amount of \nuk{Ni}{56} ejected and/or the final kinetic energy of the ejecta. The choice of the calibration is crucial because it directly affects the location of the mass cut and hence the yields of many nuclei (basically those produced by the explosive burning). In our previous set of models (CL13), the yields were computed by assuming that all stars eject 0.1\ \msun\ of \nuk{Ni}{56}. In this paper, vice versa, we computed three different set of yields obtained for different choices of the mass cut in order to show which is the consequence of one or another choice (other calibrations may be provided upon request). The first set of yields (set F) is obtained as in our previous paper, i.e. assuming that the stars  eject a fixed amount of  \nuk{Ni}{56} - 0.07 \msun\ in this case.  The second one (set M) is obtained by adopting the mixing and fall back (MFB hereinafter) scheme \citep{UN02}: for each star the inner border of the mixed region is fixed by requiring that [Ni/Fe]=0.2 and the outer one is fixed at the base of the O burning shell; the mass cut is then chosen by assuming once again that each star ejects 0.07 \msun\ of \nuk{Ni}{56}. The third set (set R), which is our recommended one, is obtained by assuming that all the stars in the range 13-25\ \msun\ behave like those in set M, while those more massive than 25\ \msun\ fully collapse to a black hole and therefore their yields include only the stellar wind. Note that also the yields of the stellar models that become PISN contain only the wind because we could not follow their evolution once they entered the pair instability regime (see above).

\begin{figure*}
\epsscale{0.8}
\plotone{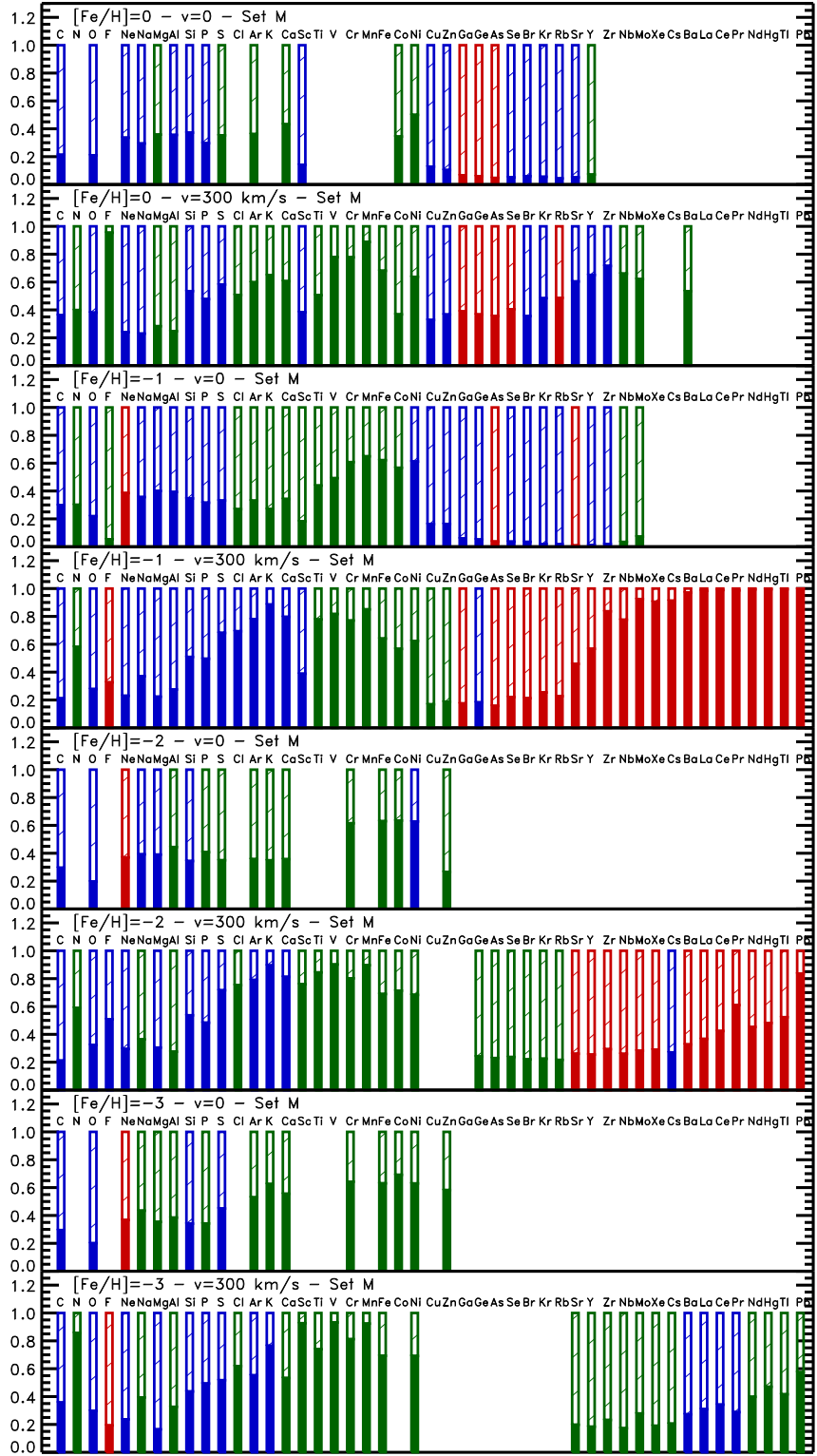}
\caption{Relative contributions of the stars in the ranges $\rm 13 \leq M/M_\odot \leq 25$ (solid bars) and $\rm 25 < M/M_\odot \leq 120$ (dashed bars) to the total yields integrated over a standard Salpeter IMF (x=1.35). The elements under produced with respect to Oxygen by a factor between 0.25 and 0.5 are marked green, those co produced (i.e. in the interval 0.5 and 2) are blue while those overproduced (i.e. more than a factor of 2) are red. The missing bars mean that the corresponding elements are largely underproduced (less then 0.25) with respect to O.}
\label{fracy}
\end{figure*}

Let us start the analysis of the yields by showing the relative contributions of the stars in the ranges $\rm 13 \leq M/M_\odot \leq 25$ (LINT) and $\rm 25 < M/M_\odot \leq 120$ (UINT) to the total yields integrated over a standard Salpeter IMF (i.e. having the slope x=1.35). Figure \ref{fracy} shows, for set M, the relative contributions of the LINT (solid bars) and UINT (hatched bars) groups to the yield of each element. The color coding identifies the elements which are under-, co- and over- produced with respect to O: those moderately under-produced with respect to O ($\rm -0.6\leq[X/O]<-0.3$) are marked green, the ones more or less co-produced with O ($\rm -0.3\leq[X/O]\leq0.3$) are in blue while those over-produced with respect to O ($\rm [X/O]>0.3$) are red. The elements which do not have any bar are those severely under produced with respect to O and therefore the ones for which it is irrelevant to determine which mass interval contributes most to their yields. O is adopted as the leading element because it is either the most abundant element after H and He and it is also almost exclusively produced by massive stars.

Let us remind that if we define $\rm PF_X$ as the production factor of a given element "X" averaged over a Salpeter IMF between 13 and $\rm 120~M_\odot$, the [X/O] ratios are simply the $\rm Log_{10}(PF_X)$ vertically shifted by $\rm Log_{10}(PF_O)$ (with a minor correction due to the enhancement of the alpha elements) and hence that the [X/O] basically share the same properties of the PFs: in particular they directly show which elements are produced by massive stars and which don't (clearly all the elements that have [X/O] close to zero are co-produced with O and hence are produced mainly by massive stars). Obviously a flat distribution of the PFs produced by a generation of stars would imply that their ejecta preserve the relative scaling of the initial composition.

In absence of rotation (first, third, fifth and seventh panel in Figure \ref{fracy}), the LINT group contributes to the synthesis of the elements C to Ni between $\sim 20-50\%$ in most cases. Elements between the Fe peak and the first neutron closure shell are produced preferentially by the UINT group down to [Fe/H]=-1, while they are basically not produced by massive stars at lower metallicities. No element beyond the first neutron closure shell is produced by non rotating massive stars. 

\begin{figure}[ht]
\center{\includegraphics[scale=0.18]{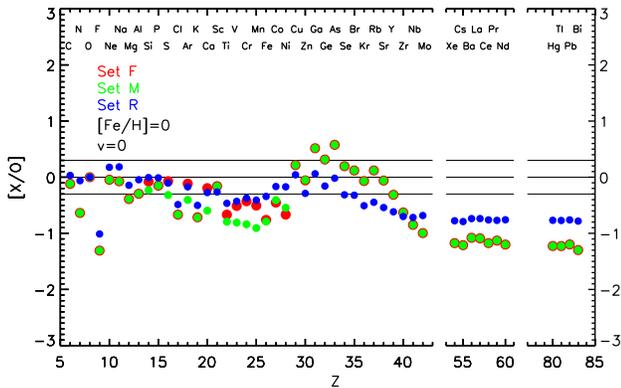}}
\caption{Comparison among the [X/O] obtained for the three different sets of yields, F (red), M (green) and R (blue), non rotating solar metallicity stars. See text for the definition of the three sets. }
\label{Ya0}
\end{figure}

\begin{figure}[ht]
\center{\includegraphics[scale=0.18]{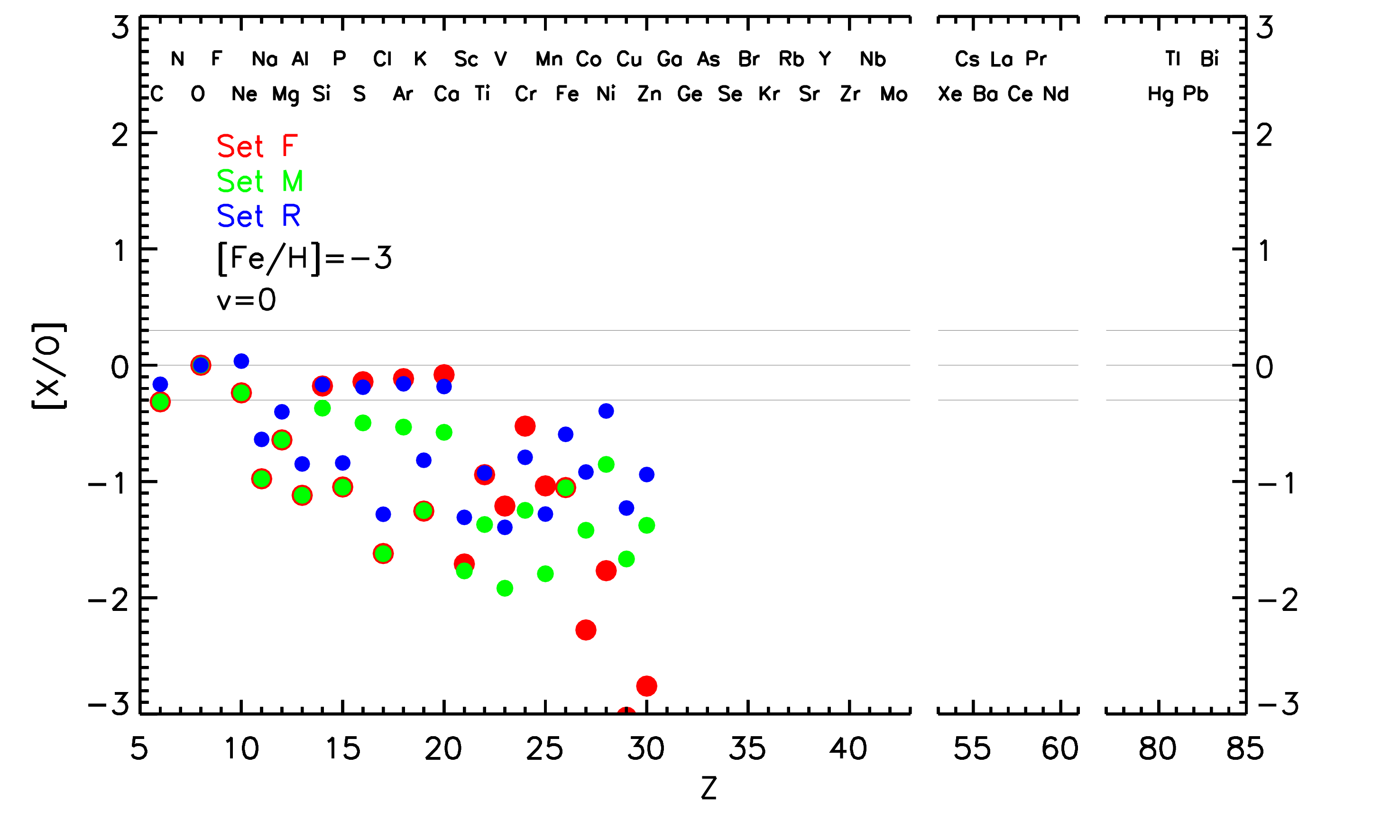}}
\caption{Same as Figure \ref{Ya0} but for [Fe/H]=-3}
\label{Yd0}
\end{figure}

Figure \ref{Ya0} shows a comparison among the [X/O] obtained in the three set, F (red) - M (green) - R (blue), for the solar metallicity non rotating models. This Figure shows that in a scenario in which all the stars eject 0.07\ \msun\ of \nuk{Ni}{56} (set F - red dots), the intermediate mass elements, i.e. elements from O to Ca, are basically co-produced with O and hence produced by massive stars. F, Mg, Cl and K are the only ones significantly under produced with respect to O but, while F may be probably produced by the intermediate mass stars, Mg, Cl and K could in principle constitute a problem because no other source for their production has been identified so far. The block of the Fe peak nuclei (Sc-Zn) are significantly (and correctly) under produced with respect to O, since they are mainly produced by the SN Ia. The [X/O] of the elements Ga to Zr, the so called weak component, are more or less co-produced with O, the only exceptions being Ga and As that are slightly overproduced, this result confirming the general belief that massive stars contribute significantly to  the synthesis of these elements. The [X/O] of all the elements heavier than Zr drop quickly well below zero, which means that they are not produced by massive stars. 

A comparison between set F (red) and M (green) in Figure \ref{Ya0} shows the effect of the MFB, at least in the framework of the parameters (inner and outer borders of the mixed zone and final mass cut) described above. The differences between the two set are obviously confined to the elements mainly produced in the more internal zones by the explosive burning. In particular the [X/O] of Si (\nuk{Si}{28}), S (\nuk{S}{32}), Ar (\nuk{Ar}{36}), Ca (\nuk{Ca}{40}), Ti (\nuk{Cr}{48}), V (\nuk{Mn}{51}), Cr (\nuk{Fe}{52}) and Mn (\nuk{Co}{55}) decrease in set M. By the way, in parenthesis we report either the most abundant isotope of that element if it is synthesized directly, or its parent isotope if it is fed from the decay of another isotope. The reason for such a decrease is that these elements are mostly synthesized by the incomplete explosive Si burning and/or explosive O burning, in regions more external than those where \nuk{Ni}{56} is produced. Therefore, the MFB mechanism spread them back in a region that remains locked in the remnant. Ni, on the contrary, shows an opposite behavior, its [X/O] slightly increasing in set M. The reason is that Ni (\nuk{Ni}{58}) is basically produced by the complete explosive Si burning, in zones more internal than those where \nuk{Ni}{56} is synthesized, and in this case the effect of the MFB is to mix some \nuk{Ni}{58} from the region where it would be locked in the remnant into the one that is then ejected, raising therefore its yield. It is worth noting that the odd-even effect, usually interpreted in terms of initial metallicity, is actually significantly affected by the possible presence of MFB. In fact, while the yields of the $\alpha$ elements Si, S, Ar and Ca are lowered by the MFB (see above), the ones of P, Cl (\nuk{Cl}{35}) and K (\nuk{K}{39}) remain basically constant because they are mainly synthesized (at solar metallicity) in the C convective shell which is not affected by the MFB. For sake of completeness let us remind that the other, much less abundant, isotopes of Cl and K, namely \nuk{Cl}{37}, \nuk{K}{40} and \nuk{K}{41}, are mainly produced by the He burning. Sc and Co deserve a specific comment because they are usually considered elements produced by the explosive burning (and hence they should be affected by the MFB) while, on the contrary, both elements have an important (secondary) contribution from the hydrostatic burning. Sc is produced either as \nuk{Ti}{45} by the explosive O burning, and as \nuk{Sc}{45} and \nuk{Ca}{45} in the shell C burning (plus a minor contribution from the He shell) via neutron captures. Also Co has a double production site, being synthesized either as \nuk{Cu}{59} by the complete explosive Si burning and directly as \nuk{Co}{59} by the hydrostatic He and C burning (again via n captures). At solar metallicity, the presence of a strong neutron flux favors the hydrostatic production (in the C and He shells) of both these nuclei with respect to the explosive one. Since the C and He burning shells are not affected by the MFB, this explains why these elements are not significantly affected by the presence or not of the MFB. A comparison between set M (green) and R (blue) shows the influence of the stars more massive than  25\ \msun\ on the distribution of the [X/O]. The most evident effect of the choice that all the stars more massive than 25\ \msun\ fully collapse to a remnant contributing to the yields only through the wind (set R) is that the overproduction of some elements of the weak component disappears. The reason is clearly due to the fact that these nuclei, mainly produced in the C convective shell of the more massive stars, remain locked in the remnant in this case. Vice versa most of the intermediate mass elements have [X/O] systematically higher (and closer to 0) than those of set "M" because the more massive stars contribute more to the yield of the O than to the yields of the intermediate mass nuclei.

Figure \ref{Yd0} is the analogous of Figure \ref{Ya0} but for [Fe/H]=-3. As expected, set F (red) shows a quite flat distribution of the even nuclei between C and Ca because of their primary origin and a consistent odd-even effect due to the low metallicity; elements beyond the Fe peak are not produced at all because of the negligible neutron flux present at this low metallicity \citep{cl04,prantzos90,raiteri92,raiteri93}. The effect of the MFB (set M, green color) on the [X/O] is qualitatively similar to that seen at solar metallicity. In this case, however, we find also a substantial increase of the abundances of Co, Ni (\nuk{Ni}{58}), Cu (\nuk{Cu}{63}) and Zn (\nuk{Ge}{64}). The reason is that these nuclei have a composite production, either explosive and hydrostatic. The hydrostatic component is mainly due to the core He burning, while the explosive one is due to the complete explosive Si burning (as it has already discussed above in the case of Co). As the initial metallicity reduces, the hydrostatic component, basically of secondary origin, progressively vanishes leaving therefore full visibility to the explosive one, which is mainly of primary origin. The effect of the MFB on these elements is therefore similar to the one already discussed for Ni at [Fe/H]=0. The lack of the contribution of the more massive stars (set R, blue) systematically raises, also at sub solar metallicity, the [X/O] of all the elements with respect to those of set M because of the steeper scaling of the yield of O with the mass than that of most of the other elements.

\begin{figure}[ht]
\gridline{\fig{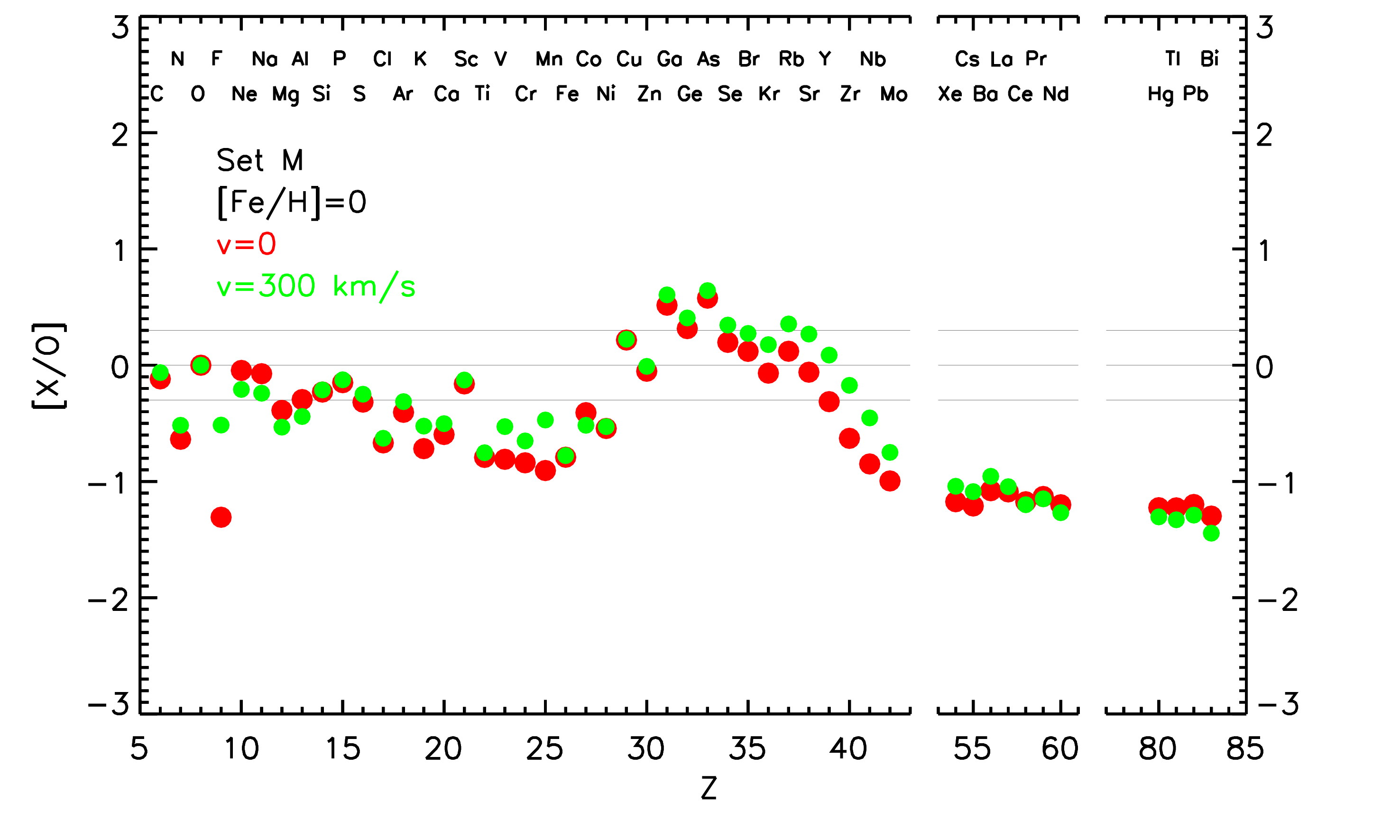}{0.5\textwidth}{}}
\gridline{\fig{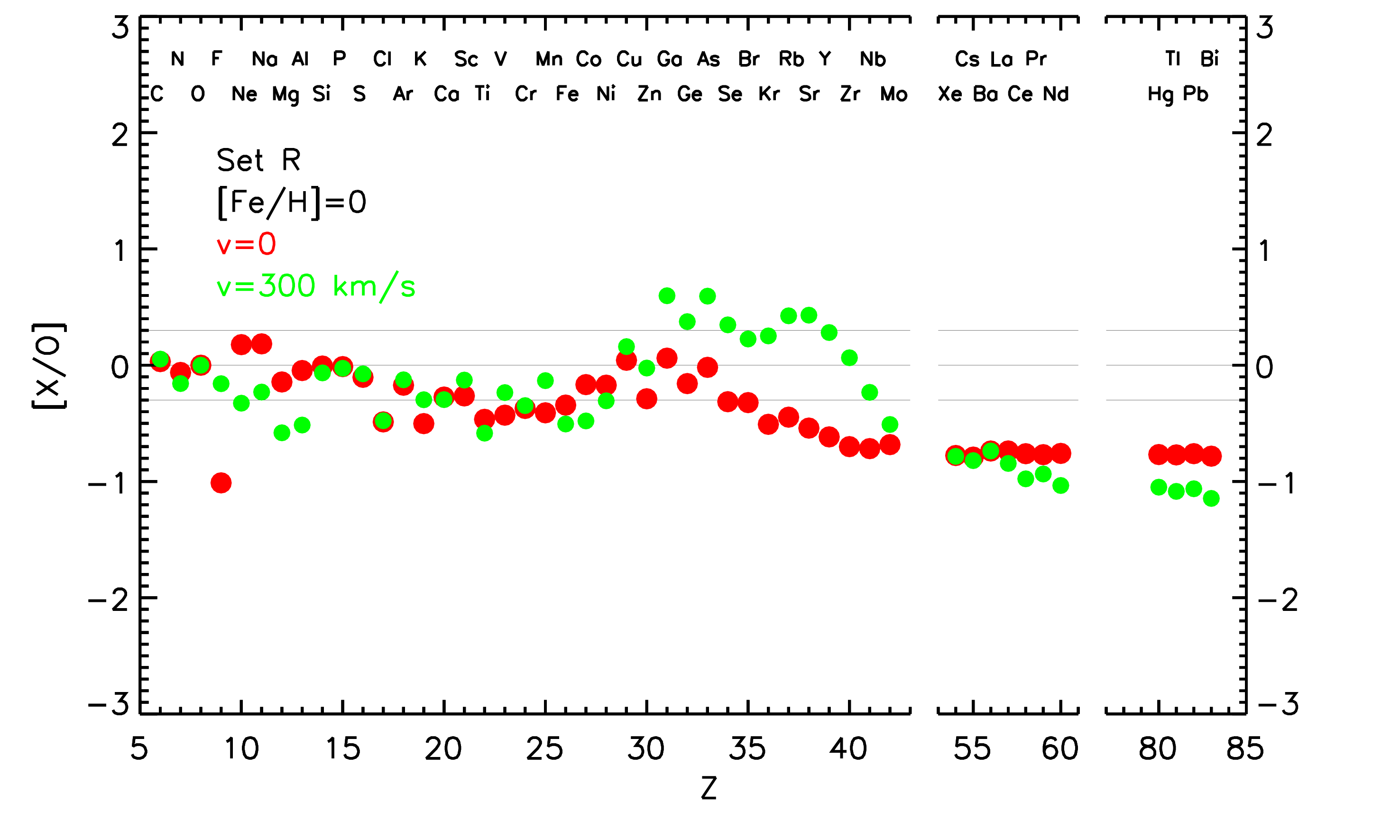}{0.5\textwidth}{}}
\caption{Upper panel: comparison between the [X/O] obtained for non rotating (red dots) and rotating (300 km/s, green dots) models, for set M and solar metallicity; lower panel same as the upper panel but for set R}
\label{pfa300}
\end{figure}

\begin{figure}[ht]
\gridline{\fig{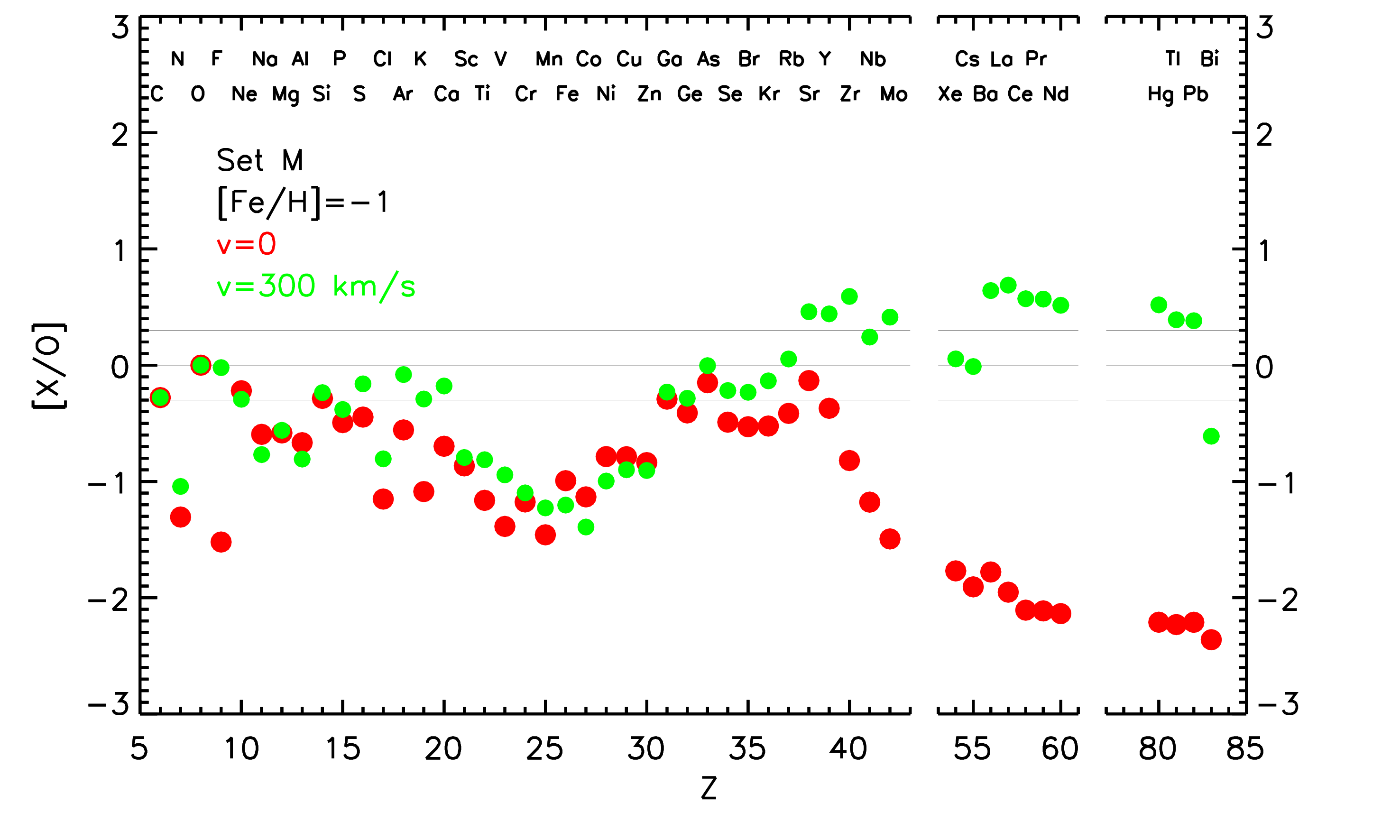}{0.5\textwidth}{}}
\gridline{\fig{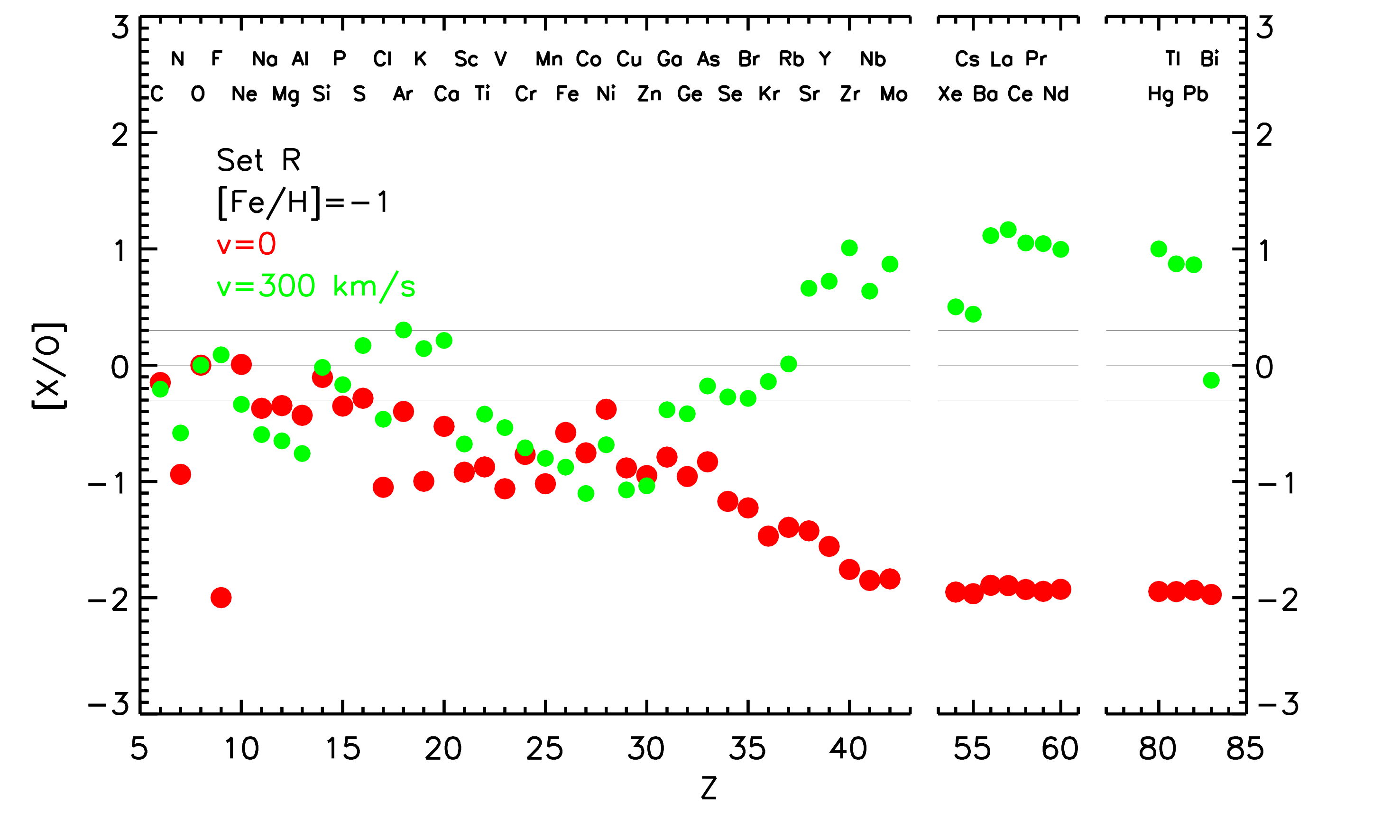}{0.5\textwidth}{}}
\caption{Same as Figure \ref{pfa300} but for [Fe/H]=-1}
\label{pfb300}
\end{figure}

\begin{figure}[ht]
\gridline{\fig{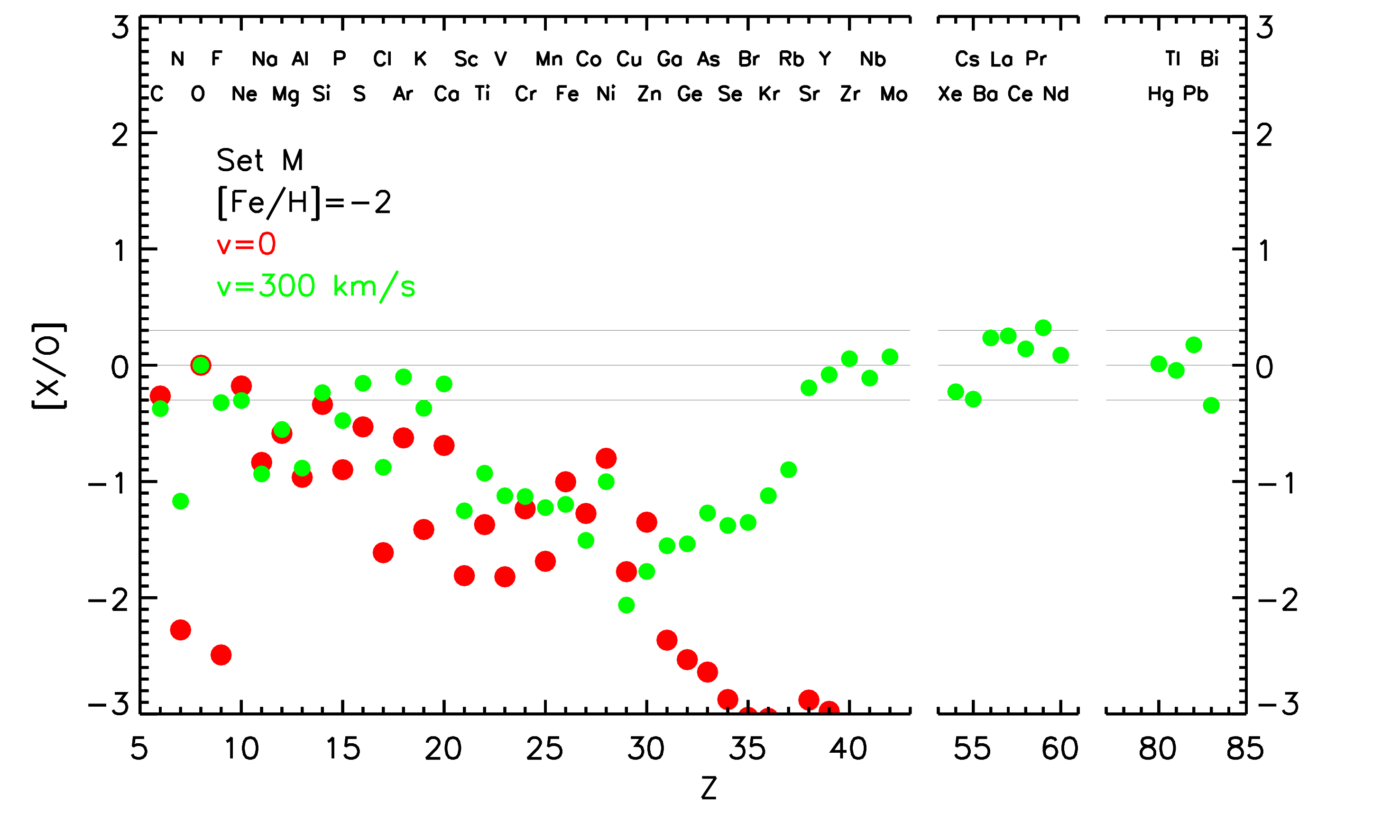}{0.5\textwidth}{}}
\gridline{\fig{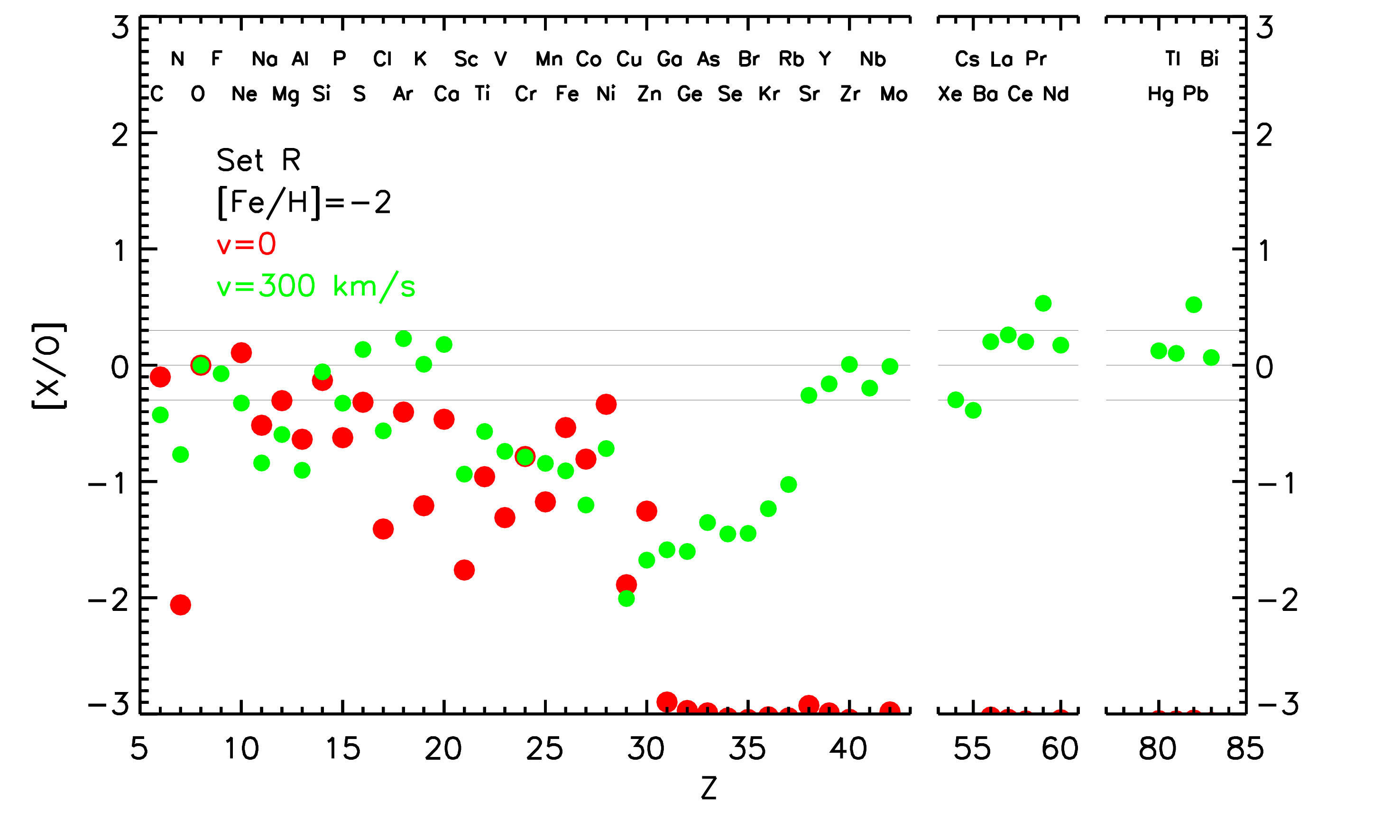}{0.5\textwidth}{}}
\caption{Same as Figure \ref{pfa300} but for [Fe/H]=-2}
\label{pfc300}
\end{figure}

\begin{figure}[ht]
\gridline{\fig{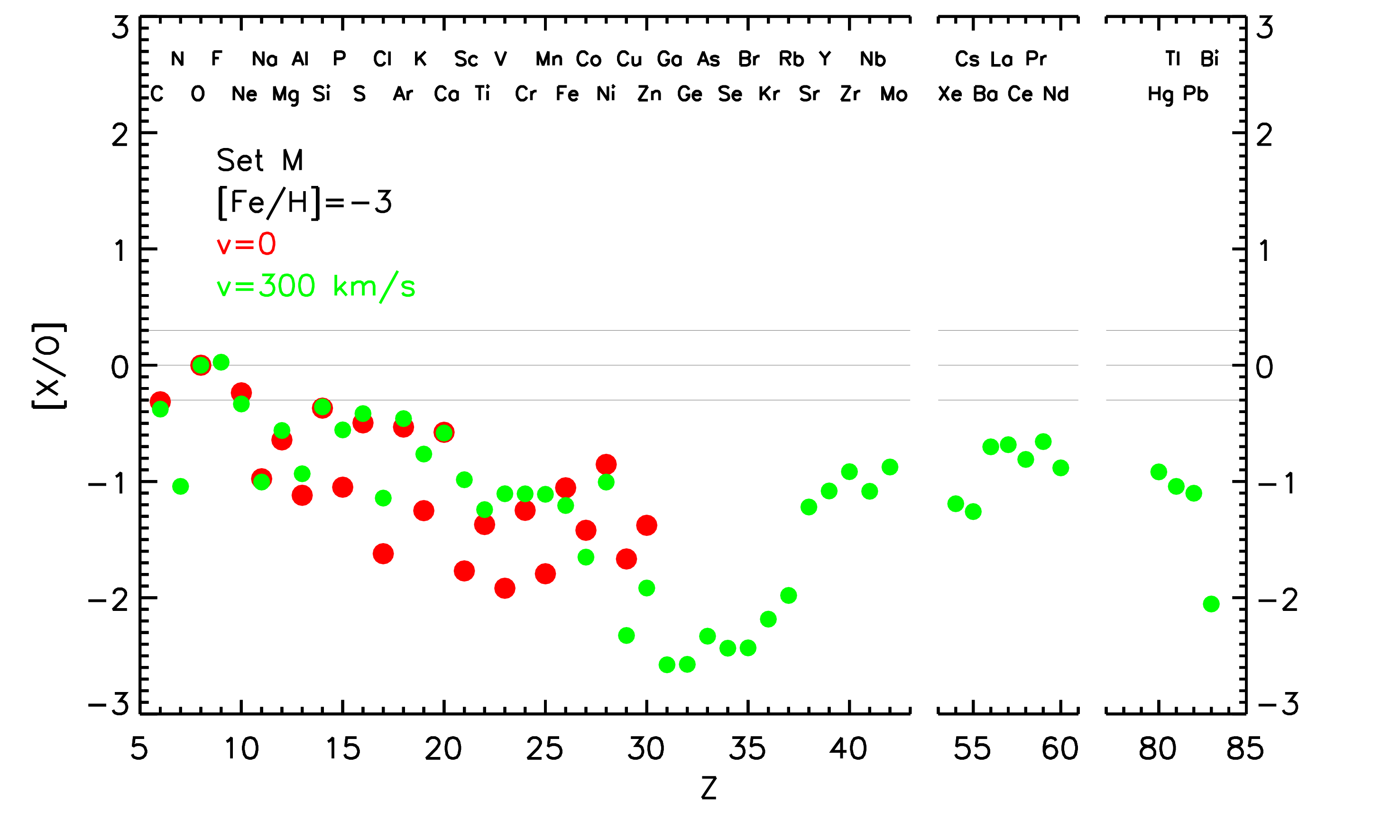}{0.5\textwidth}{}}
\gridline{\fig{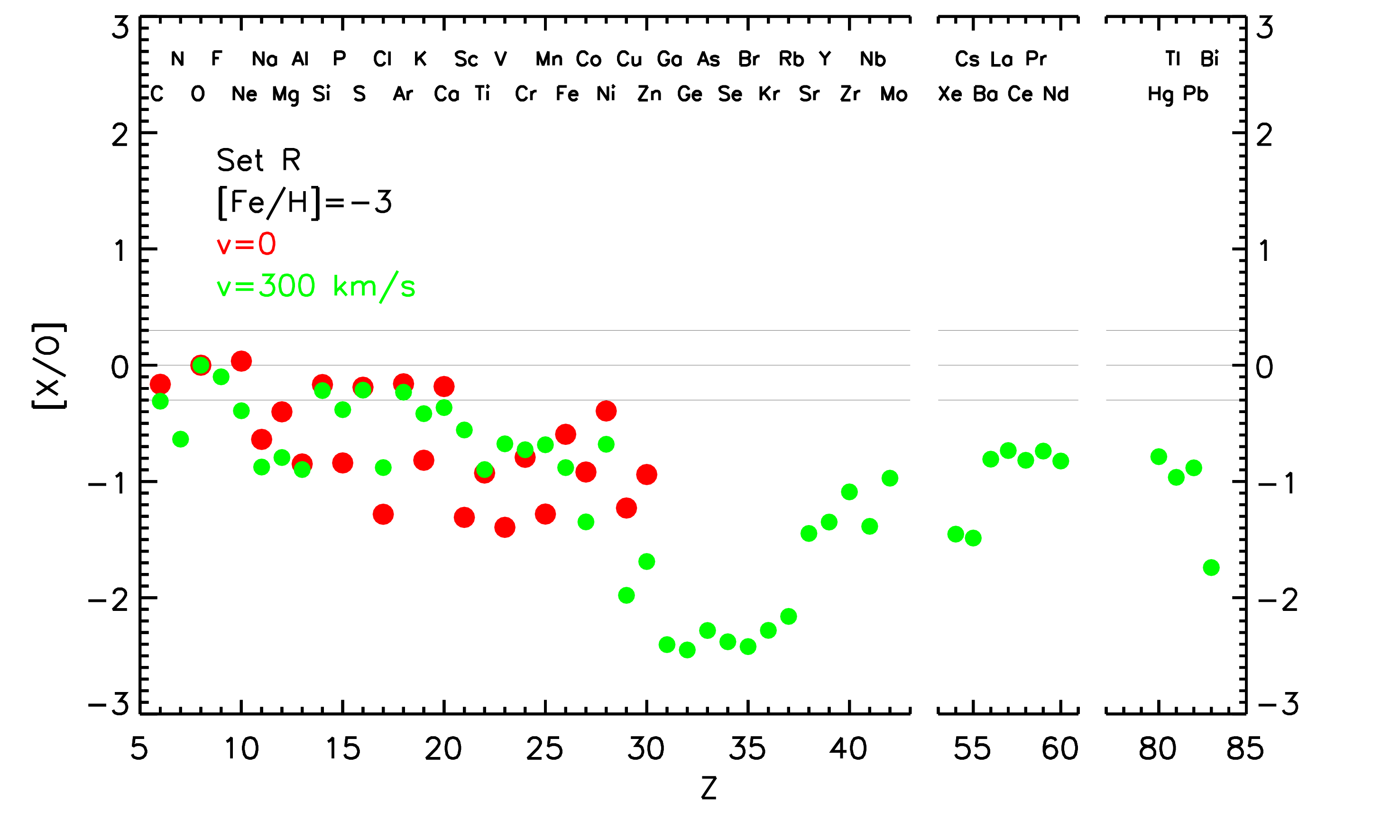}{0.5\textwidth}{}}
\caption{Same as Figure \ref{pfa300} but for [Fe/H]=-3}
\label{pfd300}
\end{figure}

Rotation changes the scenario depicted above. The first thing worth noting is that the contribution of the LINT group to the total yield of most elements increases in rotating models at any metallicity. This is clearly visible in Figure \ref{fracy} by comparing the four pairs of panels. This result is the consequence of the fact that we chose to adopt the same initial rotation velocity for all the masses of each given generation of stars. In fact, as it has already {\color{black} been} mentioned above, since the effects of rotation scale directly with $\rm v/v_{\rm crit}$ and this parameter scales inversely with the initial mass (Figure \ref{osuocritini300}), the effects of rotation scale inversely with the initial mass. If we now consider that rotating models (on average) have larger yields than their non rotating counterparts (either because they develop larger He core masses and because of the formation of a primary neutron source), it turns out that the LINT group is more affected by rotation than the UINT group and therefore it tends to prevail in rotating models. 

Keeping this in mind, Figure \ref{pfa300} shows in the upper panel a comparison between the [X/O] obtained for non rotating (red dots) and rotating (300 km/s, green dots) models, for set M and solar metallicity. In this case most of the [X/O] show a very modest dependence on the rotation velocity, with the exception of F and the weak component that are enhanced because of the presence of a primary source of neutrons. It is worth noting that Ne, Na, Mg and Al tend to have systematically smaller [X/O] in rotating models because their yield depend on the amount of C left by the He burning that, in turn, scales inversely with the rotation velocity (see Figure \ref{c12mco} and the discussion above). The lower panel in Figure \ref{pfa300} shows the same comparison but for set R. It is quite evident that now the differences between rotating and non rotating models is larger but concerns basically the weak component plus the group formed by Ne, Na, Mg and Al. The reason is simply that the UINT group contributes more to the yields in the non rotating models than in the rotating ones, so that their substantial removal in set R has a major effect in the non rotating models while is quite marginal in the rotating ones.

Figure \ref{pfb300} shows the same comparisons shown in Figure \ref{pfa300} but for [Fe/H]=-1. The striking difference between the solar and the sub solar metallicity is the large production of F and heavy nuclei up to Pb in the rotating models. This result is due to the large increase of the neutron/seed ratio {\color{black} (i.e., the ration between the number densities of neutrons and $\rm Fe$ nuclei)} in rotating models as the initial metallicity decreases. In fact, while the primary neutron source - produced by the conversion of fresh \nuk{C}{12} in \nuk{N}{14} first and \nuk{Ne}{22} later - increases or remains roughly constant as the initial metallicity reduces (this is clearly visible in Figure \ref{ntomg} that shows the trend with the metallicity of the primary \nuk{N}{14} for models rotating at 300 km/s), the global abundance of the seed nuclei (basically Fe) obviously scales directly with the initial metallicity. This large increase of the neutron/seed ratio as the initial metallicity decreases favors a substantial flux of matter through the bottlenecks corresponding to the two neutron magic numbers N=50 and N=82 and hence a consistent flux of matter up to Pb. The capability of a large neutron/seed ratio to push matter up to Pb {\color{black} has been shown by \citet[Fig.7-22]{clayton68} and has been studied for the first time by \citet{Ga98} in low metallicity low mass AGB stars}. Both panels in Figure \ref{pfb300} show a large production of the heavier nuclei, set R providing again a larger overproduction because of the dominant contribution of the LINT interval to the yields of the various elements (see the forth panel in Figure \ref{fracy}). A reduction of the initial metallicity below [Fe/H]=-1 does not lead to a further increase of the [X/O] of the heavier nuclei but actually to their progressive reduction. Figures \ref{pfc300} and \ref{pfd300} show in fact that the very high [X/O] obtained at [Fe/H]=-1 of the elements beyond Zn, reduce progressively as the metallicity drops. There are two con causes that contribute to such a trend. The first one is that the increase of the abundances of the heaviest nuclei is obtained at the expenses of the global abundance of the Fe peak nuclei that therefore constitutes the maximum buffer that can be used to build up the heavier nuclei (independently on the neutron flux). In other words the lower the metallicity the lower the total amount of matter that can be used to build up the heaviest nuclei. The second reason is that O mildly increases as the metallicity decreases and therefore this helps in pull down the [X/O] ratios.

\subsection{N, F and s-processes}\label{nfsprocess}
\begin{figure}[ht]
\gridline{\fig{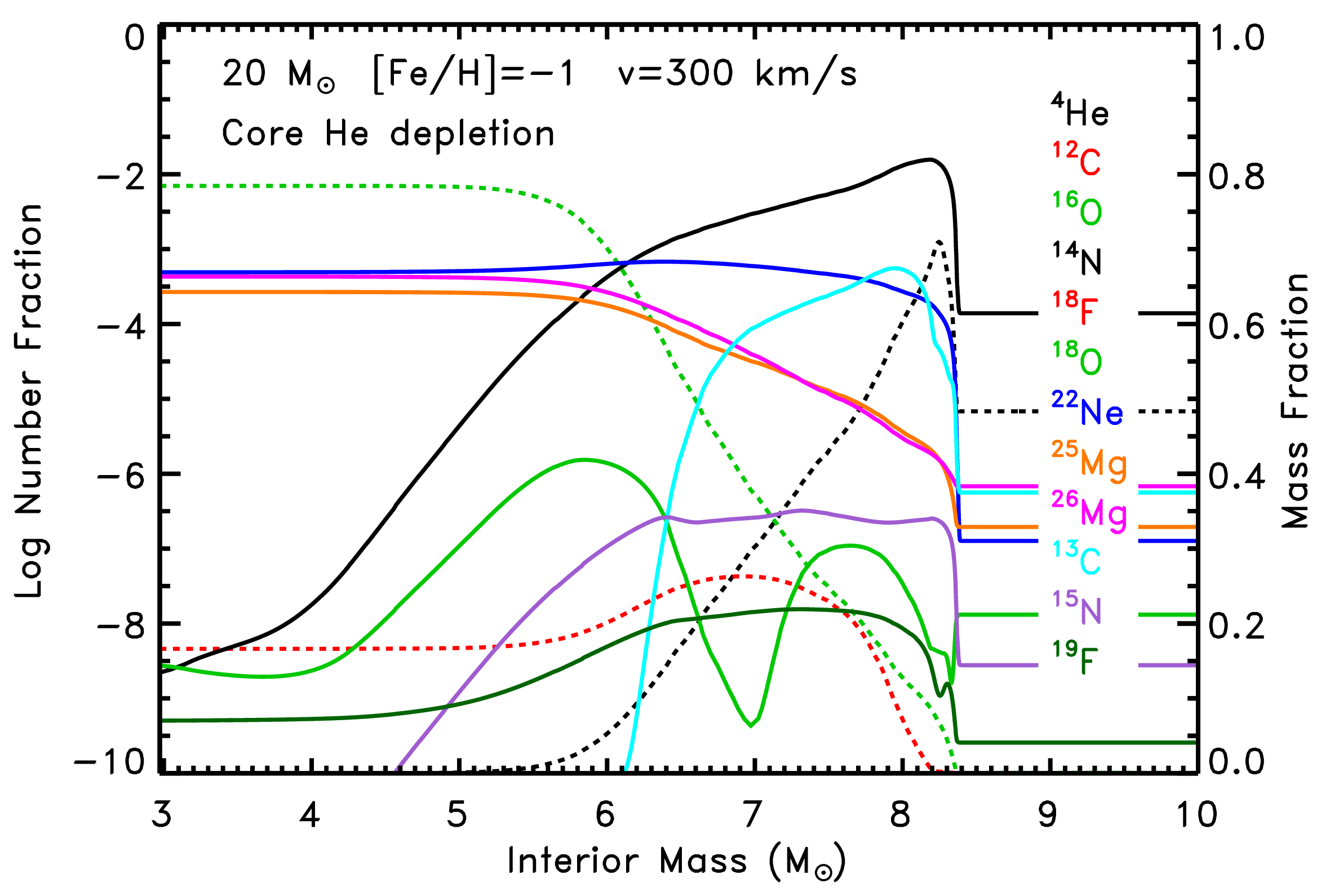}{0.47\textwidth}{}}
\gridline{\fig{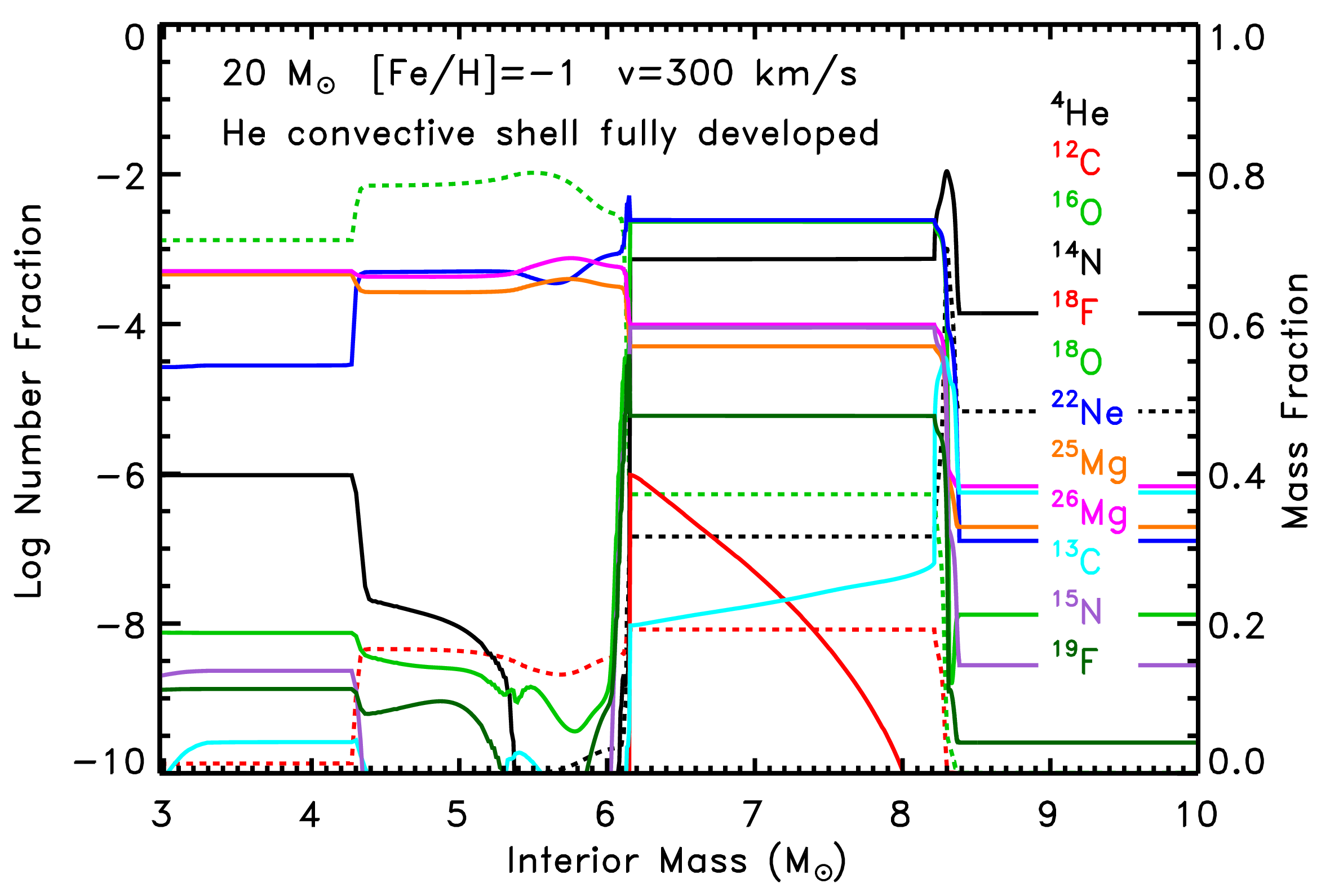}{0.47\textwidth}{}}
\caption{Distribution of the CNO nuclei plus those involved in the synthesis of \nuk{F}{19} in the He core of a $\rm 20~M_\odot$, [Fe/H]=-1 and $\rm v_{\rm ini}=300$ km/s. The dashed lines show the abundances in mass fraction of $\rm ^{4}He$ (black), \nuk{C}{12} (red) and \nuk{O}{16} (green) (right y-axis), while the solid lines show the logarithmic abundances (by number fraction) of all the other nuclei (left y-axis). The upper panel shows the internal run of these nuclei at the central He exhaustion while the lower panel shows the same profiles once the He convective shell has fully developed.}
\label{f1920b300}
\end{figure}

In the previous section we showed that the formation of a conspicuous primary neutron flux within the He core leads to the synthesis of large amounts of elements like N, F and heavy nuclei (i.e. those beyond Zn) at subsolar metallicities. It is therefore important to provide additional details about the synthesis of these nuclei in presence of rotation. Let us repeat here that the diffusion of matter between the H and He burning  zones largely increases the abundances of all the CNO nuclei in the He core. Though the most abundant isotope is obviously \nuk{N}{14}, also \nuk{C}{13}, \nuk{N}{15} and \nuk{O}{17} are largely increased. Part of these nuclei flow in the convective core, but part of them remains locked in the radiative region of the He core. Figure \ref{f1920b300} shows, in the upper panel, a typical distribution of the CNO nuclei in the He core of a $\rm 20~M_\odot$ with [Fe/H]=-1, initially rotating at 300 km/s: the dashed lines refer to $\rm ^{4}He$ (black), \nuk{C}{12} (red) and \nuk{O}{16} (green) (right y-axis), while the solid lines refer to all the other nuclei (left y-axis). The figure shows clearly that the radiative part of the He core is largely enhanced in all the CNO nuclei.   

This is an ideal environment for the synthesis of \nuk{F}{19}. To understand why, let us start by reminding that the sequence of reactions that leads to the production of $\rm ^{19}F$ is \nuk{N}{14}$\rm (\alpha,\gamma)$\nuk{F}{18}$\rm (\beta^+)$\nuk{O}{18}$\rm (p,\alpha)$\nuk{N}{15}$\rm (\alpha,\gamma)$\nuk{F}{19} \citep{gor90}. The activation of this chain requires a hot environment enriched in \nuk{N}{14}, $\alpha$ particles and protons. While both $\rm ^{14}N$ and $\alpha$ particles are simultaneously present in the He core, the abundance of protons is negligible and the most efficient way to produce them in these condition is through the \nuk{N}{14}(n,p)\nuk{C}{14} reaction that, of course, requires a neutron flux. The \nuk{Ne}{22}($\alpha$,n)\nuk{Mg}{25} must be excluded because, at the temperatures at which this nuclear reaction occurs, \nuk{N}{14}, \nuk{F}{18} and \nuk{O}{18}, are already fully converted into \nuk{Ne}{22}. The \nuk{C}{13}($\alpha$,n)\nuk{O}{16} activates at much lower temperatures and is therefore suitable as a neutron source in a condition in which \nuk{N}{14} is still abundant. The natural place in which all these requirements meet together is the He convective shell. In fact when the He convective shell forms, it engulfs both \nuk{N}{14} and \nuk{C}{13} (present in the outer part of the He core) so that they burn simultaneously: \nuk{C}{13} captures $\alpha$ particles and emits neutrons while \nuk{N}{14} captures either $\alpha$ particles and the just produced neutrons, synthesizing therefore both the \nuk{O}{18} and the protons whose further reaction eventually provides the \nuk{N}{15} necessary to synthesize \nuk{F}{19}. This sequence of events obviously occurs also in non rotating models but in this case the available amount of CNO nuclei is too small to raise significantly the \nuk{F}{19} abundance. Vice versa in presence of rotation there is a large abundance of CNO nuclei in the region where the He convective shell develops (upper panel in Figure \ref{f1920b300}). The lower panel of Figure \ref{f1920b300} shows the profiles of the same isotopes reported in the upper panel, after the He convective shell is fully developed. Note that the region between 6.1 and 8.3 \msun, i.e. the one in which the He convective shell develops, is the one in which the peaks of \nuk{N}{14} and \nuk{C}{13} are present. The ingestion of both these nuclei leads to a convective shell eventually greatly enriched in $\rm ^{19}F$, \nuk{O}{18}, \nuk{N}{15} and \nuk{O}{17} while \nuk{N}{14} and and \nuk{C}{13} are partially and fully destroyed, respectively. There are few additional things worth noting in Figure \ref{f1920b300}: a) rotational mixing already raises the number fraction of \nuk{N}{15} at a level of a few times $\rm 10^{-7}$ in the He rich zone (upper panel of Figure \ref{f1920b300}), therefore, even without a neutron source, it would be possible to raise substantially the $\rm ^{19}F$ abundance; b) \nuk{O}{17} is also very abundant and contributes to the neutron flux via the \nuk{O}{17}$\rm (\alpha,n)$\nuk{Ne}{20} reaction; c) at the onset of the core collapse the He convective shell is still largely enriched in \nuk{N}{15}, which means that additional production of $\rm ^{19}F$ would have been possible if the He shell where not frozen by the very fast evolution of the stellar core in the advanced burning phases. Of course this complex interplay among the various physical phenomena contributing to the synthesis of $\rm ^{19}F$ implies that we do not expect a strict monotonic dependence of the \nuk{F}{19} yield on the mass, metallicity and rotational velocity, but we certainly we do expect a large primary production of both \nuk{N}{14} and \nuk{F}{19} in rotating models. 


\begin{figure}[ht]
\gridline{\fig{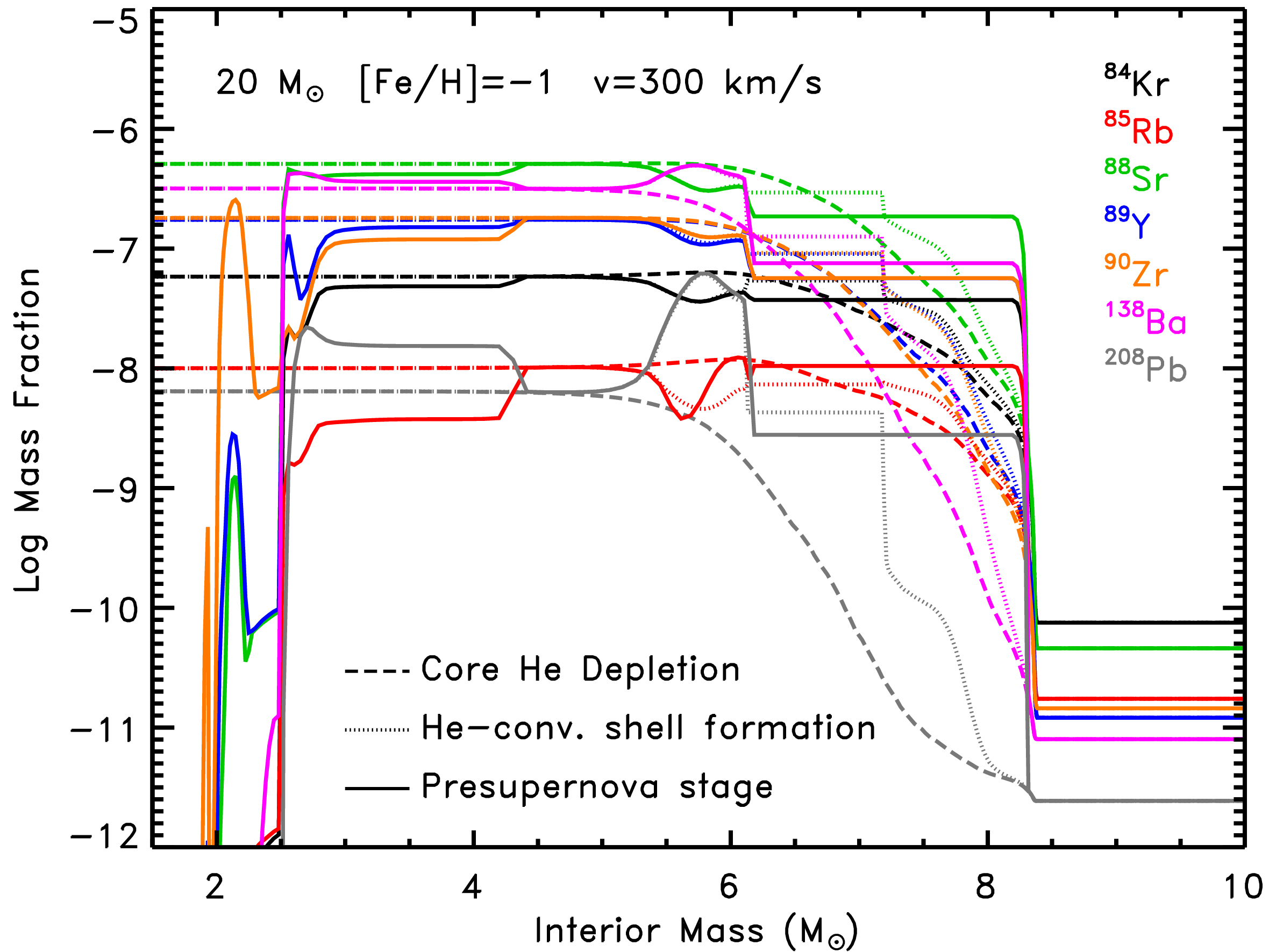}{0.45\textwidth}{}}
\gridline{\fig{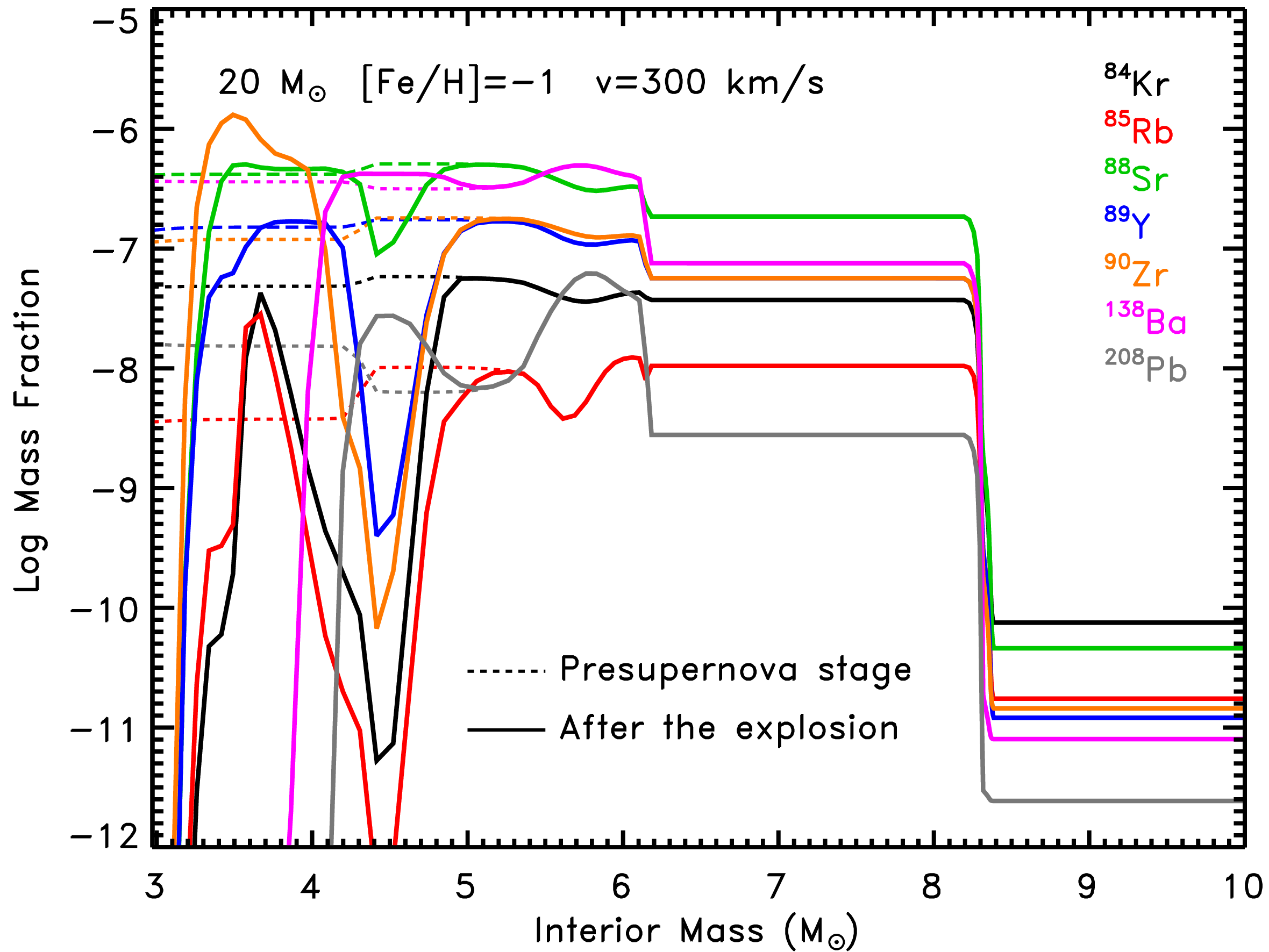}{0.45\textwidth}{}}
\caption{Upper panel: internal profiles of a subset of heavy nuclei (logarithmic abundances in number fraction) at the central He exhaustion (dashed lines), at the birth of the He convective shell (dotted lines) and at the presupernova stage (solid lines) for a 20\ \msun\ of [Fe/H]=-1 and $\rm v_{\rm ini}=300~km/s$). Lower panel: comparison between the pre explosive (dashed) and explosive abundances (solid) of the same nuclei and the same model shown in the upper panel.}
\label{s_last}
\end{figure}

While the synthesis of \nuk{F}{19} is controlled by the \nuk{C}{13}($\alpha$,n)\nuk{O}{16} and occurs in the He convective shell, the large abundances of the elements beyond Zn at sub solar metallicity are due to the activation of the \nuk{Ne}{22}($\alpha$,n)\nuk{Mg}{25} during core He-, radiative shell He- and convective shell C-burning. Figure \ref{s_last} shows in the upper panel (again for a 20\ \msun\ of [Fe/H]=-1 and $\rm v_{\rm ini}=300~km/s$) the profiles of a subset of key nuclei at the central He exhaustion (dashed lines), at the formation of the He convective shell (dotted lines) and at the presupernova stage (solid lines). These isotopes have been chosen because they mark the passage of the matter through the neutron closure shells at n=50 (Kr, Rb, Sr, Y and Zr) and at n=82 (Ba) and therefore its capability to reach the end point (Pb) of the neutron capture nucleosynthesis. The abundances of these nuclei in the He convective shell at the presupernova stage (between $\sim 6.1$ and $\rm \sim 8.3~M_\odot$) are not the result of a He convective shell itself, but mainly the result of the mixing of matter previously synthesized either by the radiative He shell and/or the He convective core. This is clearly shown by a comparison between the dotted and the solid lines (note that the y-scale is logarithmic). Another thing worth noting in the same panel is the contribution of the C convective shell (extending between $\sim 2.8$ and $\sim 4.2~M_\odot$). It is clear that there are isotopes that are destroyed, untouched or produced by the C convective shell. This result is however highly misleading if one forgets that the explosive nucleosynthesis may alter, even significantly, the chemical composition within the CO core. The lower panel in Figure \ref{s_last} shows the same nuclei, reported in the upper panel, before (dashed) and after (solid) the explosion (the minimum value of the x-axis corresponds now to the mass of the remnant). 
The first thing worth noting is that the explosive nuclesynthesis alters significantly the abundances of all the nuclei plotted in the figure, within a substantial fraction of the CO core.
In particular, \nuk{Kr}{84}, \nuk{Rb}{85}, \nuk{Y}{89}, \nuk{Ba}{138} and \nuk{Pb}{208} are destroyed while \nuk{Zr}{90} is produced; \nuk{Sr}{88}, on the contrary, is only mildly affected. Of course each nuclear species will have its own behavior and also the amount of reprocessing will depend on the energy of the explosion. It is not the purpose of this paper to discuss each isotope in detail, but what it is important to keep in mind is that, at variance with what is largely trusted, the explosive burning cannot be neglected when computing the yields of almost all the nuclear species. In this case, for example, only the zones more external to $\rm \sim 5.3~M_\odot$, which is close to the outer border of the CO core, are not affected by the explosion.



Summarizing the results discussed so far, the nuclei produced by the neutron captures are synthesized predominantly in He convective core plus a contribution from the radiative He burning shell and the C convective shell. Their presupernova abundances can be affected by the explosive nucleosynthesis in a non negligible way, therefore this must be taken into account for a correct determination of the final yields of all these elements.




\begin{figure}[ht]
\gridline{\fig{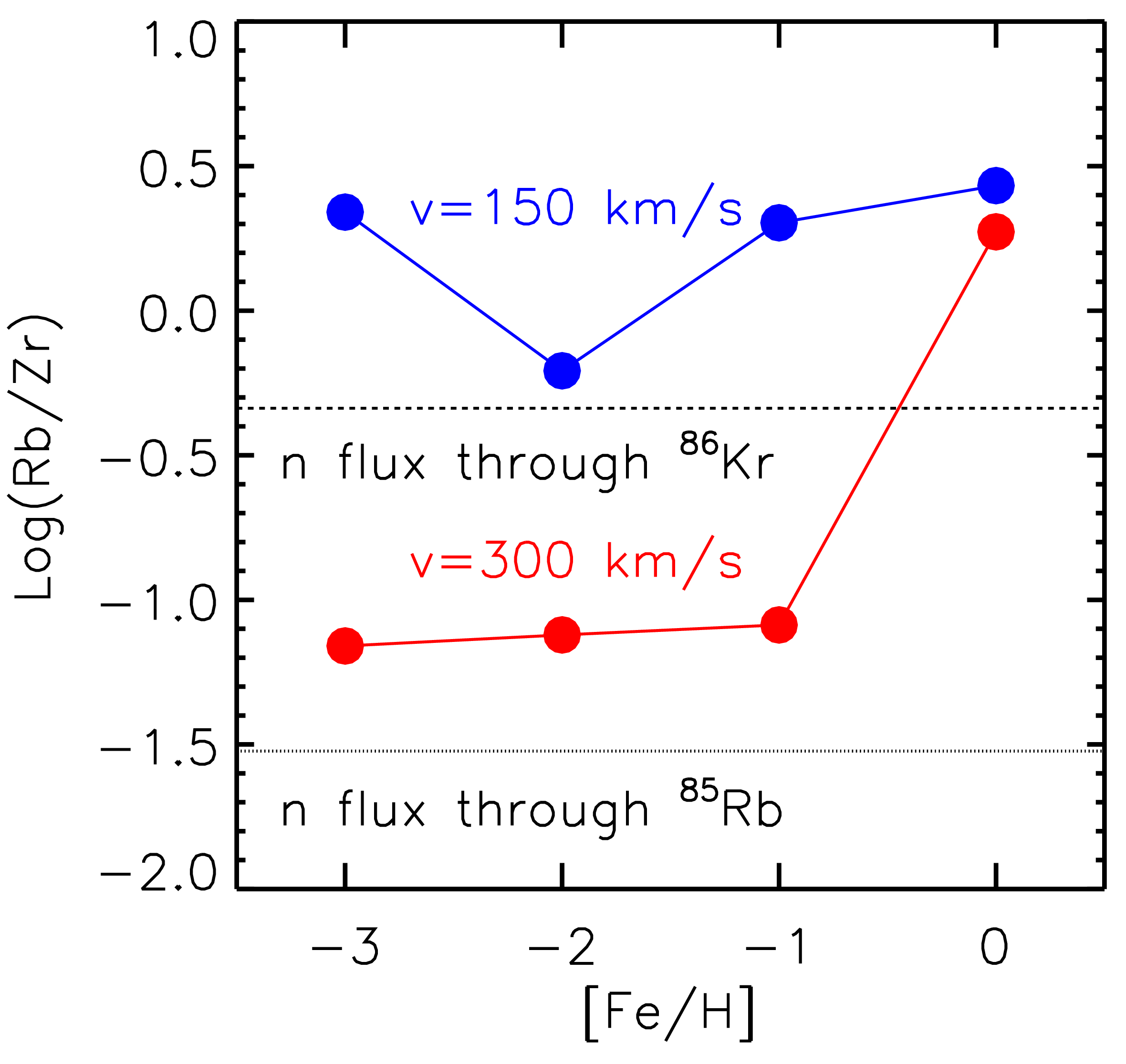}{0.45\textwidth}{}}
\gridline{\fig{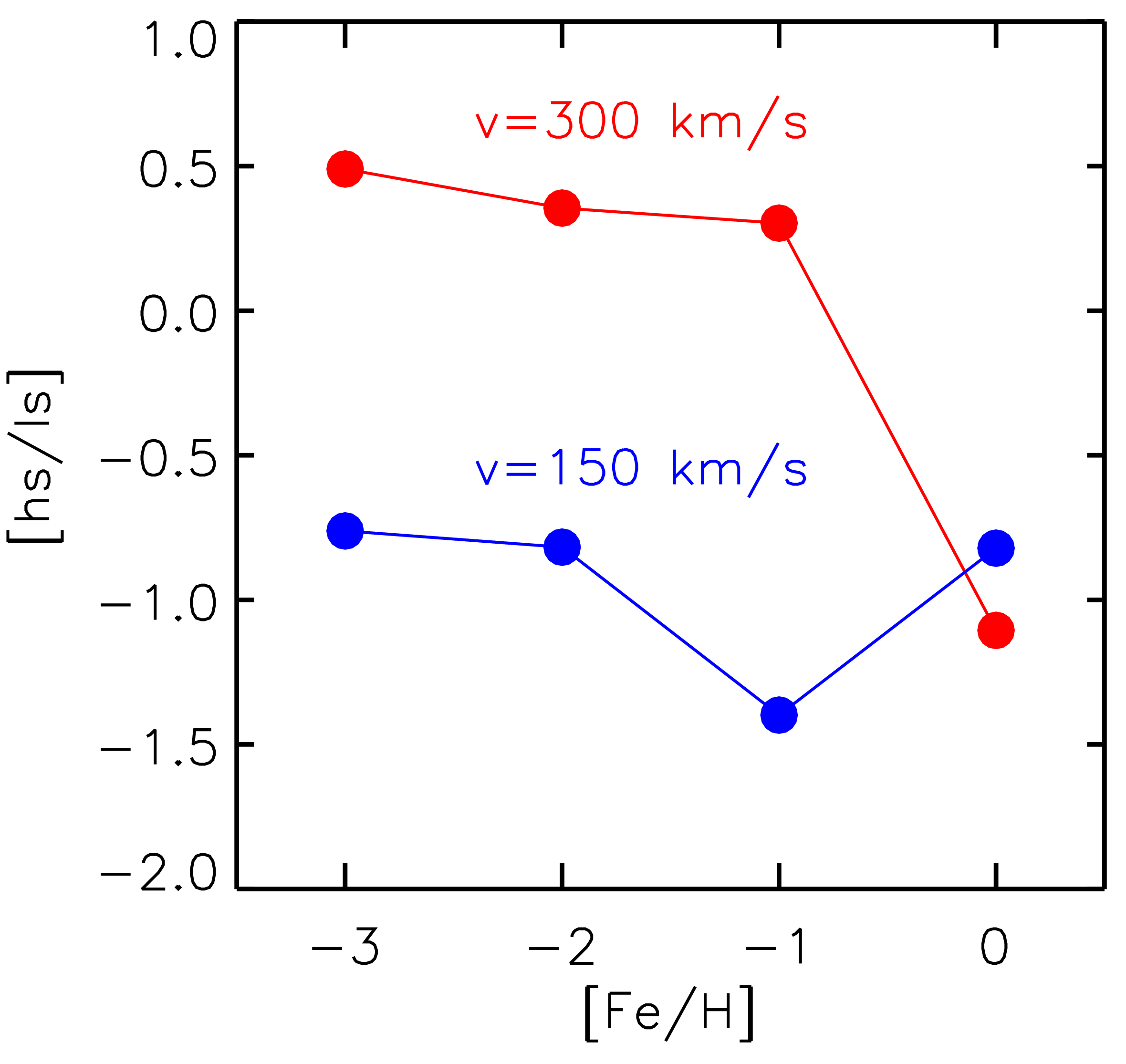}{0.45\textwidth}{}}
\caption{Left panel: Dependence of the (Rb/Zr) ratio produced by a generation of massive stars (set R) on the initial metallicity. The blue and red lines refer to the two initial rotation velocities. The lower thin dashed line shows the limiting ratio in which \nuk{Kr}{85} completely decays in \nuk{Rb}{85} while the thick dashed line shows the limiting ratio obtained by assuming that the \nuk{Kr}{85} formed in the ground state fully goes in \nuk{Kr}{86}. All cross sections have been evaluated at $\rm 3\times10^8\ K$. Right panel: trend with the metallicity of the ratio between nuclei located on the neutron closure shell n=82 (hs = average abundance of Ba, La and Ce) and those located on the neutron closure shell at n=50 (ls = average abundance of Sr, Y and Zr). Both lines refer to set R.}  
\label{trendratio}
\end{figure}


A routinely way to infer the strength of the neutron flux responsible for the neutron capture nucleosynthesis in a given environment is the analysis of the so called "branching points". A branching point occurs when there is a competition between the beta decay and the neutron capture on an unstable nucleus. For each fixed temperature there is a threshold neutron density above which the neutron capture is favored with respect to the decay and below which the opposite occurs. An important example of branching point is the one at \nuk{Kr}{85}. If the neutron density is lower than the threshold value of $\rm \sim 10^{8}~n/cm^3$, the s-process path follows the sequence \nuk{Kr}{85}($\beta^-$)\nuk{Rb}{85}(n,$\gamma$)\nuk{Rb}{86}($\beta^-$)\nuk{Sr}{86}. Since the neutron capture cross section on \nuk{Rb}{85} is quite large its equilibrium abundance is quite low and therefore the Rb abundance is quite low as well. Vice versa, if the neutron density is higher than the threshold value of $\rm \sim 10^8~n/cm^3$ the path followed by the matter goes through the sequence \nuk{Kr}{85}(n,$\gamma$)\nuk{Kr}{86}(n,$\gamma$)\nuk{Kr}{87}($\beta^-$)\nuk{Rb}{87} that continues then through the s-process path merging with the other chain at \nuk{Sr}{88}. By the way, let us remind that a fraction of the \nuk{Kr}{85} is produced in a very short living metastable state and hence this component always decays in \nuk{Rb}{85}. Anyway, since \nuk{Rb}{87} lies on the neutron magic number line n=50, its neutron capture cross section is quite low and hence if the neutron density is larger than the threshold value the Rb abundance is rather high. Hence the ratio between Rb and another element whose abundance does not depend on any branching point would provide clues about the neutron density present when the two elements were synthesized. An element largely used in combination with Rb is Zr: a high Rb/Zr value (of the order of 0.45 by number) will imply a high neutron density while a low Rb/Zr ratio (of the order of $3\times 10^{-2}$ by number) would imply a low neutron density (the critical n density being in this case $\rm \sim 10^8~n/cm^3$). Figure \ref{trendratio} shows, in the upper panel, the Rb/Zr ratio produced by a generation of massive stars (set R) as a function of the metallicity, for the two initial rotation velocities v=150 and v=300 km/s. The two approximate limiting values for low (dotted) and high (dashed) neutron densities, corresponding to the full conversion of $\rm ^{85}Kr$$_{Ground}$ into $\rm ^{85}Rb$ or into $\rm ^{86}Kr$, are also reported for reference. None of the rotating models actually reaches the threshold neutron density necessary to raise the Rb/Zr ratio, so one would expect a similar (low) ratio for both initial rotation velocity. Vice versa Figure \ref{trendratio} shows that the two sets of rotating models are systematically shifted, at subsolar metallicities, one respect to the other. Moreover the set of models rotating at 150 km/s shows a higher ratio, like if matter would have passed through \nuk{Rb}{87}, in spite of the fact that the neutron density in thse models is lower than that obtained for the models rotating faster. To understand these results it must be reminded that the above discussion about the relative equilibrum abundances assumes, obviously, that a star has enough time for all these nuclei to go to the relative equilibrium. Unfortunately the achievement of the relative equilibrium abundances of Rb and Zr requires the passage of the matter through \nuk{Sr}{88}, the nucleus that has the lowest neutron capture cross section and hence the one that requires the largest amount of time to propagate matter towards heavier nuclei. What happens in the set rotating at 150 km/s is that matter has not time to proceed beyond Sr and hence the Rb/Zr is high just because the bottleneck of \nuk{Sr}{88} does not allow the build up of Zr. Vice versa in models rotating at 300 km/s the flux of matter through the bottleneck is much larger because either the rotation driven instabilities lead to a larger primary neutron source and also because the phase in which the release of neutrons occurs starts earlier and lasts longer, providing more time for the matter to flow beyond \nuk{Sr}{88} and bring Zr to the equilibrium. Since the neutron density is in any case lower than the threshold value, Rb/Zr tends towards the low equilibrum value. Hence it is very important to realize that, at least in the frame of the massive stars, a low or high Rb/Zr ratio does not necessarily implies a low/high neutron density but also just the capability to reach or not the equilibrium abundances. At solar metallicity in both cases matter does not overturn efficiently the bottleneck at \nuk{Sr}{88}. Note that a possible comparison with the solar ratio must take into account the fact that the contribution of the r-process may alter significantly the abundances of some of these nuclei, like e.g. \nuk{Rb}{87}. For a comprehensive, quantitative and extended discussion of all the branching points we refer the reader to \cite{luchi11}. 

A widely adopted way to determine the relative efficiencies of the weak and main components of the s process nucleosynthesis is the hs/ls ratio, where hs and ls stand for the average abundance of the elements that lie on the neutron closure shell at n=82 (Ba, La and Ce in the present definition) and at n=50 (Sr, Y and Zr in this case), respectively. The capability of the matter to cross (or not) the n=50 neutron closure shell and populate the stability valley beyond this barrier is well represented by this ratio: if the neutron/seed ratio is not large enough to overcome the n=50 neutron closure shell, matter will accumulate on the ls bottleneck (and the hs/ls will be low) while in the opposite case matter can proceed further on and populate the second neutron closure shell, raising therefore the hs/ls ratio. The lower panel in Figure \ref{trendratio} shows our predictions for the trend of [hs/ls] as a function of [Fe/H], for both initial rotation velocities (set R): as expected the [hs/ls] increases in models with a higher initial rotation velocity because of the increasing neutron/seed ratio (as discussed above). 

\section{Comparison with other sets of models}\label{compa}

The comparison among yields obtained by different groups and, more specifically, the understanding of the source(s) of the differences is, in our opinion, an extremely difficult task, sometime hopeless. Over the decades we compared many times our results with those obtained by other authors, e.g. \cite{lsc00}, without being really capable of understanding where the differences came from. To really understand the source of the differences it would be necessary to look into the other evolutionary codes, effort that, even if it were possible, it would be too time consuming. In spite of this, it is certainly important to compare different sets of yields, at least to be aware of the existing differences. For these reasons in this section we show a few comparisons between our yields and those provided by other groups without pretending to really understand quantitatively where the differences come from but giving, whenever possible, some hints.

\begin{figure}[ht]
\gridline{\fig{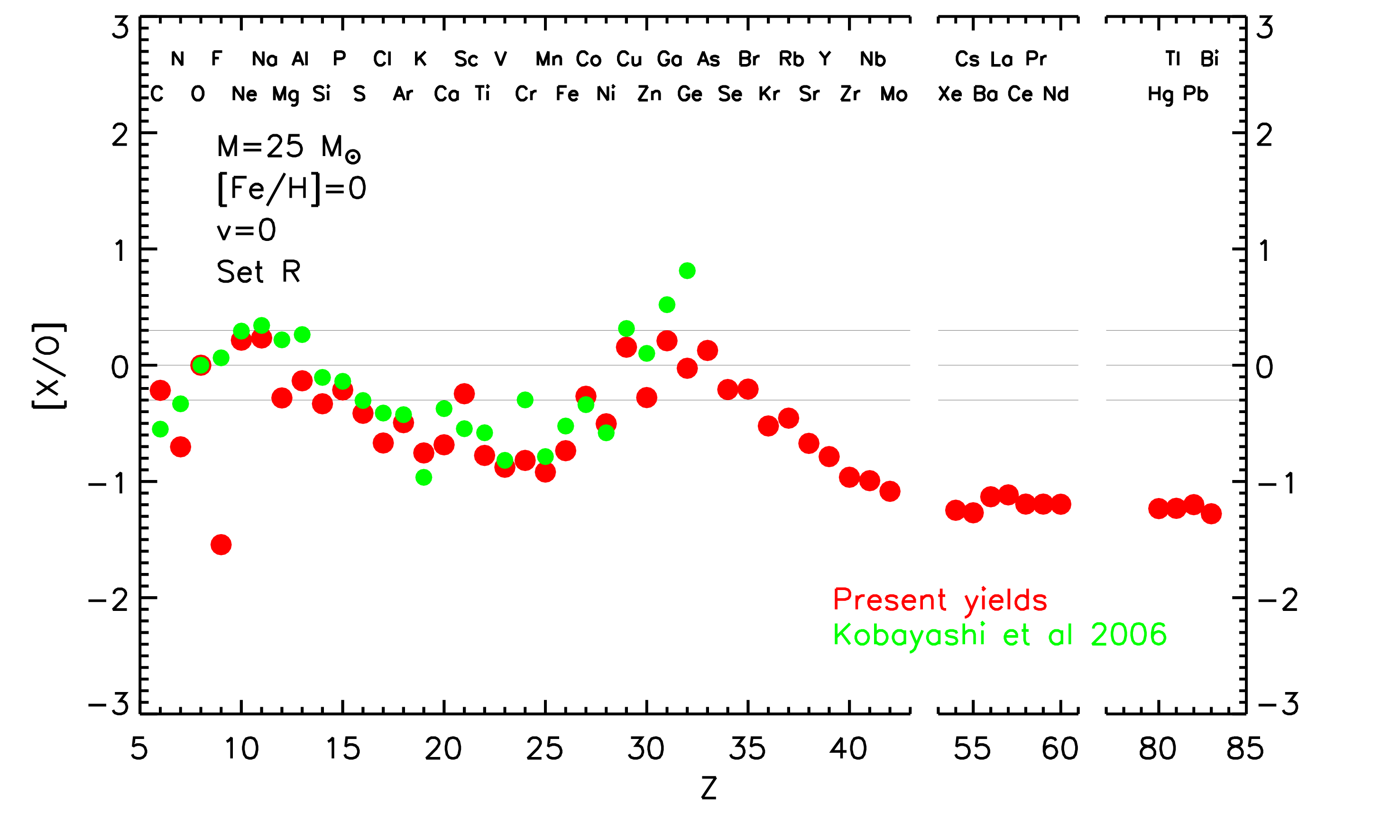}{0.45\textwidth}{}}
\gridline{\fig{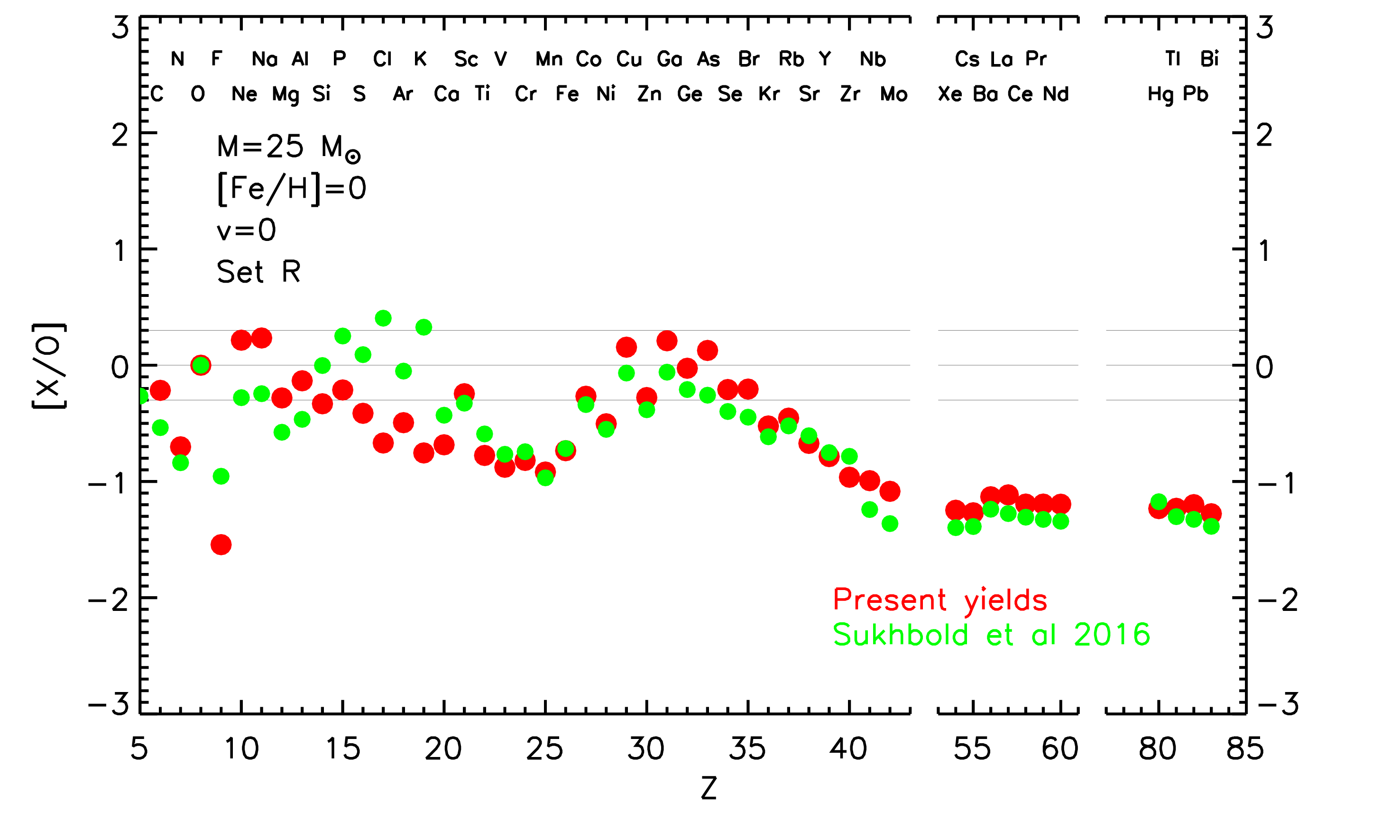}{0.45\textwidth}{}}
\gridline{\fig{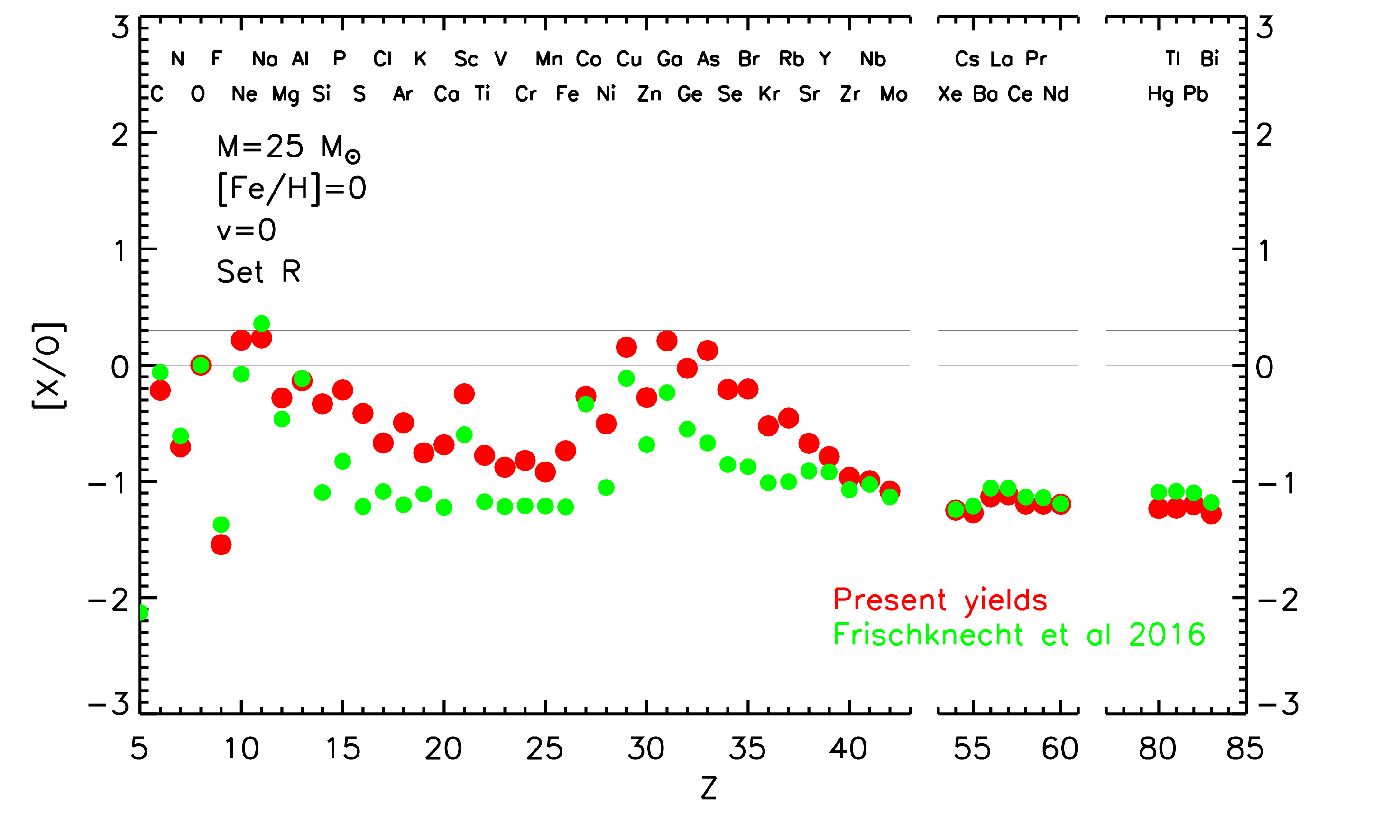}{0.45\textwidth}{}}
\caption{Comparison between the [X/O] produced by a non rotating 25\ \msun \ of solar metallicity  and analogous yields provided by \cite{koba06} (upper panel), \cite{sukh16} (middle panel) and \cite{fresco16} (lower panel).}
\label{compa_nr}
\end{figure}

Figure \ref{compa_nr} shows the comparison between the [X/O] produced by a non rotating 25\ \msun \ of solar metallicity  and the analogous yields provided by \cite{koba06}, \cite{sukh16} and \cite{fresco16}. The yields provided by \cite{koba06} extend up to Ge and show an overall agreement with the present ones (upper panel), the only extremely large difference being F that is much more abundant in their yields (their [F/O]$\sim$0 implies that massive stars, in their scenario are the main F producers at solar metallicity). Their high [Ge/O] value, vice versa, is quite certainly due to the fact that their nuclear network ends at Ge and hence matter can not flow onward and accumulates there. The middle panel shows a comparison with a model computed by \cite{sukh16}. In this case there is a basic overall agreement even if there are not negligible differences in the region P to K. Since the yields of these nuclei are largely of explosive origin, it is probable that the differences reflect the different techniques adopted to simulate the explosion and/or different mass-radius relations at the onset of the collapse. C, Ne, Na, Mg and Al are systematically lower in their model with respect to our: these differences could be understood if they would have a lower C abundance at the end of the central He burning phase. In fact, in this case, the C yield would be obviously lower and also Ne, Na, Mg and Al would be lower as well because they are products of the C burning and hence scale inversely with the amount of C left by the He burning. The amount of C left by the He burning influences also the number, and extension, of the C convective shell(s) as well as their speed in advancing in mass. It must also be noted that their CO core mass is much larger than our (7.13\ \msun \ versus 6.2\ \msun) and obviously such a large difference (certainly due to a different extension of the H convective core) may play a relevant role in the production of the final yields. The bottom panel in Figure \ref{compa_nr} shows a similar comparison with a model published by \cite{fresco16}. In principle this comparison is not much pertinent because their abundances (not really yields) do not include the explosive burning. However, we think it is worthwhile because this is the only set of models that includes the computation of abundances up to Pb for rotating models (see below). As a consequence of what we have just said, elements between Si and Zn must not be considered at all because they are an outcome of the explosive burning. We pointed out above that also other nuclei (not of explosive origin) may be affected even significantly by the explosive nucleosynthesis, but in this comparison we ignore such an occurrence. Elements C to Al are in quite good agreement, while those produced by the weak component are substantially underproduced by \cite{fresco16}.

\begin{figure}[ht]
\gridline{\fig{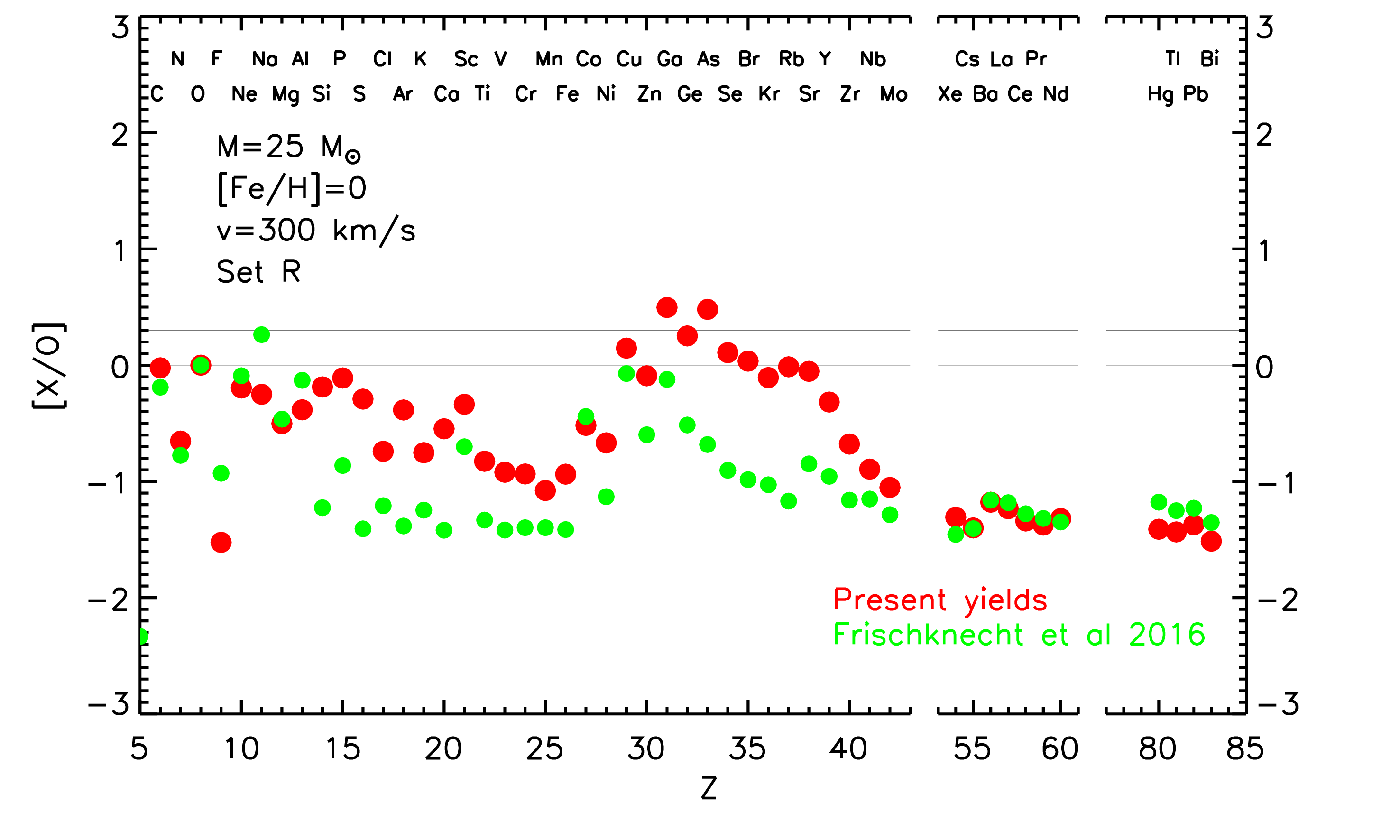}{0.45\textwidth}{}}
\gridline{\fig{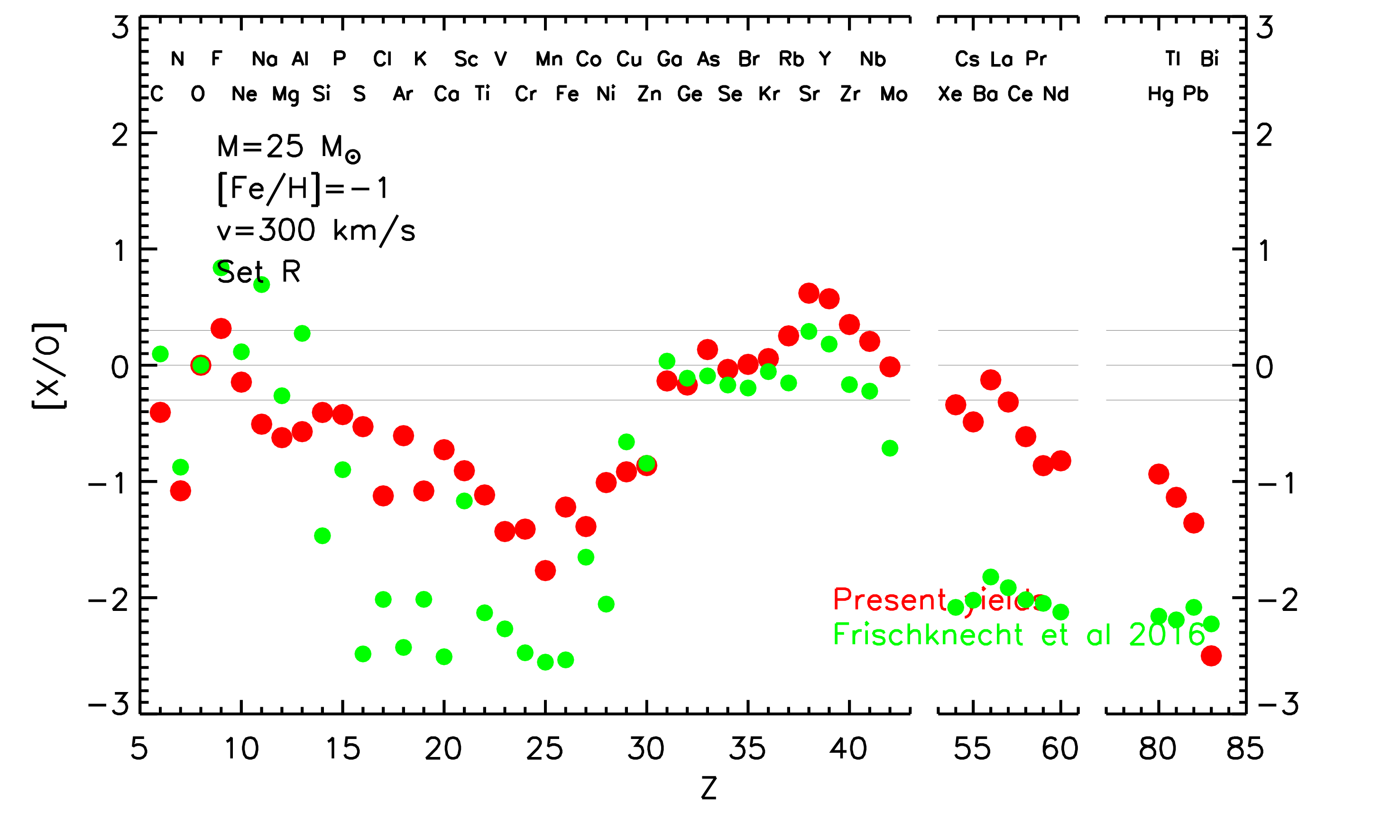}{0.45\textwidth}{}}
\caption{Upper panel: comparison between the [X/O] obtained after the explosion of a 25\ \msun \ of solar metallicity rotating at 300 km/s and the pre explosive yields obtained for the same mass by \cite{fresco16}. Lower panel: same comparison but for [Fe/H]=-1}
\label{compa_r}
\end{figure}

The comparison between our yields and the abundances published by \cite{fresco16} for rotating models is shown in Figure \ref{compa_r}. It is important to remind that the two models have not been computed with the same initial rotation velocity. In fact, while we fix an initial velocity (e.g. 150 or 300 km/s) independent of the initial mass, \cite{fresco16} assume a given value of $v/v_{\rm crit}$ (e.g., 0.4 or 0.5) which obviously implies a different initial velocity for the various models. 
Nevertheless, our 25\ \msun \ computed with $\rm v_{\rm ini}=300~km/s$ roughly corresponds to their model computed for $v/v_{\rm crit}$=0.4. The upper panel refers to the solar metallicity while the lower one to [Fe/H]=-1. In the solar metallicity case, the main differences between the two models concern essentially the light s-process elements (the weak component) up to the neutron closure shell at n=50. In particular while these elements are co-produced with O in our models, they are substantially underproduced by \cite{fresco16}. At low metallicity, on the contrary, the differences mainly concern the heavy elements above n=50. In fact, while these elements are produced in our model, although the production decreases progressively with the atomic number (note, e.g., that Ba and La are co-produced with O), they are not produced at all by \cite{fresco16}.
In addition to this, also F, Na and Al are largely overproduced with respect to O in their model. Again we do not have any real explanation for these differences but we want to remind again that the inclusion of rotation in an evolutionary code {\it requires} a calibration, 
and that different calibrations may lead to even very different results. It is worth noting, however, that even in the non rotating models, at solar metallicity, where it is generally accepted that massive stars produce the weak component, the model of \cite{fresco16} shows a substantial underproduction of all these elements. 

\section{Initial mass - remnant mass relation}\label{inifi}

\begin{figure}
\epsscale{1.2}
\centering
\plotone{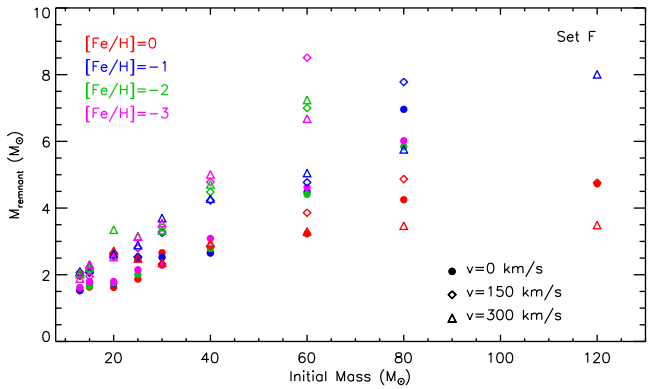}
\caption{Remnant masses obtained for set F as a function of the initial mass}
\label{mremF}
\end{figure}

\begin{figure}
\epsscale{1.2}
\centering
\plotone{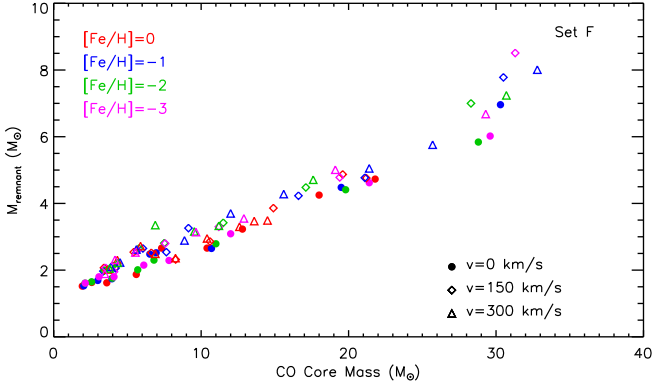}
\caption{Remnant masses obtained for set F as a function of the CO core mass}
\label{mremco}
\end{figure}

The three different choices for the calibration of the explosions that we discussed above (Sets F, M and R) obviously imply different remnant masses. Figure \ref{mremF} collects for the Set F the remnant masses for all the metallicities and all the initial rotational velocities. The figure shows clearly that the choice that all the stars eject 0.07\ \msun\ of \nuk{Ni}{56} does not allow the formation of remnants more massive than $\rm \sim 9~M_\odot$. The same would occur even for lower amounts of \nuk{Ni}{56} provided, however, that at least part of the nuclei produced by the explosive burning are ejected.
It is also worth noting that for each mass there is a spread of the remnant masses in spite of the fact that all of them are obtained by requiring the ejection of the same amount of \nuk{Ni}{56}. Such a spread largely reduces if the remnant masses are plotted versus the CO core mass (Figure \ref{mremco}). The reason is that, as already mentioned in this paper, the evolution of a star beyond the central He burning is predominantly determined by the size of the CO core, irrespective of the initial rotational velocity and metallicity. The residual spread in Figure \ref{mremco} is due to the fact that the advanced evolutionary phases are not just a function of the CO core mass but also of the amount of $\rm ^{12}C$ left by the core He burning, since it determines how much the C burning shell may advance in mass before the final collapse. Though the relation depicted in Figure \ref{mremco} may be used to have a good idea of the remnant masses, we strongly discourage any user from applying this relation to other models, different from the present ones, to infer any explosive yield. The reason is that the spread shown in this Figure, though quite tight, is larger than the region where all the explosive burning occur (which amounts to a few tens of solar masses). The remnant masses obtained for Set M do not differ significantly from those obtained for Set F because, although we apply the mixing and fall back for these models, the mixed zone is not very extended and therefore the mass cuts needed to ejected the 0.07 $\rm M_\odot$ of \nuk{Ni}{56} are in any case deep enough and not very different from the ones obtained for set F.

\begin{figure}
\epsscale{1.2}
\centering
\plotone{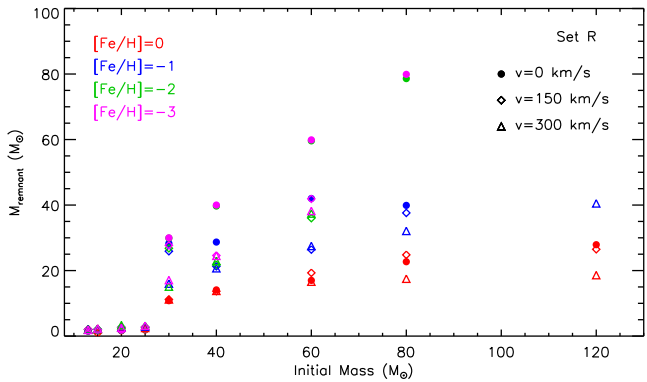}
\caption{Remnant masses obtained for set R as a function of the initial mass}
\label{mremR}
\end{figure}

Set R, on the contrary, is profoundly different from the previous ones because in this case we assume that all stars more massive than 25\ \msun\ fully collapse. Obviously in this case the remnant masses are much larger than in the previous cases (see Figure \ref{mremR}) and some of the masses larger than $\rm 25~M_\odot$ leave black hole remnants even much larger than those associated to the detections of the gravitational waves GW150914, GW151226, GW170104, GW170608 and GW170814 \citep{Abb6,Abb5,Abb4,Abb3,Abb2,Abb1}. Though this result could be considered quite trivial and obtained "by construction", Figure \ref{mremR} shows some interesting features. At solar metallicity the maximum remnant mass that can be formed is $\rm \sim 30~M_\odot$, irrespective of the initial rotational velocity (at least up to 300 km/s). The reason is that at solar metallicity mass loss is efficient enough that the final mass of all these models is lower than $\rm \sim30~M_\odot$. At lower metallicities the strong reduction of the mass loss allows the non rotating models to form much larger remnant masses. However, in this case the maximum remnant mass ranges between 80-100 $\rm M_\odot$ (at these metallicities the mass of the remnant is almost equal to the initial mass because mass loss is very inefficient) because stars above this limit enter the PISN regime and do not leave any remnant (see section \ref{advanced}). In rotating models, on the contrary, the remnants are on average much smaller because rotation pushes many stars above their Eddington luminosity where they lose much more mass than their non rotating counterparts. Because of the increase of the CO core induced by rotation, $\rm M_{PISN}$ lowers in rotating models (section \ref{advanced}) and therefore the maximum remnant mass that can be formed in this case reduces as well. 
   
\section{Summary and Conclusions}\label{conclusions}
In this paper we presented a new very extended set of models and yields of massive stars in the range 13-120 $\rm M_\odot$, initial metallicities [Fe/H]=0, -1, -2, -3 and initial rotation velocities v=0, 150, 300 km/s. All the models were followed from the pre-main sequence up to the presupernova stage. The explosion was simulated artificially by means of a hydrodynamic code in the framework of a kinetic bomb and the computation of the explosive nuclesynthesis is fully coupled to the hydrodynamics. Given the arbitrariness in the calibration of the explosion we discussed three different choices for this calibration and we showed how the yields and the masses of the remnants depend on the adopted choice. We took into account a large nuclear network, fully coupled to the stellar evolution, that extends from neutrons to $\rm ^{209}Bi$, includes explicitly 335 nuclear species and more than 3000 nuclear reactions. The efficiency of the rotation induced mixing was calibrated by requiring the fit to a subset of star (taken from the LMC samples of the FLAMES survey \citet{untetal09}) for which both the surface N abundance and the projected rotation velocity are available. Most of the properties of this grid of models, together to the final yields, are available for download from the website http://orfeo.iaps.inaf.it. More specific details about the models and/or the explosions may be obtained upon request.

\begin{figure*}
\epsscale{1.15}
\plotone{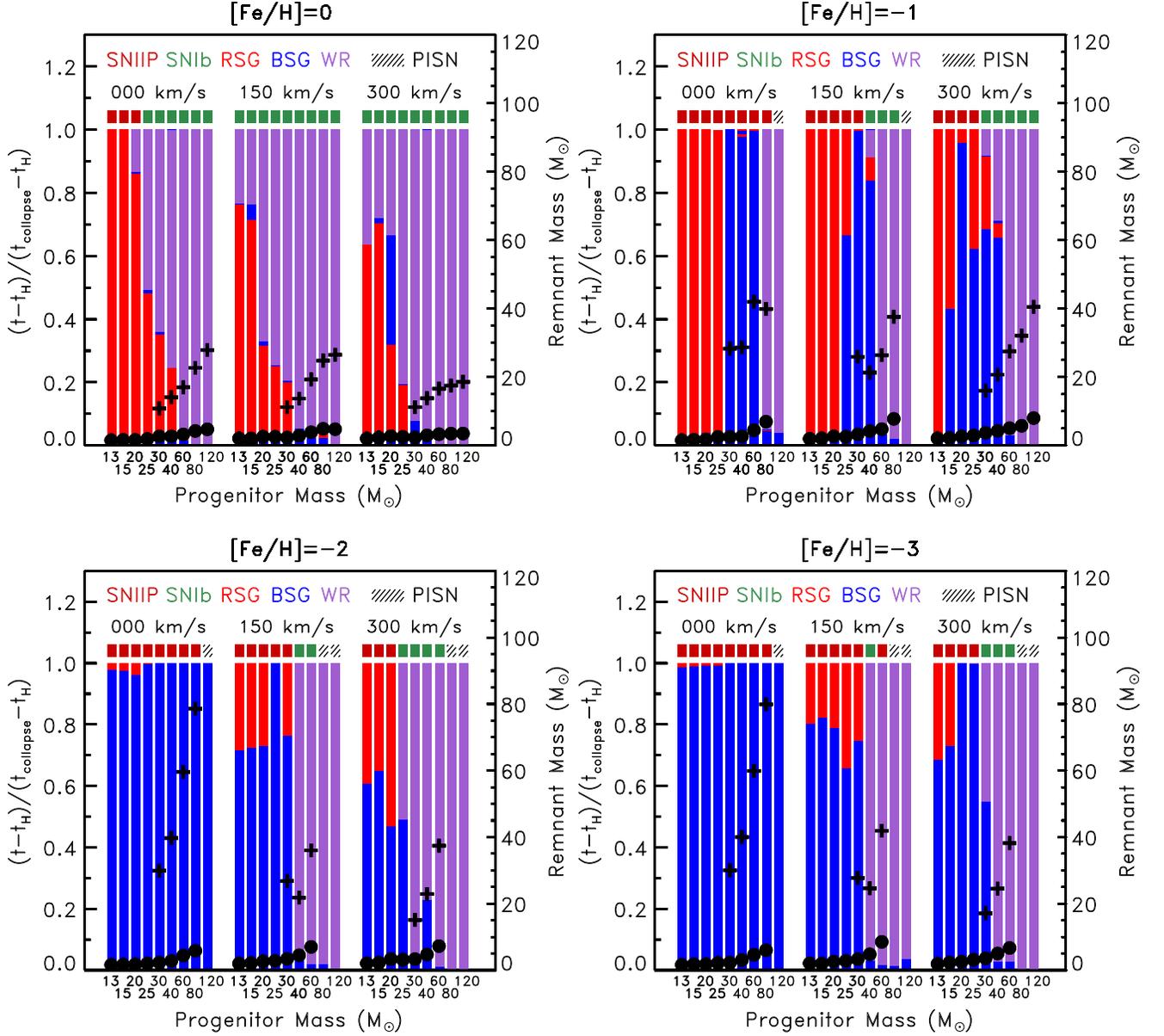}
\caption{Global properties of all the models as a function of mass, metallicity and initial rotation velocity. Each bar represents, for each model, the configuration of the star as a function the fraction of time remaining to the final collapse after core H depletion. The colors refer to the RSG (red), BSG (blue) and WR (purple) configurations, respectively. The expected SN type is reported on top of each bar: dark red refers to SNIIP while green to SNIb, respectively. Stars expected to explode as pair instability supernovae are marked by a hatched black square. For each stellar model, the final remnant masses are shown by filled circles, for the set F, and by crosses, for the set R.}
\label{summa}
\end{figure*}

A synthesis of the surface properties of all our models is shown in Figure \ref{summa} and it helps in visualizing what is described in the remaining of this section.

At solar metallicity, non rotating stars populate the RSG branch up to a luminosity $\rm Log(L/L_\odot) \sim 5.7$, which corresponds to a mass of $\rm \sim 40~M_\odot$. More massive stars lose a major fraction of their H rich envelope because they overcome their Eddington luminosity before reaching the RSG branch and hence turn back blueward where they spend all their He burning lifetime. Stars more massive than $\rm \sim 20~M_\odot$ leave the RSG branch sometime during the core He burning and evolve towards a BSG configuration, eventually becoming WR stars ($\rm M_{\rm WR}\sim 20~M_\odot$). Accordingly, the maximum mass we predict for the progenitors of SNIIP is $\rm M_{\rm IIP} \sim 17~M_\odot$ (which corresponds to a maximum luminosity of $\rm Log(L/L_\odot) \sim 5.1$). Stars with $\rm M>M_{\rm IIP}$ are expected to explode as SNIb. {\color{black} By the way, because our grid has not enough resolution in the initial mass, and therefore in the final H envelope mass, we cannot clearly define the transition SNIIP/SNIIb/SNIb according to \citet{hachietal12}, therefore we chose for the moment to consider only two limits, i.e. SNIIP and SNIb, also beacuse from the observational side the classification is not yet much clear.}
The properties of these models are in reasonable agreement with some observed properties of the massive stars: (a) the predicted distribution of the stars in the HR diagram is compatible (both in luminosity and effective temperature) with the distribution observed in a sample of the galactic RSG stars \citep{levesque05}; (b) the maximum mass exploding as SNIIP is compatible with the lack of detected progenitors (of SNIIP) with luminosities higher than $\rm Log(L/L_\odot) \sim 5.1$, as reported by \citet{smartt09, smartt15}; therefore our scenario {\color{black} naturally explains} the so called "RSG problem" \citep{smartt09} {\color{black} - note however that other hypothesis to solve this problem are provided, e.g., by \citet{we2012} and \citet{yc2010}}; (c) in a scenario in which all the stars with $\rm M>25~M_\odot$ fail to explode and collapse directly to a black hole, we do not expect SNIb at luminosities larger than $\rm Log(L/L_\odot) \sim 5.6$. This result is at least not in contradiction with the 14 SNIb progenitors with no detection reported by \citet{smartt15}; moreover, the expected amount of mass ejected by stars exploding as SNIb is of the order of $\rm M_{ejecta}\sim 6~M_\odot$, in rather good agreement with the ejected masses of $\rm 3-4~M_\odot$ estimated from the SNIb light curve fitting \citep{wl85,ew88,des11,des15,des16,Lyman16}; (d) a limiting mass between SNIIP and SNIb corresponding to $\rm M_{\rm IIP}\sim 17~M_\odot$  implies a fraction of SNIb of $\sim 26~\%$ of all the core collapse supernovae (CCSNe) (having assumed a Salpeter IMF, a minimum mass for the CCSNe $\rm \sim 9.5~M_\odot$ and an upper mass limit $\rm M_{\rm top}\sim 25~M_\odot$): this number is in very good agreement with the SN rates estimated from the volume-limited Lick Observatory Supernova Search (LOSS) \citep{smith11}. 

Rotation changes completely this scenario because it lowers both the maximum mass that spends a substantial amount of time on the RSG branch and the minimum mass that becomes WR. In particular we get that even the 13\ \msun \ rotating at either 150 and 300 km/s becomes a WR star. The maximum mass the reaches the RSG branch lowers to 30\ \msun \ for v=150 km/s and to 25\ \msun for v=300 km/s. Therefore we predict that solar metallicity models, with initial rotation velocities $\rm v\geq 150~km/s$, explode as SNIb ($\rm M_{Ib}\lesssim 13~M_\odot$) and that RSG stars are expected up to a luminosity of the order of $\rm Log(L/L_\odot) \sim 5.5$. 

At [Fe/H]=-1, in absence of rotation the strong decrease of the mass loss (coupled to the strong reduction of the opacity) raises the maximum mass that settles on the RSG branch up to 25-30\ \msun \ but it also prevents the more massive ones from reaching the RSG branch before the explosion because these stars turn towards their Hayashi track on a nuclear timescale. Hence we expect RSG SNIIP progenitors with masses as high as $\rm M_{\rm IIP}\sim 25-30~M_\odot$ and that above this value they explode as BSGs SNIIP. Given the very low amount of mass lost at this metallicity, WR stars may form only at masses $\rm M \geq 80~M_\odot$. At lower metallicities $\rm ([Fe/H] \leq -2)$ also the less massive stars turn to the red on a nuclear timescale so that basically no stars are expected to populate significantly the RSG branch at these metallicities. This does not imply that we do not expect RSG SNIIP because stars in the range 13 to 25\ \msun \ reach the RSG branch towards the end of the central He burning. Given the extremely low amount of mass lost by these stars all of them explode as SNIIP (blue or red supergiants).

The main role of rotation at these metallicities is that of pushing the stars towards the RSG branch. As a consequence, stars below 25-30\ \msun \ spend now a considerable amount of time on the RSG branch where they eventually explode as red supergiants (SNIIP) while the more massive ones approach their Eddington luminosity when their surface temperature drops to $\rm Log_T{\rm eff}\sim 3.8-4.0$, lose a huge a mount of mass and turn to the blue becoming WR stars. So at subsolar metallicities rotation populates the RSG branch up to 25-30\ \msun, leads to the formation of WR stars in a quite wide range of masses, and settles a limiting mass between SNIIP and SNIb in the range $\rm M_{\rm IIP} 25-30 M_\odot$. Though up to now no progenitor for SNIIP as massive as $\rm 25-30~M_\odot$ has been detected yet, the present results imply the existence of these massive RSG progenitors (and SNIIP) at low metallicities and we certainly expect such a finding in the next future.

Before closing this part it is important to stress that since stars do not rotate at the same rotation velocity (as seen from the available observational data) but they show a spread, the choice of an IDROV (Initial Distribution of ROtation Velocities) is mandatory (as the IMF for the mass distribution) when trying to fit the properties of a given sample of stars or a trend with the metallicity \citep[CL13]{prantzos2018}. In presence of a spread of initial rotation velocities even the definition of a limiting mass becomes much more ambiguous because it depends not just on the metallicity but also on $\rm V_{\rm ini}$. A detailed and quantitative analysis of the properties of a population of massive stars, where both the IMF and the IDROV are taken into account, are beyond the scope of the present paper and will be addressed in a future work. 


The direct detections of gravitational waves, GW150914, GW151226, GW170104, GW170608, GW170814 \citep{Abb6,Abb5,Abb4,Abb3,Abb2,Abb1} have been associated to the merger of two black holes, presumably of stellar origin, GW150914 and GW170608 being associated to the largest ($\rm \sim 29-36~M_\odot$) and the smallest ($\rm \sim 7-12~M_\odot$) black holes in a binary system, respectively. In a scenario in which all stars eject 0.07 $\rm M_\odot$ of $\rm ^{56}Ni$ the maximum mass of the remnant ranges between $\rm \sim 4~M_\odot$ and $\rm \sim 8~M_\odot$, for  $\rm -3\leq[Fe/H]\leq 0$ and $\rm 0\leq v_{ini}\leq 300~km/s$. In the alternative scenario in which all stars with $\rm M>25~M_\odot$ fail to explode and directly collapse to a black hole, the remnant masses are obviously much larger and can reach values as high as $\rm \sim 80~M_\odot$ for low metallicity non rotating models. {\color{black} By the way, in this paper we just want to show that, in principle, a black hole with a mass as large as $\rm \sim 80~M_\odot$ can be obtained as the final product of the evolution of a single massive star, without any claim on the route towards the formation of the black hole binary systems as well as their merging \citep{antonini2016,belczynski2016,mandel2016}. Of course such a result may change, even significantly, if the massive star evolves in a binary system \citep{demink2016}.}
A natural upper limit to the maximum mass a (stellar) black hole is given by the fact that stars above a given critical mass ($\rm M_{PISN}$) enter the pair instability regime and explode as pair instability supernovae without leaving any remnant. In general $\rm M_{PISN}$ decreases with increasing the initial rotation velocity because of the substantial increase of the CO core induced by the rotation driven mixing so that rotation favors the formation of PISN. More specifically we get that at subsolar metallicities $\rm M_{PISN}\sim 100~M_\odot$ for $\rm v_{\rm ini}=0$ and it decreases down to $\rm M_{PISN}\sim 70~M_\odot$ for an initial rotation velocity $\rm v_{\rm ini}=300$ km/s.

As far as the final yields are concerned, our current preferred scenario (set R) is obtained by requiring that stars in the range 13 to 25\ \msun \ experience mixing and fall back \citep{UN02} (the inner and the outer borders of the mixed region being fixed by imposing that [Ni/Fe]=0.2 and that it cannot extend beyond the base of the Oxygen shell) and then the mass cut is fixed by requiring the ejection of 0.07\ \msun \ of \nuk{Ni}{56}. Stars more massive than 25\ \msun \ are assumed to fully collapse in the remnant so that they contribute to the yields only through their wind. The same choices were assumed for all metallicities and initial rotation velocities. The overall behavior of the yields provided by the non rotating models is basically the expected one. At solar metallicity the production factors of the intermediate mass nuclei shows a reasonably flat profile, the elements between Zn and Zr are also produced by the so called weak component. Elements beyond the neutron closure shell n=50 are not produced at all. As the metallicity reduces, the production factors of the intermediate mass even particle remain quite flat (because of their primary origin) and close to that of the O while the odd nuclei show a progressively more pronounced odd-even effect. Elements beyond Zn are not produced at all, as well as N and F that are also severely underproduced. 

The most striking effect of rotation on the yields is the substantial production of N, F and all the elements heavier than Fe up to Pb in low metallicity models. Vice versa at solar metallicity the influence of rotation on the yields is modest (at least up to $\rm V_{\rm ini}=300$ km/s) . The continuous stirring of matter between the central He burning and the H burning shell leads to a large production of all the CNO cycle (and obviously \nuk{N}{14} which is the most abundant) that are then redistributed within the He core. The fraction of these nuclei that falls in the convective core is rapidly nuclearly processed and in particular \nuk{N}{14} is converted in \nuk{Ne}{22}, i.e. a powerful primary neutron source. Another fraction of them remains frozen in the radiative part of the He core and becomes the fuel that powers the synthesis of F when the He convective shell forms after the central He exhaustion. The capability of the matter to flux beyond the two neutron closure shells at n=50 and n=82 depends basically on the neutron to seed ratio because this is the quantity that controls the timescale necessary to overcome the neutron closure shells. A natural limit to the maximum overproduction of the heavy elements is obviously given by the total amount of matter that may be pushed onward, and obviously this is the total amount of Fe available. Our rotating models raise the neutron to seed ratio above 1000, {\color{black} a} value more than enough to allow the passage of the matter through the neutron closure shell in a timescale much smaller than the lifetimes of the stars in He burning {\color{black} (see, e.g., Fig7-22 in \citet{clayton68})}. Hence our models are able to produce efficiently nuclei up to Pb. The production factors of the heavy nuclei, however, reaches the maximum at [Fe/H]=-1 and then drops mildly, the obvious explanation being that the lower the metallicity the lower the availability of target nuclei, i.e. Fe.  

The final yields are the results of the computation of a large number of models from the pre main sequence to the core collapse plus the explosion. All these computations are however based on a large number of assumptions and choices that may vary from author to author. In order to check if these yields are capable to fit the solar distribution as well as their observed trend with the metallicity we have already included them in a galactic chemical evolution model \citep{prantzos2018} and we strongly refer the reader to that paper to have an idea of how these yields behave. Here we want just to point out that in that paper we calibrated the IDROV (Initial Distribution of ROtation Velocities) by requiring a primary-like behavior of $\rm ^{14}N$ in the Milky Way at low metallicity and at the same time to avoid overproduction of s-process elements at intermediate metallicities. The next step in this endless attempt to produce scenarii closer and closer to the real world will the extension of the grid to zero metallicity and super metal rich stars.

\bigskip

\noindent
M.L. warmly thanks Francesca Primas and Ferdinando Patat for thier generous hospitality during my visit at ESO and acknowledges the support by the ESO Visitor Program 2017-2018.
This work has been partially supported by the italian grants "Premiale 2015 MITiC" (P.I. B. Garilli) and "Premiale 2015 FIGARO" (P.I. G. Gemme).

\appendix

\section{The FRANEC code}\label{app:franec}

{\color{black}
In this section we recall briefly all the main features and input physics of our evolutionary code (FRANEC), already presented in detail in CL13.

The set of equations describing the physical structure of the star plus the ones governing the chemical evolution of the chemical composition (i.e., the local burning due to the nuclear reactions plus the various kinds of mixing: convection, semiconvection, and rotationally induced mixing) are coupled together and solved simultaneously by means of a relaxation technique. The angular momentum transport, together to the determination of the velocity of the meridional circulation, is solved separately, again by means of a relaxation technique.

The borders of the convective zones are defined according to the Ledoux criterion. In addition, we assume 0.2 $\rm H_{P}$ of overshooting at the outer edge of the convective core only during the core H burning phase.

The equation of state (EoS) adopted for temperatures lower than $\rm 10^{6}~K$ in the one provided by  \citet{rogers96} and \citet{rogers01} (EOS and ESOPLUS). For temperatures higher than $\rm 10^{6}~K$ we are using the EoS Tables given by \citet{straniero88}.

The radiative opacity coefficients are derived from \citet{kurucz91} for $\rm T\leq10^{4}~K$, from \citet{iglesias92} for $\rm 10^{4}< T(K) \leq 10^{8}$ (OPAL), and from the Los Alamos Opacity Library (LAOL) \citet{huebner77} for $\rm 10^{8}< T(K) \leq 10^{10}$. The opacity coefficients due to the thermal conductivity are derived from \citet{itoh83}.

Mass loss has been included following the prescriptions of \citet{val00,val01} for the blue supergiant phase ($\rm T_{eff}>12000~K$), \citet{dejager88} for the red supergiant phase ($\rm T_{eff}<12000~K$) and \citet{nl00} for the Wolf-Rayet phase.  The enhancement of the mass loss due to the formation of dust during the red supergiant phase has been included following the prescriptions of \citet{vanloonetal05}. Mass loss is enhanced, in rotating models, following the prescription of \cite{hlw00}.

The criteria adopted to define the different WR subclasses are the same adopted in \cite{lc06}, i.e., we as- sume that the star enters the WR phase when $\rm Log T_{eff} > 4$ and $\rm H_{surf}<0.4$, and we adopt the following usual definitions for the various WR phases: WNL ($\rm 10^{-5}< H_{surf}<0.4$), WNE [$\rm H_{surf}<10^{-5}$ and $\rm (C/N)_{surf}<0.1$], WNC [$\rm 0.1<(C/N)_{surf}<10$], and WCO [$\rm (C/N)_{surf}>10$].

The nuclear network and the cross section adopted have been discussed and presented in detail in the text (see Tables \ref{tabnetwork}, \ref{tabreferences} and \ref{crosspecial}). 

The effect of rotation on the structure of the star has been included following the "shellular rotation" approach \citep{z92,mm97}. The transport of the angular momentum due to meridional circulation and shear turbulence has been treated by means of the advective-diffusive equation \citep{tal97,cz92}. 
\begin{equation}
\rho r^{2}{dr^{2}\omega \over dt}={1 \over 5}{\partial \over \partial r}(\rho r^{4}U\omega )+
{\partial \over \partial r}\left(\rho r^{4}D_{s.i.}{\partial \omega \over \partial r}\right)
\label{advdiff}
\end{equation}
where $U$ is the radial component of the velocity of the meridional circulation, $D_{s.i.}$ the diffusion coefficient for the shear turbulence and all other quantities have their usual meaning. We adopted the prescription for $U$ provided by \citet{mz98} (their eq. 4.38) and the formulation of $D_{s.i.}$ proposed by \citet{tz97}, modified later by \citet{Pal03}:
\begin{equation}
D_{s.i.}=\frac{8}{5}\frac{R_{\mathit{ic}}(rd\omega/\mathit{dr})^{2}}{N_{T}^{2}/(K+D_{h})+N_{\mu }^{2}/D_{h}}
\label{dshear}
\end{equation}
where $N_{T}^{2}=\frac{g\delta }{H_{P}}(\nabla _{\mathit{ad}}-\nabla_{\mathit{rad}})$,
$N_{\mu }^{2}=\frac{g\delta }{H_{P}}(\frac{\varphi}{\delta }\nabla _{\mu })$ and
$R_{\mathit{ic}}=\frac{1}{4}$. $\rm D_h$ is the coefficient of horizontal turbulence that we assume citet{Zah92}:
\begin{equation}
D_h \simeq | r~U |
\end{equation}
where $K=\frac{4acT^3}{3C_pk\rho^2}$ is the thermal diffusivity \citep{mm00}. The adoption of the expression for the velocity of meridional circulation provided by \citet{mz98} in equation \ref{advdiff} leads to a fourth order partial differential equation that is solved by means of a relaxation technique.
In contrast, the transport of the chemical species has been treated as a pure diffusive process according to \citet{cz92}. in this case the diffusion coefficient is given by
\begin{equation}
D = D_{s.i.}+D_{m.c.}
\label{diffcoeff}
\end{equation}
where $D_{s.i.}$ is given by equation \ref{dshear} while $D_{m.c.}$ is
\begin{equation}
D_{m.c.} \simeq {| r~U |^2 \over 30 D_h}
\label{diffcoeffmc}
\end{equation}
as suggested by \citet{z92}.

The calibration of the mixing efficiency has been performed by means of two free parameters, namely, 
$f_c$ and $f_{\mu}$. The first free parameter, $f_c$, simply multiplies the total diffusion coefficient defined in Equation \ref{diffcoeff} and adopted for the mixing of the chemical composition, i.e.
\begin{equation}
D = f_c\times (D_{s.i.}+D_{m.c.})
\label{diffcoeff_calibrated}
\end{equation}
The second free parameter, $f_{mu}$, multiplies the gradient of molecular
weight, i.e., $\nabla_\mu^{\rm adopted}=f_\mu \times \nabla_\mu$, and regulates the influence of this 
quantity on the mixing of both the angular momentum and the chemical composition.
The calibration procedure is described in detail in the text (section \ref{calimix}).

As usual the initial He abundance and the mixing-length parameters have been determined by fitting the present properties of the Sun \citep{scl97}.

}

\floattable                              
\startlongtable
\begin{deluxetable*}{ccccccccccccccc}
\tablecolumns{15}
\tabletypesize{\scriptsize}
\tablecaption{Network reference matrix \label{tabreferences}}
\tablehead{
\colhead{isotope} &
\colhead{(p,$\alpha$)} & \colhead{(p,n)} & \colhead{(p,$\gamma$)} &
\colhead{($\alpha$,p)} & \colhead{($\alpha$,n)} & \colhead{($\alpha$,$\gamma$)} &
\colhead{(n,p)} & \colhead{(n,$\alpha$)} & \colhead{(n,$\gamma$)} &
\colhead{($\gamma$,p)} & \colhead{($\gamma$,$\alpha$)} & \colhead{($\gamma$,n)} &
\colhead{($\beta^+$)} & \colhead{($\beta^-$)} }
\startdata   
$\rm ^{2  }H $ &       &      & desc &      &      & nacr &      &      & ka03 &      &      &      &      &       \\
$\rm ^{3  }H $ &       & desc & cf88 &      &      &      &      &      &      &      &      &      &      & rath  \\
$\rm ^{3  }He$ &       &      & bet+ &      &      &      & desc &      & ka03 &      &      &      &      &       \\
$\rm ^{6  }Li$ &       &      & nacr & nacr &      & cf88 &      &      & mafo &      &      &      &      &       \\
$\rm ^{7  }Li$ &  desc & desc &      & wago & cf88 & nacr &      &      & ka03 &      &      &      &      &       \\
$\rm ^{7  }Be$ &       &      & nacr & nacr &      & nacr & desc & wago &      &      &      &      &      & cf88  \\
$\rm ^{9  }Be$ &  nacr &      & nacr & wago & nacr &      &      &      & ka03 &      &      &      &      &       \\
$\rm ^{10 }Be$ &  wago &      & wago &      & fkth & fkth &      &      & bb92 &      &      &      &      & rath  \\
\enddata 
\tablenotetext{a}{We treat the ground ($\rm ^{26 }Al^{\rm g}$) and isomeric ($\rm ^{26 }Al^{\rm m}$) states of $\rm ^{26}Al$ as separate species for $\rm T\leq 10^9~K$, while we assume the two states to be in statistical equilibrium (and therefore we consider just one isotope) above this temperature \citep{lc06}}
\tablenotetext{b}{$\rm ^{26}Al^{\rm g}$ ground state}
\tablenotetext{c}{$\rm ^{26}Al^{\rm m}$ isomeric state}
\tablecomments{
bet+ = beta plus decay;
desc = \citet{desc04};
ka03 = \citet{kadonis03}, KADoNiS v0.3; 
nacr = \citet{nacre}, NACRE;
cf88 = \citet{ca88};
ku02 = \citet{kunzetal02}, adopted rate;
im05 = \citet{imbriani05};
mc10 = \citet{mc10};
bb92 = \citet{bb92};
wies = M. Wiescher and collaborators;
rath = last version of the REACLIB available at the web site https://nucastro.org/reaclib.html;
wago = \citet{wago};
mafo = \citet{mafo};
fkth = \cite{fkth};
wfho = \cite{wfho};
od94 = \citet{odaetal94};
bl03 = \cite{bl03};
kart = \citet{kadonis03} in the energy range 5-100 keV and \citet{RT2000} above this limit, but rescaled to match the experimental values of \citet{kadonis03} at 100 keV;
mc11 = \citet{mc11};
rt   = \citet{RT2000};
de07 = \cite{de07};
ko97 = \cite{ko97};
ffn8 = \citet{FFN82};
og10 = \cite{og10};
ka02 = \cite{ka02}, , KADoNiS v0.2; 
il09 = \cite{il09};
nass = \cite{nass}
lp00 = \citet{lp00};
laur = \citet{laur};
re98 = \citet{re98};
taka = \citet{ty87};
tanu = terrestrial half-life.
}
\end{deluxetable*}                                                                                                                                                                                                                                                                                                                                                                                                                                                                                                  

\floattable                              
\startlongtable
\begin{deluxetable*}{lc}
\tablecolumns{2}
\tabletypesize{\scriptsize}
\tablecaption{Network reference matrix for special nuclear reactions \label{crosspecial}}
\tablehead{
\colhead{reaction} & \colhead{reference} }
\startdata   
         p             (p            ,$\rm e^+\nu$)$\rm ^{2}H$   & nacr    \\
         $\rm ^{3}He$  ($\rm ^{3}He$ ,2p          )$\rm \alpha$  & nacr    \\
         $\rm \alpha$  ($\rm ^{3}He$ ,$\rm \gamma$)$\rm ^{7}Be$  & cy08    \\
         $\rm ^{2}H$   ($\rm ^{2}H$  ,p           )$\rm ^{3}H$   & nacr    \\       
         $\rm ^{2}H$   ($\rm ^{2}H$  ,n           )$\rm ^{3}He$  & desc    \\                  
         $\rm ^{2}H$   ($\rm ^{2}H$  ,$\rm \gamma$)$\rm \alpha$  & nacr    \\
         $\rm ^{3}H$   (p            ,$\rm ^{2}H$ )$\rm ^{2}H$   & nacr    \\       
         $\rm ^{3}H$   ($\rm ^{2}H$  ,n           )$\rm \alpha$  & desc    \\
\enddata 
\tablecomments{
cy08 = \citet{cy08};
da77 = \citet{da77};
rolf = C. Rolfs and collaborators;
}
\end{deluxetable*}                                                               

\include{tab01}

\clearpage                                                                                                                                                                                                                   
\floattable                                                                                                                                                                                                                                                                                                                                                                                                                                                                                                         
\begin{deluxetable*}{ccccccccccc}                                                                                                                                                                                                                                                                                                                                                                                                                                                                                   
\tabletypesize{\scriptsize}                                                                                                                                                                                                                                                                                                                                                                                                                                                                                         
\tablecolumns{11}                                                                                                                                                                                                                                                                                                                                                                                                                                                                                                   
\tablewidth{0pc}                                                                                                                                                                                                                                                                                                                                                                                                                                                                                                    
           \tablecaption{Wolf-Rayet Lifetimes of the models with v=0   km/s and metallicity [Fe/H]=0  \label{tabwr0000}}
\tablehead{                                                                                                                                                                                                                                                                                                                                                                                                                                                                                                         
 Initial Mass   &   $t_{\rm O}$   & $t_{\rm WR}$   &   $t_{\rm WNL}$  &  $\rm H_c/He_c~(WNL)$   &  $t_{\rm WNE}$  & $\rm H_c/He_c~(WNE)$ & $t_{\rm WNC}$ & $\rm H_c/He_c~(WNC)$ & $t_{\rm WC}$ & $\rm H_c/He_c~(WC)$  \\                                                                                                                                                                                                                                                                                            
($\rm M_\odot$) &     (yr)        & (yr)            &   (yr)          &  mass fraction          &  (yr)           & mass fraction        &  (yr)           & mass fraction        &  (yr)       & mass fraction        }                                                                                                                                                                                                                                                                                            
\startdata                                                                                                                                                                                                                                                                                                                                                                                                                                                                                                          
  13   &             &             &           &               &           &               &           &               &           &              \\
  15   &  8.31(+4)   &             &           &               &           &               &           &               &           &              \\
  20   &  6.44(+6)   &  7.79(+4)   &  7.79(+4) &    He(0.05)   &           &               &           &               &           &              \\
  25   &  5.94(+6)   &  2.89(+5)   &  1.51(+5) &    He(0.45)   &  1.39(+5) &    He(0.14)   &           &               &           &              \\
  30   &  5.14(+6)   &  3.07(+5)   &  2.07(+5) &    He(0.57)   &  1.00(+5) &    He(0.11)   &           &               &           &              \\
  40   &  4.24(+6)   &  3.03(+5)   &  1.75(+5) &    He(0.73)   &  1.28(+5) &    He(0.20)   &           &               &           &              \\
  60   &  3.20(+6)   &  3.52(+5)   &  7.69(+4) &    He(0.96)   &  1.56(+5) &    He(0.67)   &  1.67(+4) &    He(0.20)   &  1.03(+5) &    He(0.16)  \\
  80   &  2.63(+6)   &  3.71(+5)   &  9.27(+4) &    H (0.01)   &  1.01(+5) &    He(0.78)   &  1.61(+4) &    He(0.40)   &  1.61(+5) &    He(0.34)  \\
\enddata
 \end{deluxetable*}
\floattable                                                                                                                                                                                                                                                                                                                                                                                                                                                                                                         
\begin{deluxetable*}{ccccccccccc}                                                                                                                                                                                                                                                                                                                                                                                                                                                                                   
\tabletypesize{\scriptsize}                                                                                                                                                                                                                                                                                                                                                                                                                                                                                         
\tablecolumns{11}                                                                                                                                                                                                                                                                                                                                                                                                                                                                                                   
\tablewidth{0pc}                                                                                                                                                                                                                                                                                                                                                                                                                                                                                                    
           \tablecaption{Wolf-Rayet Lifetimes of the models with v=150 km/s and metallicity [Fe/H]=0  \label{tabwr1500}}
\tablehead{                                                                                                                                                                                                                                                                                                                                                                                                                                                                                                         
 Initial Mass   &   $t_{\rm O}$   & $t_{\rm WR}$   &   $t_{\rm WNL}$  &  $\rm H_c/He_c~(WNL)$   &  $t_{\rm WNE}$  & $\rm H_c/He_c~(WNE)$ & $t_{\rm WNC}$ & $\rm H_c/He_c~(WNC)$ & $t_{\rm WC}$ & $\rm H_c/He_c~(WC)$  \\                                                                                                                                                                                                                                                                                            
($\rm M_\odot$) &     (yr)        & (yr)            &   (yr)          &  mass fraction          &  (yr)           & mass fraction        &  (yr)           & mass fraction        &  (yr)       & mass fraction        }                                                                                                                                                                                                                                                                                            
\startdata                                                                                                                                                                                                                                                                                                                                                                                                                                                                                                          
  13   &             &  3.59(+5)   &  3.42(+5) &    He(0.15)   &           &               &  1.71(+4) &               &           &              \\
  15   &  1.72(+4)   &  3.12(+5)   &  3.09(+5) &    He(0.14)   &  8.80(+2) &               &  2.99(+3) &               &           &              \\
  20   &  7.89(+6)   &  5.45(+5)   &  3.86(+5) &    He(0.64)   &           &               &  2.89(+4) &    He(0.13)   &  1.30(+5) &    He(0.09)  \\
  25   &  7.51(+6)   &  4.89(+5)   &  2.71(+5) &    He(0.71)   &           &               &  1.48(+4) &    He(0.24)   &  2.03(+5) &    He(0.22)  \\
  30   &  6.74(+6)   &  4.49(+5)   &  2.32(+5) &    He(0.76)   &           &               &  1.14(+4) &    He(0.29)   &  2.05(+5) &    He(0.27)  \\
  40   &  5.46(+6)   &  4.62(+5)   &  2.30(+5) &    He(0.94)   &           &               &  7.63(+3) &    He(0.37)   &  2.24(+5) &    He(0.36)  \\
  60   &  4.06(+6)   &  3.79(+5)   &  1.12(+5) &    He(0.95)   &           &               &  9.06(+3) &    He(0.57)   &  2.58(+5) &    He(0.55)  \\
  80   &  3.40(+6)   &  3.41(+5)   &  7.92(+4) &    He(0.94)   &           &               &  9.07(+3) &    He(0.64)   &  2.53(+5) &    He(0.61)  \\
\enddata
 \end{deluxetable*}
\floattable                                                                                                                                                                                                                                                                                                                                                                                                                                                                                                         
\begin{deluxetable*}{ccccccccccc}                                                                                                                                                                                                                                                                                                                                                                                                                                                                                   
\tabletypesize{\scriptsize}                                                                                                                                                                                                                                                                                                                                                                                                                                                                                         
\tablecolumns{11}                                                                                                                                                                                                                                                                                                                                                                                                                                                                                                   
\tablewidth{0pc}                                                                                                                                                                                                                                                                                                                                                                                                                                                                                                    
           \tablecaption{Wolf-Rayet Lifetimes of the models with v=300 km/s and metallicity [Fe/H]=0  \label{tabwr3000}}
\tablehead{                                                                                                                                                                                                                                                                                                                                                                                                                                                                                                         
 Initial Mass   &   $t_{\rm O}$   & $t_{\rm WR}$   &   $t_{\rm WNL}$  &  $\rm H_c/He_c~(WNL)$   &  $t_{\rm WNE}$  & $\rm H_c/He_c~(WNE)$ & $t_{\rm WNC}$ & $\rm H_c/He_c~(WNC)$ & $t_{\rm WC}$ & $\rm H_c/He_c~(WC)$  \\                                                                                                                                                                                                                                                                                            
($\rm M_\odot$) &     (yr)        & (yr)            &   (yr)          &  mass fraction          &  (yr)           & mass fraction        &  (yr)           & mass fraction        &  (yr)       & mass fraction        }                                                                                                                                                                                                                                                                                            
\startdata                                                                                                                                                                                                                                                                                                                                                                                                                                                                                                          
  13   &             &  5.23(+5)   &  4.40(+5) &    He(0.31)   &           &               &  8.29(+4) &    He(0.02)   &           &              \\
  15   &  1.14(+4)   &  3.68(+5)   &  2.93(+5) &    He(0.20)   &           &               &  7.45(+4) &    He(0.02)   &           &              \\
  20   &  6.81(+6)   &  2.85(+5)   &  1.51(+5) &    He(0.25)   &           &               &  1.34(+5) &    He(0.09)   &           &              \\
  25   &  8.92(+6)   &  5.29(+5)   &  2.19(+5) &    He(0.78)   &           &               &  2.41(+4) &    He(0.39)   &  2.86(+5) &    He(0.34)  \\
  30   &  7.79(+6)   &  5.29(+5)   &  2.57(+5) &    He(0.92)   &           &               &  1.31(+4) &    He(0.38)   &  2.59(+5) &    He(0.36)  \\
  40   &  6.04(+6)   &  4.65(+5)   &  1.60(+5) &    He(0.96)   &           &               &  1.08(+4) &    He(0.54)   &  2.95(+5) &    He(0.51)  \\
  60   &  4.17(+6)   &  1.01(+6)   &  6.88(+5) &    H (0.12)   &  4.19(+3) &    He(0.72)   &  5.05(+4) &    He(0.71)   &  2.64(+5) &    He(0.55)  \\
  80   &  3.33(+6)   &  1.09(+6)   &  7.74(+5) &    H (0.15)   &  1.98(+4) &    He(0.74)   &  4.39(+4) &    He(0.67)   &  2.50(+5) &    He(0.53)  \\
\enddata
 \end{deluxetable*}
\floattable                                                                                                                                                                                                                                                                                                                                                                                                                                                                                                         
\begin{deluxetable*}{ccccccccccc}                                                                                                                                                                                                                                                                                                                                                                                                                                                                                   
\tabletypesize{\scriptsize}                                                                                                                                                                                                                                                                                                                                                                                                                                                                                         
\tablecolumns{11}                                                                                                                                                                                                                                                                                                                                                                                                                                                                                                   
\tablewidth{0pc}                                                                                                                                                                                                                                                                                                                                                                                                                                                                                                    
           \tablecaption{Wolf-Rayet Lifetimes of the models with v=0   km/s and metallicity [Fe/H]=-1 \label{tabwr0001}}
\tablehead{                                                                                                                                                                                                                                                                                                                                                                                                                                                                                                         
 Initial Mass   &   $t_{\rm O}$   & $t_{\rm WR}$   &   $t_{\rm WNL}$  &  $\rm H_c/He_c~(WNL)$   &  $t_{\rm WNE}$  & $\rm H_c/He_c~(WNE)$ & $t_{\rm WNC}$ & $\rm H_c/He_c~(WNC)$ & $t_{\rm WC}$ & $\rm H_c/He_c~(WC)$  \\                                                                                                                                                                                                                                                                                            
($\rm M_\odot$) &     (yr)        & (yr)            &   (yr)          &  mass fraction          &  (yr)           & mass fraction        &  (yr)           & mass fraction        &  (yr)       & mass fraction        }                                                                                                                                                                                                                                                                                            
\startdata                                                                                                                                                                                                                                                                                                                                                                                                                                                                                                          
  13   &  3.66(+6)   &             &           &               &           &               &           &               &           &              \\
  15   &  8.95(+6)   &             &           &               &           &               &           &               &           &              \\
  20   &  8.25(+6)   &             &           &               &           &               &           &               &           &              \\
  25   &  6.96(+6)   &             &           &               &           &               &           &               &           &              \\
  30   &  6.09(+6)   &             &           &               &           &               &           &               &           &              \\
  40   &  4.99(+6)   &             &           &               &           &               &           &               &           &              \\
  60   &  3.82(+6)   &             &           &               &           &               &           &               &           &              \\
  80   &  3.20(+6)   &  2.80(+5)   &  2.80(+5) &    He(0.96)   &           &               &           &               &           &              \\
\enddata
 \end{deluxetable*}
\floattable                                                                                                                                                                                                                                                                                                                                                                                                                                                                                                         
\begin{deluxetable*}{ccccccccccc}                                                                                                                                                                                                                                                                                                                                                                                                                                                                                   
\tabletypesize{\scriptsize}                                                                                                                                                                                                                                                                                                                                                                                                                                                                                         
\tablecolumns{11}                                                                                                                                                                                                                                                                                                                                                                                                                                                                                                   
\tablewidth{0pc}                                                                                                                                                                                                                                                                                                                                                                                                                                                                                                    
           \tablecaption{Wolf-Rayet Lifetimes of the models with v=150 km/s and metallicity [Fe/H]=-1 \label{tabwr1501}}
\tablehead{                                                                                                                                                                                                                                                                                                                                                                                                                                                                                                         
 Initial Mass   &   $t_{\rm O}$   & $t_{\rm WR}$   &   $t_{\rm WNL}$  &  $\rm H_c/He_c~(WNL)$   &  $t_{\rm WNE}$  & $\rm H_c/He_c~(WNE)$ & $t_{\rm WNC}$ & $\rm H_c/He_c~(WNC)$ & $t_{\rm WC}$ & $\rm H_c/He_c~(WC)$  \\                                                                                                                                                                                                                                                                                            
($\rm M_\odot$) &     (yr)        & (yr)            &   (yr)          &  mass fraction          &  (yr)           & mass fraction        &  (yr)           & mass fraction        &  (yr)       & mass fraction        }                                                                                                                                                                                                                                                                                            
\startdata                                                                                                                                                                                                                                                                                                                                                                                                                                                                                                          
  13   &  6.31(+5)   &             &           &               &           &               &           &               &           &              \\
  15   &  1.03(+7)   &             &           &               &           &               &           &               &           &              \\
  20   &  9.75(+6)   &             &           &               &           &               &           &               &           &              \\
  25   &  8.10(+6)   &             &           &               &           &               &           &               &           &              \\
  30   &  6.89(+6)   &             &           &               &           &               &           &               &           &              \\
  40   &  5.62(+6)   &  3.72(+4)   &  3.72(+4) &    He(0.04)   &           &               &           &               &           &              \\
  60   &  4.19(+6)   &  3.42(+5)   &  1.07(+5) &    He(0.98)   &           &               &  2.81(+3) &    He(0.57)   &  2.32(+5) &    He(0.56)  \\
  80   &  3.42(+6)   &  3.15(+5)   &  9.55(+4) &    He(0.99)   &           &               &  5.91(+2) &    He(0.60)   &  2.19(+5) &    He(0.60)  \\
\enddata
 \end{deluxetable*}
\floattable                                                                                                                                                                                                                                                                                                                                                                                                                                                                                                         
\begin{deluxetable*}{ccccccccccc}                                                                                                                                                                                                                                                                                                                                                                                                                                                                                   
\tabletypesize{\scriptsize}                                                                                                                                                                                                                                                                                                                                                                                                                                                                                         
\tablecolumns{11}                                                                                                                                                                                                                                                                                                                                                                                                                                                                                                   
\tablewidth{0pc}                                                                                                                                                                                                                                                                                                                                                                                                                                                                                                    
           \tablecaption{Wolf-Rayet Lifetimes of the models with v=300 km/s and metallicity [Fe/H]=-1 \label{tabwr3001}}
\tablehead{                                                                                                                                                                                                                                                                                                                                                                                                                                                                                                         
 Initial Mass   &   $t_{\rm O}$   & $t_{\rm WR}$   &   $t_{\rm WNL}$  &  $\rm H_c/He_c~(WNL)$   &  $t_{\rm WNE}$  & $\rm H_c/He_c~(WNE)$ & $t_{\rm WNC}$ & $\rm H_c/He_c~(WNC)$ & $t_{\rm WC}$ & $\rm H_c/He_c~(WC)$  \\                                                                                                                                                                                                                                                                                            
($\rm M_\odot$) &     (yr)        & (yr)            &   (yr)          &  mass fraction          &  (yr)           & mass fraction        &  (yr)           & mass fraction        &  (yr)       & mass fraction        }                                                                                                                                                                                                                                                                                            
\startdata                                                                                                                                                                                                                                                                                                                                                                                                                                                                                                          
  13   &  3.46(+4)   &             &           &               &           &               &           &               &           &              \\
  15   &  7.19(+6)   &             &           &               &           &               &           &               &           &              \\
  20   &  9.78(+6)   &             &           &               &           &               &           &               &           &              \\
  25   &  8.95(+6)   &             &           &               &           &               &           &               &           &              \\
  30   &  7.63(+6)   &  4.33(+4)   &  4.03(+4) &    He(0.04)   &  2.97(+3) &               &           &               &           &              \\
  40   &  5.85(+6)   &  1.25(+5)   &  6.64(+4) &    He(0.19)   &           &               &  2.60(+4) &    He(0.07)   &  3.25(+4) &    He(0.03)  \\
  60   &  4.35(+6)   &  3.47(+5)   &  9.87(+4) &    He(0.98)   &           &               &  3.85(+3) &    He(0.61)   &  2.45(+5) &    He(0.60)  \\
  80   &  3.27(+6)   &  4.78(+5)   &  2.35(+5) &    H (0.05)   &           &               &  3.01(+3) &    He(0.66)   &  2.40(+5) &    He(0.65)  \\
\enddata
 \end{deluxetable*}

\clearpage

\floattable                                                                                                                                                                                                                                                                                                                                                                                                                                                                                                         
\begin{deluxetable*}{ccccccccccc}                                                                                                                                                                                                                                                                                                                                                                                                                                                                                   
\tabletypesize{\scriptsize}                                                                                                                                                                                                                                                                                                                                                                                                                                                                                         
\tablecolumns{11}                                                                                                                                                                                                                                                                                                                                                                                                                                                                                                   
\tablewidth{0pc}                                                                                                                                                                                                                                                                                                                                                                                                                                                                                                    
           \tablecaption{Wolf-Rayet Lifetimes of the models with v=0   km/s and metallicity [Fe/H]=-2 \label{tabwr0002}}
\tablehead{                                                                                                                                                                                                                                                                                                                                                                                                                                                                                                         
 Initial Mass   &   $t_{\rm O}$   & $t_{\rm WR}$   &   $t_{\rm WNL}$  &  $\rm H_c/He_c~(WNL)$   &  $t_{\rm WNE}$  & $\rm H_c/He_c~(WNE)$ & $t_{\rm WNC}$ & $\rm H_c/He_c~(WNC)$ & $t_{\rm WC}$ & $\rm H_c/He_c~(WC)$  \\                                                                                                                                                                                                                                                                                            
($\rm M_\odot$) &     (yr)        & (yr)            &   (yr)          &  mass fraction          &  (yr)           & mass fraction        &  (yr)           & mass fraction        &  (yr)       & mass fraction        }                                                                                                                                                                                                                                                                                            
\startdata                                                                                                                                                                                                                                                                                                                                                                                                                                                                                                          
  13   &  1.33(+7)   &             &           &               &           &               &           &               &           &              \\
  15   &  1.21(+7)   &             &           &               &           &               &           &               &           &              \\
  20   &  9.37(+6)   &             &           &               &           &               &           &               &           &              \\
  25   &  7.40(+6)   &             &           &               &           &               &           &               &           &              \\
  30   &  6.25(+6)   &             &           &               &           &               &           &               &           &              \\
  40   &  4.99(+6)   &             &           &               &           &               &           &               &           &              \\
  60   &  3.90(+6)   &             &           &               &           &               &           &               &           &              \\
  80   &  3.37(+6)   &             &           &               &           &               &           &               &           &              \\
\enddata
 \end{deluxetable*}%

\startlongtable
\floattable
\begin{deluxetable}{lccccccccc}
\tablecolumns{10}
\tablecaption{Isotopic yields - Recommended set \label{tabyR}}
\tablehead{&&&&&&&&&\\ $\rm Isotope$ & $\rm 13~M_\odot$ & $\rm 15~M_\odot$ & $\rm 20~M_\odot$ & $\rm 25~M_\odot$ & $\rm 30~M_\odot$ & $\rm 40~M_\odot$ & $\rm 60~M_\odot$ & $\rm 80~M_\odot$ & $\rm 120~M_\odot$}
\startdata
\cutinhead{v=0 km/s - [Fe/H]=0}
H    & 6.1592E+00 & 6.8845E+00 & 8.5593E+00 & 1.0135E+01 & 1.1576E+01 & 1.4539E+01 & 2.0463E+01 & 2.5220E+01 & 3.6732E+01 \\
H2   & 2.9523E-06 & 3.6682E-06 & 5.8878E-06 & 7.3598E-06 & 8.8318E-06 & 1.1777E-05 & 1.7670E-05 & 2.3570E-05 & 3.5377E-05 \\
H3   & 5.4069E-29 & 3.1300E-26 & 2.0669E-16 & 4.3294E-14 & 2.2082E-13 & 2.6141E-12 & 1.4364E-11 & 4.3895E-11 & 1.3806E-10 \\
He3  & 3.6548E-04 & 3.9439E-04 & 4.6986E-04 & 5.4836E-04 & 6.2781E-04 & 7.8911E-04 & 1.1168E-03 & 1.3396E-03 & 1.7584E-03 \\
He4  & 4.6327E+00 & 5.1226E+00 & 7.1652E+00 & 8.3288E+00 & 7.3349E+00 & 1.0974E+01 & 2.1610E+01 & 3.0098E+01 & 5.1977E+01 \\
Li6  & 5.9770E-11 & 7.4266E-11 & 1.1972E-10 & 1.5018E-10 & 1.8090E-10 & 2.4319E-10 & 3.6984E-10 & 4.9849E-10 & 7.5772E-10 \\
Li7  & 8.4900E-10 & 1.0549E-09 & 1.7985E-09 & 2.2774E-09 & 2.7630E-09 & 3.8796E-09 & 5.6519E-09 & 7.2343E-09 & 1.1083E-08 \\
Be7  & 1.4797E-18 & 4.6861E-16 & 1.1940E-14 & 4.7919E-12 & 8.4431E-12 & 1.1629E-11 & 2.0945E-11 & 1.5975E-11 & 3.0266E-12 \\
\enddata
\end{deluxetable}     

\startlongtable                                                                                                                                                                                                                                                                                                                                                                                                                                                                                             
\floattable                                                                                                                                                                                                                                                                                                                                                                                                                                                                                                         
\begin{deluxetable}{lcccc}                                                                                                                                                                                                                                                                                                                                                                                                                                                                                                                                                                                                                                                                                                                                                                                                                                                                                                                                                                
\setlength{\tabcolsep}{0.045in}                                                                                                                                                                                                                                                                                                                                                                                                                                                                                     
\tablecolumns{5}                                                                                                                                                                                                                                                                                                                                                                                                                                                                                                   
\tablecaption{Isotopic yields in the wind - Recommended set \label{tabwR}}
\tablehead{&&&& \\ Isotope   & $\rm 13~M_\odot$ & $\rm 15~M_\odot$ & $\rm 20~M_\odot$ & $\rm 25~M_\odot$ }
\startdata
\cutinhead{v=0 km/s - [Fe/H]=0}
H     &  7.9437E-01 & 1.2188E+00 & 8.5000E+00 & 1.0135E+01 \\
H2    &  2.9523E-06 & 3.6682E-06 & 5.8878E-06 & 7.3598E-06 \\
H3    &  1.7054E-30 & 3.1250E-26 & 2.0669E-16 & 4.3294E-14 \\
He3   &  4.6963E-05 & 6.9968E-05 & 4.6986E-04 & 5.4836E-04 \\
He4   &  3.1961E-01 & 4.9952E-01 & 3.7908E+00 & 6.1010E+00 \\
Li6   &  5.9770E-11 & 7.4266E-11 & 1.1972E-10 & 1.5018E-10 \\
Li7   &  8.4900E-10 & 1.0549E-09 & 1.7985E-09 & 2.2774E-09 \\
Be7   &  8.1083E-25 & 1.1721E-23 & 7.2389E-17 & 4.7919E-12 \\
\enddata                                                                                                                                                                                                                                                                                                                                                                                       
\end{deluxetable}                                                                                                                                                                                                                                                                                                                                                                                                                                                                                                   


\begin{thebibliography}{(dummy)}
\bibitem[Abbott et al.(2017c)]{Abb1} Abbott, B.~P., Abbott, R., Abbott, T.~D., et al.\ 2017, \apjl, 851, L35 
\bibitem[Abbott et al.(2017b)]{Abb2} Abbott, B.~P., Abbott, R., Abbott, T.~D., et al.\ 2017, Physical Review Letters, 119, 141101 
\bibitem[Abbott et al.(2017a)]{Abb3} Abbott, B.~P., Abbott, R., Abbott, T.~D., et al.\ 2017, Physical Review Letters, 118, 221101 
\bibitem[Abbott et al.(2016c)]{Abb4} Abbott, B.~P., Abbott, R., Abbott, T.~D., et al.\ 2016, Physical Review Letters, 116, 241103 
\bibitem[Abbott et al.(2016b)]{Abb5} Abbott, B.~P., Abbott, R., Abbott, T.~D., et al.\ 2016, \prd, 93, 122003 
\bibitem[Abbott et al.(2016a)]{Abb6} Abbott, B.~P., Abbott, R., Abbott, T.~D., et al.\ 2016, Physical Review Letters, 116, 061102 
\bibitem[Angulo et al.(1999)]{nacre} Angulo, C., et al.\ 1999, Nuclear Physics A, 656, 3 
\bibitem[Antonini \& Rasio(2016)]{antonini2016} Antonini, F., \& Rasio, F.~A.\ 2016, \apj, 831, 187
\bibitem[Asplund et al.(2009)]{agss09} Asplund, M., Grevesse, N., Sauval, A. J. \& Scott, P. 2009, \araa, 47, 481 
\bibitem[Belczynski et al.(2016)]{belczynski2016} Belczynski, K., Repetto, S., Holz, D.~E., et al.\ 2016, \apj, 819, 108
\bibitem[Blackmon et al.(2003)]{bl03} Blackmon, J.~C., Bardayan, D.~W., Bradfield-Smith, W., et al.\ 2003, Nuclear Physics A, 718, 127
\bibitem[Bragan{\c c}a et al.(2012)]{braganca12} Bragan{\c c}a, G.~A., Daflon, S., Cunha, K., et al.\ 2012, \aj, 144, 130
\bibitem[Brott et al.(2011)]{brott11} Brott, I., de Mink, S. E., Cantiello, M., Langer, N., de Koter, A., Evans, C. J., Hunter, I., Trundle, C. \& Vink, J. S. 2011, \aap, 530, 115
\bibitem[Caughlan \& Fowler (1988)]{ca88} Caughlan, G.R., and Fowler, W.D. 1988, At. Data Nucl. Data Tables, 40, 283
\bibitem[Cayrel et al. (2004)]{cayreletal04} Cayrel, R., et al. 2004, \aap, 416, 1117
\bibitem[Chaboyer \& Zahn(1992)]{cz92} Chaboyer, B. \& Zahn, J.-P. 1992, \aap, 253, 173 
\bibitem[Chatzopoulos \& Wheeler(2012)]{chaz12} Chatzopoulos, E., \& Wheeler, J.~C.\ 2012, \apj, 748, 42
\bibitem[Chieffi et al.(1998)]{cls98} Chieffi, A., Limongi, M., \& Straniero, O.\ 1998, \apj, 502, 737
\bibitem[Chieffi \& Limongi(2004)]{cl04} Chieffi, A., \& Limongi, M.\ 2004, \apj, 608, 405
\bibitem[Chieffi \& Limongi (2013)]{cl13} Chieffi, A. \& Limongi, M.\ 2013, \apj, 764, 21 (CL13)
\bibitem[Clayton(1968)]{clayton68} Clayton, D.~D.\ 1968, New York: McGraw-Hill, 1968
\bibitem[Cowan et al.(1991)]{fkth} Cowan, J.~J., Thielemann, F.-K., \& Truran, J.~W.\ 1991, \physrep, 208, 267
\bibitem[Cyburt \& Davids(2008)]{cy08} Cyburt, R.~H., \& Davids, B.\ 2008, \prc, 78, 064614
\bibitem[Dayras et al.(1977)]{da77} Dayras, R., Switkowski, Z.~E., \& Woosley, S.~E.\ 1977, Nuclear Physics A, 279, 70
\bibitem[de Mink \& Mandel(2016)]{demink2016} de Mink, S.~E., \& Mandel, I.\ 2016, \mnras, 460, 3545
\bibitem[Descouvemont et al.(2004)]{desc04} Descouvemont, P., Adahchour, A., Angulo, C., Coc, A., \& Vangioni-Flam, E.\ 2004, Atomic Data and Nuclear Data Tables, 88, 203 
\bibitem[de Smet et al.(2007)]{de07} de Smet, L., Wagemans, C., Wagemans, J., Heyse, J., \& van Gils, J.\ 2007, \prc, 76, 045804
\bibitem[Dessart et al.(2011)]{des11} Dessart, L., Hillier, D.J., Livne, E., Yoon, S.C., Woosley, S., Waldman, R. \& Langer,N.\ 2011, \mnras, 414, 2985
\bibitem[Dessart et al.(2015)]{des15} Dessart, L., Hillier, D.~J., Woosley, S., et al.\ 2015, \mnras, 453, 2189
\bibitem[Dessart et al.(2016)]{des16} Dessart, L., Hillier, D.~J., Woosley, S., et al.\ 2016, \mnras, 458, 1618
\bibitem[de Jager et al.(1988)]{dejager88} de Jager, C., Nieuwenhuijzen, H., \& van der Hucht, K.~A.\ 1988, \aaps, 72, 259
\bibitem[Dillmann et al.(2006)]{kadonis02} Dillmann, I., Heil, F., K\"{a}ppeler, F., Plag, R., Rauscher, T., \& Thielemann, F.-K.\ 2006, AIP Conf. Proc. 819, 123 
\bibitem[Dillmann et al.(2009)]{kadonis03} Dillmann, I., Plag, R., K\"{a}ppeler, F., \& Rauscher, T.\ 2009, Submitted as proceeding of the workshop "EFNUDAT Fast Neutrons - scientific workshop on neutron measurements, theory \& applications" held on April 28-30 2009 at Geel, Belgium 
\bibitem[Ensman \& Woosley(1988)]{ew88} Ensman, L.~M., \& Woosley, S.~E.\ 1988, \apj, 333, 754
\bibitem[Ertl et al.(2016)]{ertl16} Ertl, T., Janka, H.-T., Woosley, S.~E., Sukhbold, T., \& Ugliano, M.\ 2016, \apj, 818, 124
\bibitem[Ekstr\"{o}m et al.(2012)]{ekstroemetal12} Ekstr\"{o}m, S., Georgy, C., Eggenberger, P., et al. 2012, \aap, 537, 146
\bibitem[Frischknecht et al.(2012)]{fresco12} Frischknecht, U., Hirschi, R., \& Thielemann, F.-K.\ 2012, \aap, 538, L2 
\bibitem[Frischknecht et al.(2016)]{fresco16} Frischknecht, U., Hirschi, R., Pignatari, M., et al.\ 2016, \mnras, 456, 1803
\bibitem[Fuller, Fowler \& Newman(1982)]{FFN82} Fuller, G.M., Fowler, W.A., and Newman, M. 1982, \apjs, 48, 279
\bibitem[Gallino et al.(1998)]{Ga98} Gallino, R., Arlandini, C., Busso, M., et al.\ 1998, \apj, 497, 388
\bibitem[Hachinger et al.(2012)]{hachietal12} Hachinger, S., Mazzali, P.~A., Taubenberger, S., et al.\ 2012, \mnras, 422, 70
\bibitem[Heger et al.(2000)]{hlw00} Heger, A., Langer, N., \& Woosley, S. E. 2000, \apj, 528, 368
\bibitem[Heger \& Woosley(2002)]{HW02} Heger, A., \& Woosley, S.~E.\ 2002, \apj, 567, 532 
\bibitem[Heger et al.(2005)]{hws05} Heger, A., Woosley, S.~E., \& Spruit, H.~C.\ 2005, \apj, 626, 350 
\bibitem[Heger \& Woosley(2010)]{he10} Heger, A., \& Woosley, S.~E.\ 2010, \apj, 724, 341 
\bibitem[Hirschi et al.(2004)]{hirschi04} Hirschi, R., Meynet, G., \& Maeder, A.\ 2004, \aap, 425, 649
\bibitem[Huebner et al.(1977)]{huebner77} Huebner, W.~F., Merts, A.~L., Magee, N.~H., \& Argo, M.~F.\ 1977, Los Alamos Sci. La. Rep. LA-6760-M 
\bibitem[Hunter et al.(2008)]{untetal08} Hunter, I., Lennon, D.~J., Dufton, P.~L., et al.\ 2008, \aap, 479, 541
\bibitem[Hunter et al.(2009)]{untetal09} Hunter, I., Brott, I., Langer, N., et al.\ 2009, \aap, 496, 841
\bibitem[Iglesias et al.(1992)]{iglesias92} Iglesias, C.~A., Rogers, F.~J., \& Wilson, B.~G.\ 1992, \apj, 397, 717 
\bibitem[Iliadis et al.(2010)]{mc10} Iliadis, C., Longland, R., Champagne, A.~E., Coc, A., \& Fitzgerald, R.\ 2010, Nuclear Physics A, 841, 31
\bibitem[Iliadis et al.(2011)]{il09} Iliadis, C., Champagne, A., Chieffi, A., \& Limongi, M.\ 2011, \apjs, 193, 16
\bibitem[Imbriani et al.(2005)]{imbriani05} Imbriani, G., Costantini, H., Formicola, A., et al.\ 2005, European Physical Journal A, 25, 455
\bibitem[Itoh et al.(1983)]{itoh83} Itoh, N., Mitake, S., Iyetomi, H., \& Ichimaru, S.\ 1983, \apj, 273, 774
\bibitem[Koehler et al.(1997)]{ko97} Koehler, P.~E., Kavanagh, R.~W., Vogelaar, R.~B., Gledenov, Y.~M., \& Popov, Y.~P.\ 1997, \prc, 56, 1138
\bibitem[Kunz et al. (2002)]{kunzetal02} Kunz, R., Fey, M., Jaeger, M., Mayer, A., Hammer, J.W., Staudt, G., Harissopulos, S., and Paradellis, T. 2002, \apj, 567, 643
\bibitem[Kurucz(1991)]{kurucz91} Kurucz, R.~L.\ 1991, NATO Advanced Science Institutes (ASI) Series C, 341, 441
\bibitem[Georgy et al.(2017)]{georgy17} Georgy, C., Meynet, G., Ekstr{\"o}m, S., et al.\ 2017, \aap, 599, L5
\bibitem[Goriely et al.(1990)]{gor90} Goriely, S., Jorissen, A., \& Arnould, M.\ 1990, Nuclear Astrophysics, 5th Workshop, 60 
\bibitem[Kobayashi et al.(2006)]{koba06} Kobayashi, C., Umeda, H., Nomoto, K., Tominaga, N., \& Ohkubo, T.\ 2006, \apj, 653, 1145 
\bibitem[Langanke \& Mart\'inez-Pinedo(2000)]{lp00} Langanke, K.H., and Mart\'inez-Pinedo, G. 2000, Nucl. Phys. A, 673, 481
\bibitem[Levesque et al.(2005)]{levesque05} Levesque, E.M., Massey, P., Olsen, K.A.G., Pletz, B., Josselin, E., Maeder, A. \& Meynet, G.\ 2005, \apj, 628, 973 
\bibitem[Limongi \& Chieffi (2006)]{lc06} Limongi, M. \& Chieffi, A.\ 2006, \apj, 647, 483 
\bibitem[Limongi et al.(2000)]{lsc00} Limongi, M., Straniero, O., \& Chieffi, A.\ 2000, \apjs, 129, 625 
\bibitem[Lugaro \& Chieffi(2011)]{luchi11} Lugaro, M., \& Chieffi, A.\ 2011, Lecture Notes in Physics, Berlin Springer Verlag, 812, 83 
\bibitem[Lyman et al.(2016)]{Lyman16} Lyman, J.~D., Bersier, D., James, P.~A., et al.\ 2016, \mnras, 457, 328
\bibitem[Limongi \& Chieffi(2012)]{lc12} Limongi, M., \& Chieffi, A.\ 2012, \apjs, 199, 38
\bibitem[Maeder \& Meynet(2000)]{mm00araa} Maeder, A., \& Meynet, G.\ 2000, \araa, 38, 143
\bibitem[Maeder \& Meynet(2001)]{MM01} Maeder, A., \& Meynet, G.\ 2001, \aap, 373, 555 
\bibitem[Maeder \& Zahn (1998)]{mz98} Maeder and Zahn 1998, \aap 
\bibitem[Maeder \& Meynet(2012)]{mm12} Maeder, A., \& Meynet, G.\ 2012, Reviews of Modern Physics, 84, 25
\bibitem[Maeder \& Meynet(2001)]{mm01} Maeder, A. \& Meynet, G. 2001, \aap, 373, 555 
\bibitem[Mandel \& de Mink(2016)]{mandel2016} Mandel, I., \& de Mink, S.~E.\ 2016, \mnras, 458, 2634
\bibitem[Meynet \& Maeder (1997)]{mm97} Meynet \& Maeder 1997, \aap, 321, 465
\bibitem[Malaney \& Fowler(1989)]{mafo} Malaney, R.~A., \& Fowler, W.~A.\ 1989, \apjl, 345, L5
\bibitem[Meynet \& Maeder (2000)]{mm00} Meynet \& Maeder 2000, \aap 
\bibitem[Meynet \& Maeder(2002)]{MM02} Meynet, G., \& Maeder, A.\ 2002, \aap, 381, L25
\bibitem[Meynet \& Maeder (2003)]{mm03} Meynet, G. \& Maeder, A. 2003, \aap, 404, 975
\bibitem[Meynet \& Maeder(2005)]{mm05} Meynet, G. \& Maeder, A.\ 2005, \aap, 429, 581
\bibitem[Meynet et al.(2006)]{MEM06} Meynet, G., Ekstr{\"o}m, S., \& Maeder, A.\ 2006, \aap, 447, 623 
\bibitem[Nassar et al.(2006)]{nass} Nassar, H., Paul, M., Ahmad, I., et al.\ 2006, Physical Review Letters, 96, 041102 
\bibitem[Nugis \& Lamers(2000)]{nl00} Nugis, T., \& Lamers, H.~J.~G.~L.~M.\ 2000, \aap, 360, 227
\bibitem[O'Connor \& Ott(2011)]{oconnorott2011} O'Connor, E., \& Ott, C.~D.\ 2011, \apj, 730, 70
\bibitem[Oda et al.(1994)]{odaetal94} Oda, T., Hino, M., Muto, K., Takahara, M., and Sato, K. 1994, A.D.N.D.T., 56, 231
\bibitem[Oginni et al.(2011)]{og10} Oginni, B.~M., Iliadis, C., \& Champagne, A.~E.\ 2011, \prc, 83, 025802 
\bibitem[Palacios et al.(2003)]{Pal03} Palacios, A., Talon, S., Charbonnel, C., \& Forestini, M.\ 2003, \aap, 399, 603
\bibitem[Pignatari et al.(2008)]{pigna08} Pignatari, M., Gallino, R., Meynet, G., et al.\ 2008, \apjl, 687, L95
\bibitem[Prantzos et al.(1990)]{prantzos90} Prantzos, N., Hashimoto, M., \& Nomoto, K.\ 1990, \aap, 234, 211
\bibitem[Prantzos et al.(2018)]{prantzos2018} Prantzos, N., Abia, C., Limongi, M., Chieffi, A., \& Cristallo, S.\ 2018, \mnras, 476, 3432
\bibitem[Ram{\'{\i}}rez-Agudelo et al.(2017)]{ramirez17} Ram{\'{\i}}rez-Agudelo, O.~H., Sana, H., de Koter, A., et al.\ 2017, \aap, 600, A81
\bibitem[Raiteri et al.(1992)]{raiteri92} Raiteri, C.~M., Gallino, R., \& Busso, M.\ 1992, \apj, 387, 263
\bibitem[Raiteri et al.(1993)]{raiteri93} Raiteri, C.~M., Gallino, R., Busso, M., Neuberger, D., \& Kaeppeler, F.\ 1993, \apj, 419, 207 
\bibitem[Rauscher et al.(1994)]{bb92} Rauscher, T., Applegate, J.~H., Cowan, J.~J., Thielemann, F.-K., \& Wiescher, M.\ 1994, \apj, 429, 499
\bibitem[Rauscher \& Thielemann(2000)]{RT2000} Rauscher, T., and Thielemann, F.K. 2000, A.D.N.D.T., 75, 1
\bibitem[Rehm et al.(1998)]{re98} Rehm, K.~E., Borasi, F., Jiang, C.~L., et al.\ 1998, Physical Review Letters, 80, 676
\bibitem[Rogers et al.(1996)]{rogers96} Rogers, F.~J., Swenson, F.~J., \& Iglesias, C.~A.\ 1996, \apj, 456, 902 
\bibitem[Rogers(2001)]{rogers01} Rogers, F.~J.\ 2001, Contributions to Plasma Physics, 41, 179
\bibitem[Sallaska et al.(2011)]{mc11} Sallaska, A.~L., Wrede, C., Garc{\'{\i}}a, A., et al.\ 2011, \prc, 83, 034611
\bibitem[Sallaska et al (2013)]{starlib} Sallaska, A. L., Iliadis, C., Champagne, A. E., Goriely, S., Starrfield, S., \& Timmes, F. X.\ 2013, \apjs, 207, 18 
\bibitem[Smartt(2009)]{smartt09} Smartt, S.~J.\ 2009, \araa, 47, 63 
\bibitem[Smartt(2015)]{smartt15} Smartt, S.~J.\ 2015, \pasa, 32, e016
\bibitem[Smith et al.(2011)]{smith11} Smith, N., Li, W., Filippenko, A.~V., \& Chornock, R.\ 2011, \mnras, 412, 1522
\bibitem[Spite et al. (2005)]{spiteetal05} Spite, M., et al. 2005, \aap, 430, 655
\bibitem[Straniero, Chieffi \& Limongi (1997)]{scl97} Straniero. O., Chieffi, A. \& Limongi, M. 1997, \apj, 490, 425
\bibitem[Straniero(1988)]{straniero88} Straniero, O.\ 1988, \aaps, 76, 157
\bibitem[Sukhbold et al.(2016)]{sukh16} Sukhbold, T., Ertl, T., Woosley, S.~E., Brown, J.~M., \& Janka, H.-T.\ 2016, \apj, 821, 38 
\bibitem[Takahashi \& Yokoi(1987)]{ty87} Takahashi, K., and Yokoi, K. 1987, A.D.N.D.T., 36, 375
\bibitem[Talon et al (1997)]{tal97} Talon et al 1997, \aap 
\bibitem[Talon and Zahn (1997)]{tz97} Talon and Zahn 1997, \aap 
\bibitem[Tornambe' \& Chieffi (1986)]{tc86} Tornambe', A., \& Chieffi, A. 1986, \mnras, 220, 547 
\bibitem[Umeda \& Nomoto(2002)]{UN02} Umeda, H., \& Nomoto, K.\ 2002, \apj, 565, 385
\bibitem[van Loon et al.(2005)]{vanloonetal05} van Loon, J.~T., Cioni, M.-R.~L., Zijlstra, A.~A., \& Loup, C.\ 2005, \aap, 438, 273
\bibitem[Wheeler \& Levreault(1985)]{wl85} Wheeler, J.~C., \& Levreault, R.\ 1985, \apjl, 294, L17 
\bibitem[van Wormer et al.(1994)]{laur} van Wormer, L., G{\"o}rres, J., Iliadis, C., Wiescher, M., \& Thielemann, F.-K.\ 1994, \apj, 432, 326
\bibitem[Vink et al.(2000)]{val00} Vink, J.~S., de Koter, A., \& Lamers, H.~J.~G.~L.~M.\ 2000, \aap, 362, 295 
\bibitem[Vink et al.(2001)]{val01} Vink, J.~S., de Koter, A., \& Lamers, H.~J.~G.~L.~M.\ 2001, \aap, 369, 574
\bibitem[Wagoner et al.(1967)]{wfho} Wagoner, R.~V., Fowler, W.~A., \& Hoyle, F.\ 1967, \apj, 148, 3
\bibitem[Wagoner(1969)]{wago} Wagoner, R.~V.\ 1969, \apjs, 18, 247
\bibitem[Walmswell \& Eldridge(2012)]{we2012} Walmswell, J.~J., \& Eldridge, J.~J.\ 2012, \mnras, 419, 2054
\bibitem[Yoon \& Cantiello(2010)]{yc2010} Yoon, S.-C., \& Cantiello, M.\ 2010, \apjl, 717, L62
\bibitem[Yoon et al.(2012)]{yoon12} Yoon, S.-C., Dierks, A., \& Langer, N.\ 2012, \aap, 542, A113
\bibitem[Zahn (1992)]{z92} Zahn, J.~P., 1992 \aap, 265, 115





\end{thebibliography}
\end{document}